 \documentclass[pra,amsmath,amssymb,onecolumn, showpacs, superscriptaddress,10pt]{revtex4-1}

\usepackage{amsmath}
\usepackage{hyperref}
\usepackage{graphicx}
\usepackage{amsfonts}
\usepackage{amsthm}
\usepackage{cases}
\usepackage{bm}

\usepackage[normalem]{ulem}

\usepackage{color}
\definecolor{Blue}{rgb}{0.00, 0.00, 1.00}
\definecolor{Red}{rgb}{1.00, 0.00, 0.00}
\definecolor{Green}{rgb}{0.00, 1.00, 0.00}
\definecolor{Cyan}{rgb}{0.00, 0.70, 0.70}

\hypersetup{
    colorlinks=true,       % false: boxed links; true: colored links
    linkcolor=red,          % color of internal links (change box color with linkbordercolor)
    citecolor=blue,        % color of links to bibliography
    filecolor=magenta,      % color of file links
    urlcolor=cyan           % color of external links
}

\newcommand{\nn}{\nonumber}
\newcommand{\be}{\begin{equation}}
\newcommand{\ee}{\end{equation}}
\newcommand{\bea}{\begin{eqnarray}}
\newcommand{\eea}{\end{eqnarray}}

\newcommand{\beq}{\begin{equation}}
\newcommand{\eeq}{\end{equation}}
\newcommand{\beqn}{\begin{eqnarray}}
\newcommand{\eeqn}{\end{eqnarray}}
\begin{document}

\title{Kernels for noninteracting fermions via a Green's function approach with applications to step potentials}

\author{David S. \surname{Dean}}
\affiliation{Univ. Bordeaux and CNRS, Laboratoire Ondes et Mati\`ere  d'Aquitaine
(LOMA), UMR 5798, F-33400 Talence, France}
\author{Pierre Le Doussal}
\affiliation{CNRS-Laboratoire de Physique Th\'eorique de l'Ecole Normale Sup\'erieure, 24 rue Lhomond, 75231 Paris Cedex, France}
\author{Satya N. \surname{Majumdar}}
\affiliation{Universit{\'e} Paris-Saclay, CNRS, LPTMS, 91405, Orsay, France}
\author{Gr\'egory \surname{Schehr}}
\affiliation{Universit{\'e} Paris-Saclay, CNRS, LPTMS, 91405, Orsay, France}
\author{Naftali R. \surname{Smith}}
\affiliation{Universit{\'e} Paris-Saclay, CNRS, LPTMS, 91405, Orsay, France}

\date{\today}

\begin{abstract}

The quantum correlations of $N$ noninteracting spinless fermions in their ground state can be 
expressed in terms of  a two-point function called the kernel. Here we develop a
general and compact method for computing the kernel 
in a general trapping potential in terms of the Green's function for the corresponding single particle Schr\"odinger equation. For smooth potentials the method allows a simple alternative derivation of the local density approximation for the density and of the sine kernel in the bulk part of the trap in the large $N$ limit. It also recovers the density and the kernel of the so-called {\em Airy gas} at the edge. 
This method allows to analyse the quantum correlations in the ground state when the potential has a singular part
with a fast variation in space. For the square step barrier of height $V_0$, we derive explicit expressions for the density and for the kernel. For large Fermi energy $\mu>V_0$ it describes the interpolation between two regions of different densities in a Fermi gas, each described by a different sine kernel. Of particular interest is the {\em critical point} of the square well potential when $\mu=V_0$. 
In this critical case, while there
is a macroscopic number of fermions in the lower part of the step potential, there
is only a finite $O(1)$ number of fermions on the shoulder, and moreover this number is independent of $\mu$. In particular, the density exhibits an algebraic decay $\sim 1/x^2$, where $x$ is the distance from the jump. Furthermore, we show
that the critical behaviour around $\mu = V_0$ exhibits universality with respect with the shape of the barrier. 
This is established (i) by an exact solution for a
smooth barrier (the Woods-Saxon potential) and (ii) by establishing a general relation between the large distance behavior of the kernel
and the scattering amplitudes of the single-particle wave-function. 
 \end{abstract}

\maketitle

\tableofcontents

\section{Introduction}
Recent developments in trapping techniques for cold atomic  systems  \cite{blo08}, along with the possibility of visualizing individual atoms using quantum Fermi microscopes \cite{che15,hal15,par15} have lead to a resurgence of theoretical interest in the equilibrium and dynamical properties of confined systems of fermions. 
%In particular the case of spinless noninteracting fermions can also be realised experimentally by suitably tuning Feshbach resonances. 
%Note that 
In particular the rather idealised case of spinless noninteracting fermions is experimentally relevant as magnetic traps are based on  polarising the spin degrees of freedom of cold atoms and furthermore in the spin polarised state interactions are weak due to the suppression of s-wave scattering. The interactions can further be suppressed experimentally by suitably tuning Feshbach resonances.
This noninteracting case, at zero temperature, still presents interesting theoretical challenges since nontrivial statistical behavior arises because of  
{the Pauli} exclusion principle. 

The statistics of spinless non interacting fermionic systems can be encoded in a two point kernel from which the local density, as well as all the higher order correlation functions, can be expressed, in essence via  Wick's theorem for fermions. For bulk systems, where a large number of fermions are confined by an external potential, the regions where the local fermionic density is large can be studied using the local density approximation or LDA \cite{gio08,cas06} which is based on the approximation that the trapping potential can be treated locally as constant in space. The LDA allows the calculation of  the  bulk density.  
Together with more controlled approaches, the LDA also predicts 
that, in the bulk and at scales of the order of the typical inter-particle distance, the kernel takes a universal form, independent of the details of the potential, given by the sine-kernel \cite{gio08,cas06,eis13,dea15,dea15b,dea16}. The LDA can be used to predict its own downfall in regions where the density of fermions becomes small. This occurs, by definition, at the edge of the trapped atomic cloud. Here, the form of the density and kernel is modified and one finds that the {\em edge} physics is described by fermions in a linear potential, as the trapping potential can in general be expanded as a locally linear potential near the edge \cite{koh98,eis13,dea15,dea15b,dea16}. The associated kernel near a locally linear edge is called the Airy kernel and the  fermions in this region referred to as the {\em Airy gas}    \cite{koh98,eis13,dea15,dea15b,dea16}. It can be shown to be universal for a broad class of smooth potentials. However, other edge regimes and edge universality classes exist, notably when the trap has an infinite hard wall or a continuous but divergent wall potential \cite{cal11,lac17}, but also when the Fermi energy coincides with the maximum of a double well potential \cite{smi20}. Interestingly, while the kernels appearing in the aforementioned problems arise from quantum mechanical problems,  most of them also arise in the context of random matrix theory, where they describe the eigenvalue statistics of certain random matrix ensembles \cite{eis13,dea15,dea15b,dea16,dea19}.

\begin{figure}[t!]
\begin{center}
   \includegraphics[width=0.8 \linewidth]{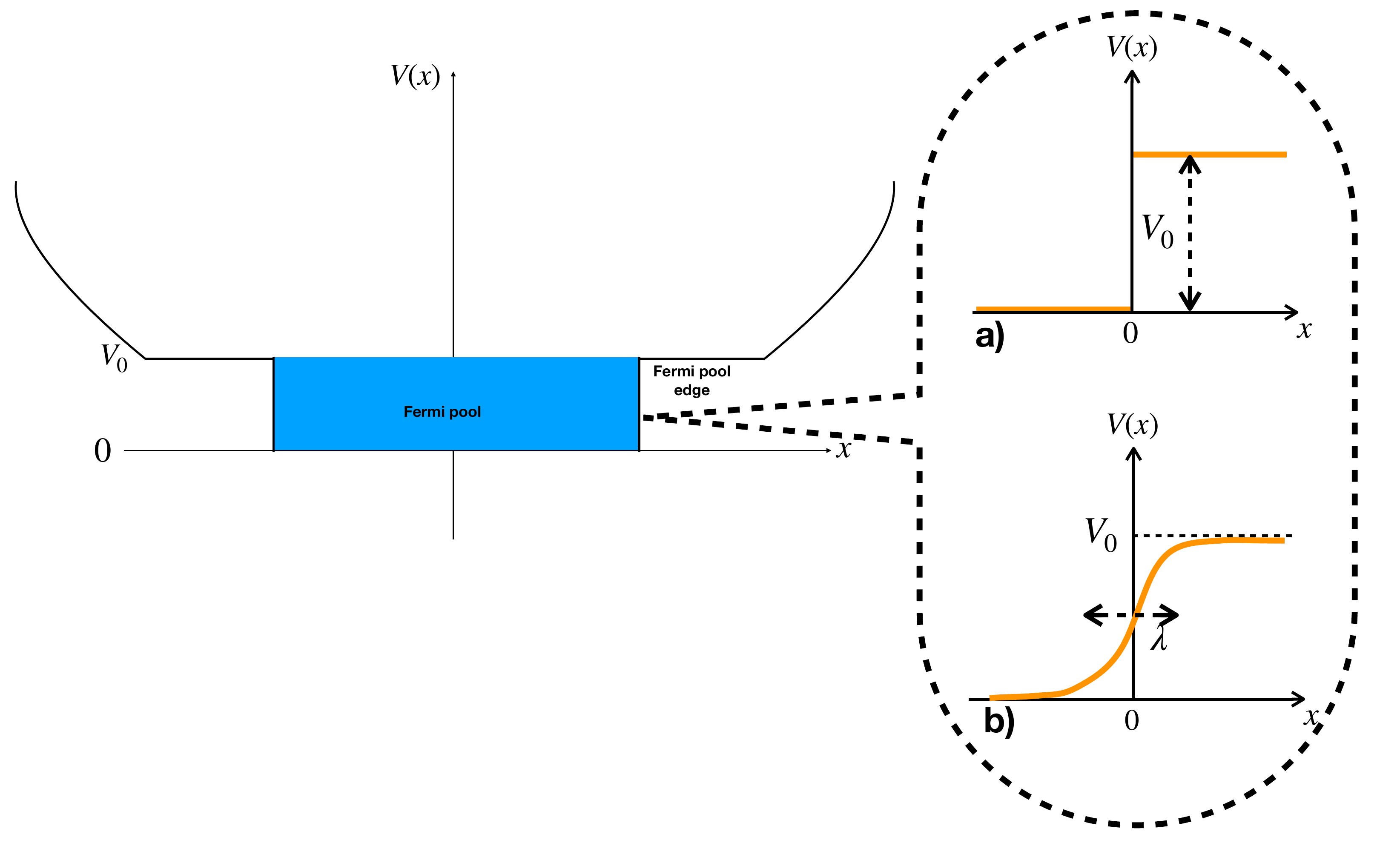}
 \caption{Square well potential of height $V_0$ within an overall confining potential $V(x)$. The critical regime where the Fermi energy $\mu$, the height of blue region, coincides with $V_0$ is shown.
The physics near the edge can be described by zooming at the edge, as shown on the right side of the figure. 
Two examples of barrier potentials are shown: {\bf a)} The square step barrier studied in Section \ref{sec:step}, {\bf b)} a smooth barrier, varying over a scale $\lambda$, such as studied in Section \ref{smooth}.}\label{schema}
\end{center}
\end{figure}

The LDA is based on the assumption that the trapping potential is locally constant, i.e. that it varies very slowly on the scale of the typical inter-particle distance. If the potential has fast variations on this scale, e.g. if it is discontinuous, the LDA will fail, even if the fermion density is 
large. Similarly, we can expect the edge universality classes to be modified. The main goal of this paper,
and of a companion paper \cite{inprep}, is to analyse the statistical properties of non interacting fermions when the trapping potential exhibits a local singularity on top of an overall smooth confining shape. An important question is what replaces the sine-kernel in the bulk near the singularity, and how far the effect of this singularity can be felt. A related question is what effect does it have on the counting statistics, such as the fluctuations of the number of fermions in a given region. 

The main goal of this paper is to address these questions. For this, we consider $N$ noninteracting fermions in a globally confining
trapping potential and focus on the large $N$ limit. On top of this, we assume that the potential has a singularity at some point in space (see Fig. \ref{schema}). We zoom in near the singularity and study a class of fast-varying potentials (see Fig. \ref{schema}), which include the square step barrier as well as other continuous barrier potentials, such as the Woods-Saxon potential \cite{woo54}, well known in nuclear physics. Our goal is to describe how the quantum correlations in the ground state are modified in the vicinity of the singularity -- as opposed to the smooth potential case. In order to carry out the analysis of such fast-varying potentials we develop a method based on the single particle Green's function associated to the Schr\"odinger equation in the presence of a general trapping potential. As a preliminary benchmark, we first show how this method can be used to derive well known properties of smooth potentials such as the LDA  and the Airy gas physics. This already allows us to discuss the limitation of the LDA which, as we find, fails when the potential varies too fast in space.

We then apply this Green's function method to obtain the exact form of the kernel, and the statistics of fermions, near a square step barrier of height $V_0$, as shown in Fig. \ref{schema}. For Fermi energies $\mu>V_0$ it describes the interpolation between two regions of different densities in a Fermi gas, each described by a differently-scaled sine kernel. We examine in particular the case where the Fermi energy coincides with the top of the step potential, $\mu=V_0$. This mimics a macroscopic system of fermions confined in a finite square well potential within an overall trapping potential - an everyday analogy being that of a swimming pool of fermions which is full to the edge - see Fig. \ref{schema}. Said otherwise, the swimming pool is on the point of overflowing a bit like in an {\em infinity pool}. The statistics of the number of particles in the region outside the square well potential (on the pool edges) have rather interesting properties when $V_0=\mu$. In particular, we analyse the mean, variance and third cumulant of the number of particles $N_{\rm out}$ to the right outside the pool is independent of $\mu$ and can be computed exactly.

\begin{figure}[t]
\includegraphics[width = 0.9\linewidth]{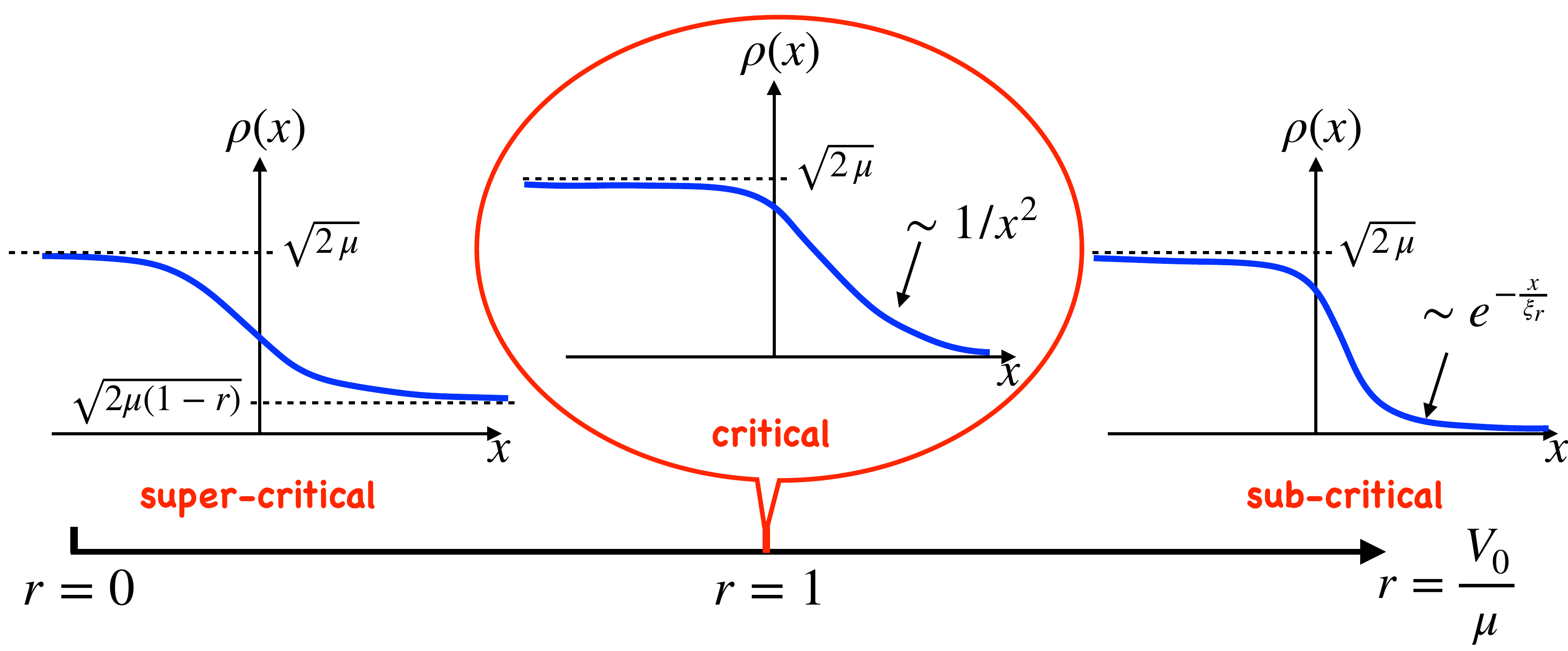}
\caption{Density profile $\rho(x)$ vs $x$ for different values of the dimensionless parameter $r = \frac{V_0}{\mu}$. For $r>1$, in the sub-critical phase, $\rho(x) \sim e^{-x/\xi_r}$ vanishes exponentially as $x \to + \infty$ on a scale $\xi_r \sim 1/\sqrt{r-1}$. For $r < 1$, the super-critical phase, the density approaches a constant $\rho(x) \sim \sqrt{2\mu(1-r)}$ as $x \to + \infty$. Exactly at the critical point $r=1$, $\rho(x)$ vanishes algebraically as $1/x^2$, as $x \to +\infty$.}\label{ph_diag}
\end{figure}

In fact, we find that a critical behavior emerges as a function of the dimensionless control parameter $r=V_0/\mu > 0$ which is summarised in Fig. \ref{ph_diag}. We find a sub-critical behavior for $r>1$ where the local density profile decays exponentially $\rho(x)\sim e^{-x/\xi_r}$ for $x \to +\infty$. The decay length diverges as $\xi_r \sim 1/\sqrt{r-1} $ as $r$ approaches the critical point $r=1$ from above. Exactly at the critical point we find an algebraic decay $\rho(x) \sim 1/x^2$ for $x \to +\infty$. On the super-critical side $r < 1$ the density approaches a non-zero constant $\sqrt{2\mu(1-r)}$ for $x \to + \infty$. 

To explore the universality of this critical behaviour with respect to the shape of the barrier, we explore more general smooth barriers of the type $V(x) = V_0\,v(x/\lambda)$ where $v(z)$ smoothly interpolates between $0$ and $1$ as $z$ increases from $-\infty$ to $+\infty$. Here $\lambda$ is the characteristic length scale describing how fast the barrier varies in space (see Fig. \ref{schema} b)).  We find that indeed this critical behavior near $r=1$ is universal and sets in whenever $\xi_r \gg \lambda$. For example, the density profile in the vicinity of the transition is described by a universal scaling function, up to a non-universal amplitude that depends on $\lambda$. This result is obtained by an exact solution in the case of the Woods-Saxon potential as well as through an analysis valid for  
more general barrier potentials. This analysis unveils a general relation between the scattering amplitudes of the single particle wave functions and the asymptotic behaviour of the kernel at large distances {from the barrier.} %{\red I added these words to make it clear that we don't mean here $K(x,y)$ for large $|x-y|$.}

%\begin{figure}[t!]
%\begin{center}
%  \includegraphics[width=0.8 \linewidth]{Fig_potentials.pdf}
% \caption{Two examples of barrier potentials: {\bf a)} The square step barrier studied in Section \ref{sec:step}, {\bf b)} a smooth barrier, fast varying over a scale $\lambda$, such as studied in Section \ref{smooth}.}\label{fig_potential}
%\end{center}
%\end{figure}

The rest of the paper is organised as follows. In Section \ref{sec:green} we introduce the Green's function method to compute the kernel for arbitrary potentials. In Section \ref{sec:smooth_pot} we study the case of a smooth trapping potential (without singularity) and show how to recover the standard scaling forms for the kernel both in the bulk (LDA and sine-kernel) and {at} {near} the edge of the Fermi gas (Airy-kernel). In Section \ref{sec:step}, we obtain the exact expression of the kernel for the square step barrier and discuss the critical behaviour. We also obtain the first three cumulants of the total number of particles to the right of the barrier. In Section \ref{smooth} A., we obtain asymptotic formula for the kernel for {a} general barrier potential in terms of the corresponding scattering amplitudes. In Section \ref{smooth} B., we study the Woods-Saxon potential, for which exact formula can be derived, before we conclude in Section \ref{sec:conclusion}. Some technical aspects are presented in Appendices.

\section{Green's function formalism to compute the kernel}
\label{sec:green} 
\subsection{The kernel - basic definitions}

We consider $N$ non interacting spinless fermions confined by a potential $V(x)$.
The  single particle Hamiltonian is $H = \frac{p^2}{2 m} + V(x)$, where $m$ is the particle mass. We denote by $\psi_k(x)$ the eigenstates of $H$ and $\epsilon_k$ the associated energies.
The zero temperature kernel can be written in terms of the Fermi energy $\mu$ as
\begin{equation}
K_\mu(x,y) = \sum_k\theta(\mu - \epsilon_k) \psi_k^*(x)\psi_k(y).\label{k1}
\end{equation}
Here $\theta(x)$ denotes the Heaviside function, and $\mu$ is considered as a continuous parameter, the total number of fermions being related to $\mu$ as $N=\sum_k \theta(\mu-\epsilon_k)$ \cite{com1}. 
By construction, the kernel \eqref{k1} is real and symmetric. %\sout{in one dimensional systems}
, {\em i.e.} $K_\mu(x,y) = K_\mu(y,x)$ \cite{comreal}.
We also consider below cases of non confining potentials $V(x)$, i.e. fermions on the whole line with a continuous spectrum for $H$, as limiting cases of
\eqref{k1} for large system sizes. In this case $N$ is infinite and $\mu$ is the control parameter. 

The kernel encodes all of the statistical properties of an $N$ body system as all $n$-point correlation functions can be constructed from it \cite{dea16,dea19}. These correlations can be computed using Wick's theorem for fermionic fields or equivalently by noting that the particle positions are described by a determinantal point process \cite{bor11}. In particular, for the purposes of the current work, we note that the number density of fermions is given by
\begin{equation}
\rho(x)=K_{\mu}(x,x),
\end{equation}
which means that the number of fermions in a region ${\cal I}$, denoted by $N_{\cal I}$, has the  average value
\begin{equation}
\langle N_{\cal I}\rangle =\int_{\cal I} dx\  K_\mu(x,x),\label{ns}
\end{equation}
while the variance of $N_{\cal I}$ is given by \cite{dea16}
\begin{equation}
{\rm{Var}}( N_{\cal I}) = \langle N_{\cal I}\rangle-\int_{\cal I}\int_{\cal I} dxdy\ [K_\mu(x,y)]^2.\label{vs}
\end{equation}

In what follows we explain, formally, how the single particle Green's function can be used
to compute the kernel. Some of these results were previously derived by other methods (such as by direct evaluation of the kernel by computing and summing the wave functions or by using the 
%\sout{short time behavior of the propagator} {\blue 
Euclidean propagator associated with $H$ \cite{dea16}), but here we use a new Green's function method that turns out to be technically advantageous compared to other methods, in particular in computing the limiting kernels for discontinuous potentials, as demonstrated later in the paper.
\subsection{Kernels via Green's function}
%The zero temperature kernel can be written in terms of the Fermi energy $\mu$ as
%\begin{equation}
%K_\mu(x,y) = \sum_k^{\epsilon_k=\mu} \psi_k^*(x)\psi_k(y)= \sum_k\theta(\mu - \epsilon_k) \psi_k^*(x)\psi_k(y).\label{k1}
%\end{equation}
%An important thing to notice is that if $\psi^*\neq \exp(i\theta)\psi$ where $\theta$ is a constant phase shift, 
%$\psi$ and $\psi^*$ do not correspond to the same  quantum  state, i.e. mathematically they are linearly independent solutions. As a consequence $K_\mu(x,y)$ is not obviously symmetric in $x$ and $y$. It is however obvious that the kernel is hermitian
%\begin{equation}
%K_\mu(x,y) = K_\mu^*(y,x).\label{adk}
%\end{equation}

Differentiating Eq. ({\ref{k1}) with respect to $\mu$ gives
\begin{equation}
{\frac{\partial}{\partial \mu}}K_\mu(x,y) =  \sum_k\delta(\mu - \epsilon_k) \psi_k^*(x)\psi_k(y),
\end{equation}
whose diagonal part is the local density of states of $H$ at energy $\mu$.
% i.e.
%$\rho(\epsilon=\mu,x)=\partial_\mu K_\mu(x,x)$.
%{\red Naftali: I don't understand what is meant here by $\rho(\epsilon=\mu,x)$. It seems to me that it should be $\partial_\mu \rho(x)$.}
Now we use the well known formula %and much used result
\begin{equation}
{ \frac{1}{z-i 0^+} := \lim_{\varepsilon\to 0^+} } \frac{1}{z-i\varepsilon}  = \pi i \delta(z) + P \frac{1}{z},
\end{equation}
%which is to be 
interpreted in terms of distributions, where $P$ indicates that one should use 
the Cauchy principle part in any integrals. { We will use everywhere below the notation $z - i 0^+$ to denote
the limit $\varepsilon\to 0^+$ at the end of the calculation.} We thus find
{
\begin{equation}
\delta(\mu-\epsilon) =\frac{1}{\pi}  \rm {Im} \frac{1}{\mu-i 0^+ -\epsilon} . 
\end{equation}
}
Now, as the terms involving  the wave function are real, we can write
\begin{equation}
\frac{\partial}{\partial \mu} K_\mu(x,y) =\frac{1}{\pi} %{\rm Im}\lim_{\varepsilon\to 0} 
 {\rm Im}  
\sum_k  \frac{1}{\mu -i 0^+ -\epsilon_k} \psi_k^*(x)\psi_k(y).
\end{equation}
This gives
\begin{equation}
\frac{\partial}{\partial \mu} K_\mu(x,y) = \frac{1}{\pi} { \rm Im} \, G_\mu(x,y)\label{dmu}
\end{equation}
where $G_\mu$ is the resolvent of the operator $H$, evaluated at $\mu - i 0^+$ just below the real axis,
in operator notation
\begin{equation}
G_\mu = (\mu -i 0^+ -H)^{-1} .
\end{equation}
Hence it is in general a complex quantity. The imaginary part
of its diagonal component gives the local density of states of $H$ at energy $\mu$.
Equivalently, $G_\mu(x,y)$ is the solution of 
\begin{equation}
(\mu -i 0^+ -H)G_\mu (x,y) = \delta(x-y),
\end{equation}
with proper decay at infinity. Hence $G_\mu$ is the Green's function corresponding to the single particle Schr\"odinger equation of Hamiltonian 
\begin{equation}
 H= -\frac{\hbar^2}{2m}\frac{\partial^2}{\partial x^2} +V(x),
\end{equation}
with $V(x)$  the trapping potential.
In other words, $G_\mu(x,y)$ is  the solution of
\begin{equation}
\frac{\hbar^2}{2m}\frac{\partial^2}{\partial x^2}G_\mu(x,y) +(\mu-i0^+ -V(x))G_\mu(x,y)=
\delta(x-y).\label{gfunction}
\end{equation}
It is important to note that when integrating Eq. (\ref{dmu}) to recover the kernel we have the boundary condition, or completeness condition, 
\begin{equation}
\lim_{\mu\to\infty} K_\mu(x,y) =\delta(x-y), 
\end{equation}
as in this limit the sum in Eq. (\ref{k1}) is over a complete set of states. We also note 
the trivial identity
\begin{equation}
K_\mu(x,y) =\delta(x-y) -  \sum_k\theta(\epsilon_k-\mu)  \psi_k^*(x)\psi_k(y),
\end{equation}
which yields 
\begin{equation}
K_\mu(x,y) = \delta(x-y) -\int_{\mu}^\infty \frac{\partial}{\partial \mu'} K_{\mu'}(x,y) d\mu'
= \delta(x-y) -\int_{\mu}^\infty d\mu'  \frac{1}{\pi} { \rm Im} \, G_{\mu'}(x,y)
\label{bigmu}
\end{equation}
which will be useful in what follows. An alternative integration formula is
 \begin{equation}
K_\mu(x,y)= \int_{-\infty}^\mu \frac{\partial}{\partial \mu'} K_{\mu'}(x,y) d\mu'=\int_{-\infty}^\mu d\mu'  \frac{1}{\pi} { \rm Im} \, G_{\mu'}(x,y)
 \label{bigmu2}
\end{equation}
which obviously holds as long as the ground state of the system is bounded from below. In what follows we will always denote by $\mu$ the Fermi energy (which we assume is fixed) and denote by $\mu'$ the running Fermi energy used in the integrands of the representations given in Eq. (\ref{bigmu}) and (\ref{bigmu2}). These two representations can also be used to represent the kernel in terms of a kernel corresponding to a locally constant potential (so exact far away from  the step) plus a term due to the variation of the potential. The derivation is rather technical and is relegated to appendix \ref{contsec}.
%which is clearly the case for physical systems. 

\section{Smooth potentials}\label{sec:smooth_pot}

Before applying {the Green's function} method to obtain new exact solutions (for any $\mu$) 
for discontinuous potentials in the next section, we show how the method can be applied to
analyse the well studied case of smooth potentials. In the bulk we recover the prediction
of the LDA (which is exact for potentials constant in space) and identify the validity of the LDA via this method. We then examine, the again well known, edge {\em Airy gas} behavior, based on a local linear approximation to the potential
(the method being again exact for purely linear potentials).

\subsection{The bulk regime and the local density approximation}
\label{sec:LDA} 
 
Here we use  the Green's function to derive the LDA or Thomas-Fermi approximation which is the standard theoretical tool used to study the bulk thermodynamics behavior of free fermionic systems. In our derivation  we identify the two key regimes where  the approximation fails, the first regime is where the conditions for bulk behavior do not apply, notably near the edge of the system in continuous potential where the density becomes small. The second case occurs  when the potential is not continuous, or varies too fast in space.

The basic approximation consists of computing the Green's function at two points
$x=x_0+z$ and $y=x_0+z'$. {We focus here on the bulk, hence one has $V(x_0) < \mu$.}
Assuming that $z$ and $z'$ are small, we can make the approximation $V(x) \approx V(x_0)$ in Eq. (\ref{gfunction}) and write
\begin{equation} \label{s1} 
\frac{1}{2}\frac{\partial^2}{\partial z^2} G_{\mu'}(x_0+z,x_0+z') +[{ \mu' - i 0^+} - V(x_0)]G_{\mu'}(x_0+z,x_0+z')=\delta(z-z'),
\end{equation}
where we have introduced the running Fermi energy $\mu'$ which will be integrated over.
To simplify notation we have set $\hbar=1$ and $m=1$. The dependence on $\hbar^2/m$ can be reintroduced by making the rescalings $G\to m/\hbar^2 \times G$ and $(\mu-V)\to  m/\hbar^2\times(\mu-V)$. 

{Let us consider $\mu' > V(x_0)$.} For $z<z'$ we have 
\begin{equation}
G_{\mu'}(x_0+z,x_0+z') = A_- \exp\left(i\sqrt{2\mu' - i 0^+ -2V(x_0)}\ z\right) \;.
\end{equation}
We have used that for $a>0$, $\sqrt{a - i 0^+} \equiv \sqrt{a} - i 0^+$, hence 
the r.h.s. tends to zero as $z\to -\infty$ as required, due to the small imaginary part regulator $- i 0^+$. Similarly
for $z>z'$ 
\begin{equation}
G_{\mu'}(x_0+z,x_0+z') = A_+ \exp\left(-i\sqrt{2\mu' - i 0^+ - 2V(x_0)}\ z\right).
\end{equation}
Matching at $z=z'$, {taking into account the delta function on the r.h.s. of \eqref{s1},}
we obtain
\begin{equation}
G_{\mu'}(x_0+z,x_0+z') = \frac{i\exp\left(-i\sqrt{2\mu' - i 0^+ -2V(x_0)}|z-z'|)\right)}{\sqrt{2\mu' -2V(x_0)} } \, ,
\label{gflocal}
\end{equation}
where the small imaginary part {has been} neglected in the denominator.
This result is sufficient %enable us 
to derive the kernel in the bulk. 

For the computation of kernels away from the bulk it is useful to derive an equivalent representation for the Green's function.
For $y>x$, (so $z'>z$) Eq. (\ref{gflocal}) can be rewritten in terms of $x$ and $y$ for $|z-z'|$ small
as
\begin{eqnarray}
G_{\mu'}(x_0+z,x_0+z') 
&=&\frac{i\exp\left(-i\sqrt{2\mu' - i 0^+ -2V(x_0)}(z'-z)\right)}{\sqrt{2\mu' -2V(x_0)} }\nonumber \\
&\approx& \frac{i\exp\left(-i\int_{z}^{z'}\sqrt{2\mu' - i 0^+ -2V(x_0+s)}\ ds\right)}{\sqrt{2\mu'-2V(x_0)}} \nonumber\\
&\approx& \frac{i\exp\left(-i\int_{z}^{z'}\sqrt{2\mu' - i 0^+ -2V(x_0)- 2V'(x_0)s}\ ds\right)}{\sqrt{2\mu' -2V(x_0)} }\nonumber \\
&=& \frac{i\exp\left(i\left[\frac{1}{3V'(x_{0})}\left(2\mu'-i0^{+}-2V(x_{0})-2V'(x_{0})s\right)^{3/2}\right]_{z}^{z'}\right)}{\sqrt{2\mu'-2V(x_{0})}}\,,
\end{eqnarray}
and thus to the same order this can be written as
\begin{equation}
G_{\mu'}(x_0+z,x_0+z') = Y_-(z)Y_+(z'),
\end{equation}
where
\begin{equation}
Y_-(z)=\frac{\sqrt{i}\exp\left(-i\frac{1}{3V'(x_{0})}\left(2\mu'-i0^{+}-2V(x_{0})-2V'(x_{0})z\right)^{3/2}\right)}{\left(2\mu'-2V(x_{0})-2V'(x_{0})z\right)^{1/4}} \, ,
\label{bcag}
\end{equation}
and 
\begin{equation}
Y_+(z)=\frac{\sqrt{i}\exp\left(i\frac{1}{3V'(x_{0})}\left(2\mu'-i0^{+}-2V(x_{0})-2V'(x_{0})z\right)^{3/2}\right)}{\left(2\mu'-2V(x_{0})-2V'(x_{0})z\right)^{1/4}} \, .
\label{bcad}
\end{equation}
Adding the extra $z$ dependence in the numerator above does not change the basic small $z$ approximation and extends the basic plane wave approximation to the more general WKB form which will be useful when we will discuss the behaviors of kernels near a linear edge and matching plane wave solutions with Airy functions.

Returning to Eq. (\ref{gflocal}) we find that the kernel in the bulk can be written as 
%{\red Naftali: here I combined what was previously two equations into a single line}
\begin{equation}
\frac{\partial}{\partial\mu'}K_{\mu'}\left(x_{0}+z,x_{0}+z'\right)=\frac{1}{\pi}{\rm Im}\frac{i\exp\left(-i\sqrt{2\mu'-i0^{+}-2V(x_{0})}|z-z'|\right)}{\sqrt{2\mu'-2V(x_{0})}} {= \frac{\cos(\sqrt{2\mu'-2V(x_0)}|z-z'|)}{\pi\sqrt{2\mu'- 2V(x_0)} } \, .}
\end{equation}
%and so
%\begin{equation}
%\frac{\partial}{\partial \mu'} K_{\mu'}(x_0+z,x_0+z') = \frac{\cos(\sqrt{2\mu'-2V(x_0)}|z-z'|)}{\pi\sqrt{2\mu'- 2V(x_0)} }.
%\end{equation}
Integrating over $\mu'$ then gives
\begin{equation}
K_\mu(x_0+z,x_0+z') = \frac{1}{\pi |z-z'|}\sin(\sqrt{2\mu-2V(x_0)}|z-z'|) \, .
\end{equation}
The constant of integration has been determined by using the identity
\begin{equation}
\lim_{k\to\infty}\frac{1}{\pi x}\sin(kx) = \delta(x) \, .
\end{equation}
Note that strictly speaking the analysis is only valid in the case where $\mu'>\mu$ (assuming that already $\mu-V(x_0)$ is large enough to justify the analysis), however one sees that the analysis can be restricted to the case $\mu'>\mu$ by using the representation of $K_\mu$ given in Eq. (\ref{bigmu}) as opposed to Eq. (\ref{bigmu2}).
We have thus recovered the well known sine kernel for the
kernel in the bulk. It is often written as 
\begin{equation} \label{sk} 
K_\mu(x_0+z,x_0+z') = \frac{1}{\pi |z-z'|}\sin(p_\mu(x_0)|z-z'|),
\end{equation}
where 
\begin{equation}
p_\mu(x_0)= \sqrt{2\mu-2V(x_0)} \, ,\label{fmom}
\end{equation}
 is the local Fermi momentum at $x_0$. Finally the density of fermions at the point $x_0$ is obtained as
\begin{equation}
\label{eq:densityLDA}
\rho_\mu(x_0) = K_\mu(x_0,x_0)= \frac{p_\mu(x_0)}{\pi} \, .
\end{equation}\\

Let us now discuss the validity of the LDA. Its validity depends on the accuracy of the approximation where the spatial dependence of the potential $V(x)$ is neglected about the point $x_0$. The correction to this potential to give the exact one is $\Delta V(x) = V(x_0 +z)-V(x_0) \approx 
zV'(x_0)$ for small $z$. The exact Green's function is given by
\begin{equation}
G_\mu = (\mu - H_{LDA}- \Delta V)^{-1} 
\end{equation}
where 
$H_{LDA} = -\frac{1}{2}\frac{\partial^2}{\partial x^2} + V(x_0)$ is the approximate form of the Hamiltonian used in the LDA and so $G_\mu^{LDA} = (\mu-H_{LDA})^{-1}$ is the corresponding Green's function. The first order correction to the LDA Green's function is thus (in operator product notation)
\begin{equation}
\Delta G_\mu = G_\mu^{LDA} \Delta V G_\mu^{LDA}.
\end{equation}
Now making the approximation $\Delta V(x) \simeq z V'(x_0)$, so assuming that the derivative of $V(x)$ exists, we find
\begin{equation}
\Delta G_\mu(z,z') = V'(x_0) \int du \, u   \ G_\mu^{LDA}(z,u)G_\mu^{LDA}(u,z').
\end{equation}
From this we obtain the estimate $|\Delta G(z,z')| \approx |V'(x_0)|/ p_\mu(x_0)^4$. However we have seen that $G_\mu^{LDA}\approx 1/p_\mu(x_0)$ and so the condition that the  relative error  in using the  LDA is small can be written as \cite{comment} 
\begin{equation}
 \frac{|\Delta G(z,z')|}{|G_{LDA}(z,z')|} \approx \frac{|V'(x_0)|}{p_\mu(x_0)^3 }\ll1 
\label{ldaval}
\end{equation}
Note that this condition agrees with the more heuristic argument that the 
potential should vary little on the scale of the inter-particle distance, which is $1/{p_\mu}(x_0)$,
i.e. that $|V'(x_0)/{p_\mu}(x_0)| \ll V(x_0) \sim \mu = \frac{{p_\mu}(x_0)^2}{2}$ (since in the bulk
$V$ is of the order of $\mu$).

We thus see that the LDA is valid in the bulk where the density $\rho(x_0) = p_\mu(x_0)/\pi$ is large. The LDA fails at the {\em edge} where the bulk density vanishes. The behavior of the kernel near this edge has been derived in several works by various methods \cite{dea16,eis13,dea15}.
In the following section we will show how the edge behavior can be derived using the Green's function method. As one might anticipate, we will see that the LDA also fails  when the potential is discontinuous.
What happens in this case has been much less studied, and in Section \ref{sec:step}  we will provide an analysis of the kernel for the cases of the step like potentials.
This analysis is exact, as we can obtain exact expressions for the Green's function.

\subsection{The edge regime and the Airy kernel}
\label{sec:airy} 

Let us now study the Green's function near the edge points $x_e$  which are defined via the vanishing of the LDA prediction of the density, {\em i.e.} as the solutions of the equation
\begin{equation}
\mu - V(x_e) = 0,
\end{equation}
and we note that in general the function $V(x)-V(x_e)$ will vanish linearly near $x_e$. Thus the Green's function will locally obey the equation
\begin{equation}
\frac{1}{2}\frac{\partial^2}{\partial z^2} G_{\mu'}(z,z') +(\mu'-V(x_e)-z V'(x_e) )G_{\mu'}(z,z')=\delta(z-z'),\label{gfe}
\end{equation}
for $z$ and $z'$ in the neighborhood of {$x_0$} {$x_e$}.
Here, for notational simplicity we denoted $\mu' - i 0^+$ by $\mu'$ hence we
temporarily assume that $\mu'$ has a small negative imaginary part.
Without loss of generality we assume that $V'(x_e)>0$ (so we are considering 
a right edge). From Eq. (\ref{gfe}) we see that when $\mu'\gg V(x_e)$ we can ignore the term linear in $z$ and repeat the bulk calculation.  However when $\mu'\approx \mu$ the conditions necessary for the validity of the bulk calculation no longer hold and so we keep the linear correction which can be larger than or of the same order as the constant term.
We define the variable 
\begin{equation} \label{cv} 
\mu' - V(x_e) -z V'(x_e) = -\alpha \zeta,
\end{equation}
where $\alpha$ is a constant determined below. The equation \eqref{gfe} becomes
\begin{equation}
\frac{\partial^2}{\partial \zeta^2}G_{\mu'}(\zeta,\zeta) - \frac{2\alpha^3}{V'^2(x_e)}\zeta G_{\mu'}(\zeta,\zeta) =
2\frac{\alpha}{V'(x_e)}\delta(\zeta-\zeta').
\end{equation}
We choose $\frac{2\alpha^3}{V'^2(x_e)} =1$, which fixes $\alpha$ as
\begin{equation} \label{defalpha}
%\alpha= (V'^2(x_e)/2)^{\frac{1}{3}}.
\alpha=\left[V'(x_{e})\right]^{2/3} \! / \, 2^{1/3}\,.
\end{equation}
This yields
\begin{equation}
\frac{\partial^2}{\partial \zeta^2}G_{\mu'}(\zeta,\zeta) - \zeta G_{\mu'}(\zeta,\zeta) =
\frac{2\alpha}{V'(x_e)}\delta(\zeta-\zeta').
\end{equation}
We therefore have
\begin{equation} \label{Gs}
G_{\mu'}(\zeta,\zeta') = \frac{2\alpha}{V'(x_e)} g(\zeta,\zeta')= \frac{V'(x_e)}{\alpha^2} g(\zeta,\zeta'),
\end{equation}
where $g(\zeta,\zeta')$ is the solution of 
\begin{equation}
\frac{\partial^2}{\partial \zeta^2}g(\zeta,\zeta') - \zeta g(\zeta,\zeta') =
\delta(\zeta-\zeta'), \label{airyeq} 
\end{equation}
i.e. it is the Green's function of the Airy operator. Note that the resolvent of
the Airy operator has a branch cut on the real axis, and here we consider its
value, $g(\zeta,\zeta')$,
for infinitesimal negative imaginary part (corresponding to $\mu' \to \mu' - i 0^+$ above).
The derivation is given in Appendix \ref{app:airygreen}. The final result is
%\begin{eqnarray} \label{resg} 
%g(\zeta,\zeta') &=& -\pi{\rm Ai}(\zeta)[-i{\rm Ai}(\zeta') +{\rm  Bi}(\zeta')] 
%  \ {\rm for } \ \zeta>\zeta' \\
%&=&-\pi{\rm Ai}(\zeta')[-i{\rm Ai}(\zeta) +{\rm  Bi}(\zeta)]   \ {\rm for } \ \zeta<\zeta'.
%\end{eqnarray}
\be
\label{resg} 
g(\zeta,\zeta')=\begin{cases}
-\pi{\rm Ai}(\zeta)[-i{\rm Ai}(\zeta')+{\rm Bi}(\zeta')] & {\rm for}\ \zeta>\zeta'\\[0.1cm]
-\pi{\rm Ai}(\zeta')[-i{\rm Ai}(\zeta)+{\rm Bi}(\zeta)] & {\rm for}\ \zeta<\zeta'.
\end{cases}
\ee

Hence, for any $\zeta,\zeta'$ we have
\begin{equation}
{\rm Im}\left[g(\zeta,\zeta') \right]=\pi {\rm Ai}(\zeta){\rm Ai}(\zeta').
\end{equation}
Using the general relation \eqref{dmu} between the kernel and the Green's function,
together with \eqref{cv}, \eqref{defalpha}, \eqref{Gs},  and the result \eqref{resg}, we obtain
\begin{equation}
\frac{\partial}{\partial \mu'} K_{\mu'}(x,y) =  \frac{V'(x_e)}{\alpha^2}{\rm Ai}(\zeta) {\rm Ai}(\zeta') = \frac{V'(x_e)}{\alpha^2}
 {\rm Ai} \left(\frac{(x-x_e) V'(x_e)-\mu' + V(x_e)}{\alpha}\right){\rm Ai} \left(\frac{(y-x_e) V'(x_e)-\mu' + V(x_e)}{\alpha}\right),
 \end{equation}
 where here $\mu'$ is real. 
Using Eq. (\ref{bigmu}) now gives
\begin{equation}
K_\mu(x,y)= \delta(x-y) - 
{\frac{V'(x_e)}{\alpha^2} } \int_{\mu}^\infty d\mu' {\rm Ai} \left(\frac{(x-x_e) V'(x_e)-\mu' + V(x_e)}{\alpha}\right){\rm Ai} \left(\frac{(y-x_e) V'(x_e)-\mu' + V(x_e)}{\alpha}\right).\end{equation}
Recalling that $x_e$ is the edge, such that $\mu = V(x_e)$, we can change variables to 
$\mu' = \mu + \alpha u$ and obtain
\begin{equation}
K_\mu(x,y) = \delta(x-y)-\frac{1}{w_\mu}\int_{-\infty}^0 du\ {\rm Ai}\left(\frac{x-x_e}{w_\mu}+ u\right){\rm Ai}\left(\frac{y-x_e}{w_\mu}+u\right), \nonumber \\
\end{equation}
where we have introduced the width of the edge regime \cite{dea16}
\be
w_\mu = \alpha/V'(x_0) = (2 V'(x_0))^{-1/3} \;.
\ee
Finally, using the completeness identity for Airy functions
\begin{equation}
\int_{-\infty}^\infty du \ {\rm Ai}(u+x){\rm Ai}(u+y)=\delta(x-y),
\end{equation}
we recover that the kernel near the edge takes the following scaling form in terms of the Airy kernel 
$K_{\rm Ai}$ 
\begin{equation} \label{Kedge} 
K_{\mu}(x,y)=\frac{1}{w_{\mu}}K_{{\rm Ai}}\left(\frac{x-x_{e}}{w_{\mu}},\frac{y-x_{e}}{w_{\mu}}\right)\quad,\quad K_{{\rm Ai}}(a,b)=\int_{0}^{+\infty}du\ {\rm Ai}(a+u){\rm Ai}(b+u).
\end{equation}

\subsection{General fast varying potential: perturbation theory}

One way of treating rapidly varying potentials which cause a breakdown of the LDA is via perturbation theory, valid when they are rapidly varying but weak. We take $H=H_0 + \delta V(x)$ where $\delta V(x) \ll H_0$ is a rapidly varying potential of a general form. If it varies notably on scales of the inter-particle distance its effect will be in general difficult to calculate. However we can study its effect via perturbation theory for the Green's function assuming that  $\delta V(x)$ is sufficiently weak. Here we find to first order
\be
G_\mu(x,y) = G_{0\mu}(x,y) +  \int dx' G_{0\mu}(x,x') \delta V(x') G_{0\mu}(x',y) + O(\delta V^2) .
\ee 
Hence  the change in the kernel due to the perturbation is given by
\be
\Delta K_\mu(x,y) = - \frac{1}{\pi} \int_\mu^{+\infty} d\mu' \,  {\rm Im} \int dx' G_{0\mu'}(x,x') \delta V(x') G_{0\mu'}(x',y) + O(\delta V^2) .
\ee
Let us study it in the bulk, as in Section \ref{sec:LDA}. Using \eqref{gflocal} and $\frac{d\mu}{d p_\mu(x_0)}= p_\mu(x_0)$, we obtain for $x,y$ near {a given} $x_0$, the general formula
\be
\Delta K_\mu(x,y) = - \frac{1}{\pi} \int_{p_\mu(x_0)}^{+\infty} \frac{dp}{p} 
\int dx' \delta V(x') \exp\left(- i p |x-x'| - i p |y-x'|\right)
+ O(\delta V^2) .
\ee 
We thus see that  the result is quantitatively significant if $\delta V(x)$ varies on scales of order or shorter than $1/p_\mu(x_0)$.
%and using this in Eq. (\ref{wron}) then yields
%\begin{equation}
%{C_1(k_1,i\kappa_2)C_1^*(k_1,-i\kappa_2)}-{C^*_1(k_1,i\kappa_2)C_1(k_1,-i\kappa_2)}=i\frac{\kappa_2}{k_1}
%\end{equation}

%
%which yields 
%\begin{equation}
%K_\mu(x,y)= \delta(x-y)-\frac{V'(x_0)}{\alpha^2}\int_{\mu}^{\infty} d\mu' {\rm Ai} (\frac{(x-x_0) V'(x_0)-\mu' +\mu }{\alpha}){\rm Ai} (\frac{(y-x_0) V'(x_0)-\mu' + \mu}{\alpha}) \nonumber \\
%\end{equation}
%and changing variables we find
%\begin{equation}
%K_\mu(x,y) = \delta(x-y)-\frac{V'(x_0)}{\alpha}\int_{-\infty}^0 du\ {\rm Ai}(\frac{(x-x_0) V'(x_0)}{\alpha}+ u){\rm Ai}(\frac{(y-x_0) V'(x_0)}{\alpha}+u).\nonumber \\
%\end{equation}
%Finally, using the completeness identity for Airy functions
%\begin{equation}
%\int_{-\infty}^\infty du \ {\rm Ai}(u+x){\rm Ai}(u+y)=\delta(x-y),
%\end{equation}
%we can write
%\begin{equation}
%K_\mu(x,y) = \frac{V'(x_0)}{\alpha}\int_{0}^\infty du\ {\rm Ai}(\frac{(x-x_0) V'(x_0)}{\alpha}+ u){\rm Ai}(\frac{(y-x_0) V'(x_0)}{\alpha}+u).\nonumber \\
%\end{equation}
%We thus recover the Airy kernel for the edge in free fermion systems, which can also be written as
%\begin{equation}
%K_\mu(x,y) = \frac{1}{(x-y)}\left[{\rm Ai}(\frac{(x-x_0) V'(x_0)}{\alpha}){\rm Ai}'(\frac{(y-x_0) V'(x_0)}{\alpha})-{\rm Ai}'(\frac{(x-x_0) V'(x_0)}{\alpha}){\rm Ai}(\frac{(y-x_0) V'(x_0)}{\alpha})\right].
%\end{equation}

\section{Square step barrier, exact results}
\label{sec:step}

\subsection{General setting} 

Here we apply the Green's function method presented above to analyse fermion statistics %at the edge 
 in the presence of a finite square step potential located at $x=0$ which can be written  as
\begin{equation} \label{step1}
V(x) = 0\  {\rm for}\ x<0;\ V(x)= V_0 \ {\rm for}\  x>0, 
\end{equation}
and is shown in Fig. \ref{square}. For the potential \eqref{step1} the basic formalism of  section \ref{sec:green}
can be implemented without any approximation for any $\mu$ without resorting to any approximation, because the Green's function can be computed exactly. 

Let us introduce 
\be
p_F^L = p_F^L(\mu) = \sqrt{2 \mu} \quad , \quad p_F^R = p_F^R(\mu) = \sqrt{(2 \mu- 2 V_0)_+} 
\ee
the Fermi momentum in the regions $x<0$ and $x>0$ respectively, where $(x)_+=\max(x,0)$,
and we work here at fixed $\mu$.

There are two main cases:

(i) $\mu > V_0$, which we will refer to as supercritical in what follows. Here
very far from the barrier, at distances much greater that $ 1/p_F^{L,R}$, the
Fermi gas has two different uniform mean densities for $x<0$ (left -L) and $x>0$ (right -R), given by 
\be \label{densLR} 
\rho_L=\frac{p_F^L}{\pi} = \frac{\sqrt{2 \mu}}{\pi} \quad , \quad \rho_R = \frac{p_F^R}{\pi} = \frac{\sqrt{2 (\mu-V_0)}}{\pi} < \rho_L.
\ee

(ii) $0< \mu < V_0$, which we refer to as sub-critical, in which case only the left half space is filled for all $x$,
with, far from the barrier, the uniform mean density $\rho_L$. The mean density
vanishes for $x>0$ at distances $x \gg 1/\sqrt{2 V_0- 2\mu}$.

These are however only the behavior of the system far from the step
and we wish to calculate the behavior at distances $\sim 1/p_F^{L, R}$ from the
step. Hence we are interested in the crossover between the behaviors 
of the two bulk Fermi gases. 

The cases (i) and (ii) are separated by a transition, for $\mu=V_0$, which we call the critical case, where the Fermi gas on the left is at the brink of overflowing. It has a very interesting behavior, that we analyse below.

The general scaling form that the kernel takes can be obtained by using simple dimensional analysis. The units of the kernel are $\text{length}^{-1}$. The most general way to construct such a quantity in this system is
\begin{equation}
\label{eq:kernel_scaling}
K_{\mu}(x,y)=\frac{1}{\ell}\kappa_{r}\left(\frac{x}{\ell},\frac{y}{\ell}\right)\quad,\quad\frac{1}{\ell}=\rho_{L}=\frac{\sqrt{2\mu}}{\pi} \quad , \quad r=V_0 / \mu 
\end{equation}
where %$r=V_0 / \mu > 0$ and 
$\kappa_r$ is a dimensionless function. The cases $r<1$, $r=1$ and $r>1$ correspond to the supercritical, critical and subcritical cases respectively. As a result, the density takes the scaling form 
\begin{equation}
\label{eq:density_scaling}
\rho(x)=\frac{1}{\ell}\,n_{r}\left(\frac{x}{\ell}\right)\quad,\quad\frac{1}{\ell}=\rho_{L}=\frac{\sqrt{2\mu}}{\pi}\quad,\quad n_{r}(a)=\kappa_{r}(a,a).%\label{denscal2}
\end{equation}

\begin{figure}[t!]
\begin{center}
  \includegraphics[scale=0.30]{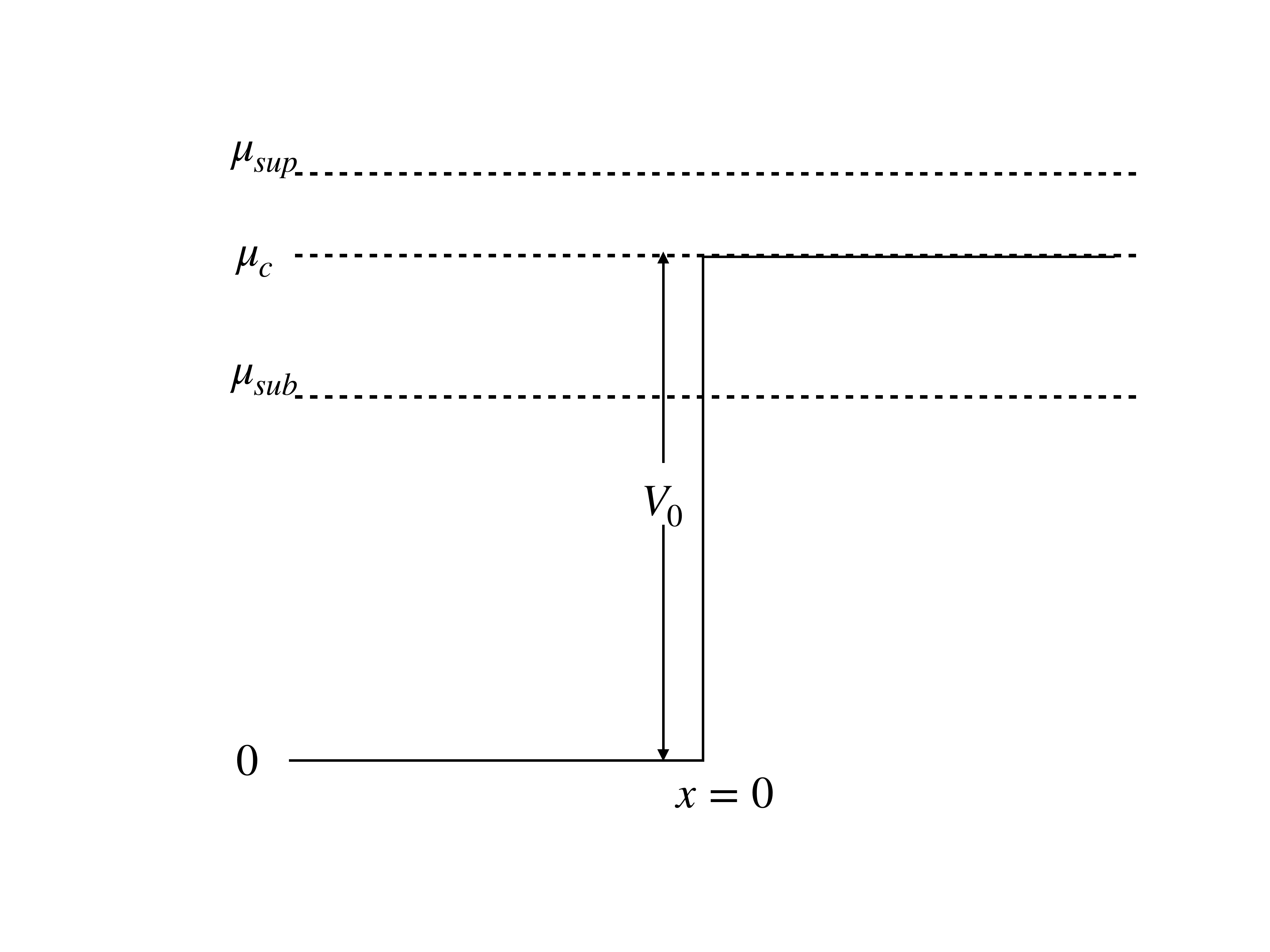} 
 \caption{Locally square well potential with barrier of height $V_0$ situated at $x=0$. The Fermi energy $\mu$ for  $\mu=\mu_c=V_0$ as well as the sub-critical case $\mu=\mu_{sub}<V_0$ and the super-critical case $\mu= \mu_{sup}>V_0$.}\label{square}
\end{center}
\end{figure}

\subsection{Green's function} 

The computation of the Green's function is long but straightforward. 
It is performed in Appendix \ref{app:green} for completeness 
and also to detail the needed analytic continuations. One can also
find a formula in the reference book on path integrals \cite{gro98}. Here we give
the results for the imaginary part, which is what we need to compute the kernel. 

For $\mu'<0$ one has ${\rm Im}\ G_{\mu'}(x,y) =0$, which is natural 
since there are no energy eigenstates for negative energy. 
Next there are two cases, either $0< \mu'<V_0$ or $\mu'> V_0$.
 
We start with  the case where $0< \mu'< V_0$. In the region $x,y<0$ we find 
\begin{equation}
{\rm Im}\,G_{\mu'}(x,y)=\frac{1}{V_{0}\sqrt{2\mu'}}\!\left[V_{0}\cos\left(\!\sqrt{2\mu'}\ (x-y)\right)+(2\mu'-V_{0})\cos\left(\!\sqrt{2\mu'}\ (x+y)\right)-2\sqrt{\mu'(V_{0}-\mu')}\sin\left(\!\sqrt{2\mu'}\ (x+y)\right)\right], \label{vbigleft} 
\end{equation}
which obeys the symmetry  $G(x,y)=G(y,x)$ of the Green's function. For $x>0>y$ we find 
\begin{equation}
{\rm Im}\ G_{\mu'}(x,y)=\frac{\sqrt{2}\exp\left(-\sqrt{2V_{0}-2\mu'}\ x\right)}{V_{0}}\left[\sqrt{\mu'}\cos\left(\sqrt{2\mu'}\ y\right)-\sqrt{V_{0}-\mu'}\sin\left(\sqrt{2\mu'}\ y\right)\right], \label{vbigright}
\end{equation}
and the  behavior for  $y>0>x$ is obtained from the symmetry of the Green's function.
In region $x,y>0$ one finds the decaying solution 
\begin{equation}
{\rm Im}\ G_{\mu'}(x,y) = \frac{\sqrt{2\mu'}}{V_0}\exp\left(-\sqrt{2V_0-2\mu'}\ (x+y)\right).
\label{g2}
\end{equation}

We now consider the case where  $\mu'> V_0$. For $x,y<0$ we find
\begin{equation}
{\rm Im}\ G_{\mu'}(x,y)=\frac{1}{V_{0}\sqrt{2\mu'}}\left[V_{0}\cos\left(\sqrt{2\mu'}\ (x-y)\right)+\left(\sqrt{\mu'}-\sqrt{\mu'-V_{0}}\right)^{2}\cos\left(\sqrt{2\mu'}\ (x+y)\right)\right],
\label{e1}
\end{equation}
while for $y<0<x$ we find
\begin{equation}
{\rm Im}\ G_{\mu'}(x,y) =
 \frac{\sqrt{2\mu'}-\sqrt{2\mu'-2 V_0}}{V_0}\cos\left(\sqrt{2\mu'-2V_0}\ x- \sqrt{2\mu'}y\right).
 \label{e2}
\end{equation}
Finally, for  $x,y>0$ we find
\begin{equation}
{\rm Im}\ G_{\mu'}(x,y)=\frac{1}{V_{0}\sqrt{2\mu'-2V_{0}}}\left[V_{0}\cos\left(\sqrt{2\mu'-2V_{0}}\ (x-y)\right)-\left(\sqrt{\mu'}-\sqrt{\mu'-V_{0}}\right)^{2}\cos\left(\sqrt{2\mu'-2V_{0}}\ (x+y)\right)\right].
\label{e3}
\end{equation}
These forms for $G_{\mu'}$ can now be used to compute the kernel.
\subsection{Infinite barrier and the hard wall kernel}

Let us start with the simplest case of an infinite barrier $V_0\to\infty $. In this limit we see
that when both points are to the left of the wall, $x,y<0$, one obtains from Eq. (\ref{vbigleft})
\begin{equation}
{\rm Im}\ G_{\mu'}(x,y) = \frac{1}{\sqrt{2\mu'}}
\left[\cos(\sqrt{2\mu'}\ (x-y))-\cos(\sqrt{2\mu'}\ (x+y))\right].
\end{equation}
By integrating Eq. (\ref{e3}) over $\mu'$ one finds from Eq. (\ref{dmu}) 
\begin{equation} \label{reflectedkernel} 
K_{\mu}(x,y) = \frac{\sin(\sqrt{2\mu}\ (x-y))}{\pi (x-y)}-\frac{\sin(\sqrt{2\mu}\ (x+y))}{\pi (x+y)}, 
\end{equation}
which is the reflected sine-kernel that describes the infinite, or hard wall, potential barrier. 
This result is well known and has been derived in the literature using different methods \cite{cal11,lac17}.

\subsection{{Super-critical case (or overflow) $\mu>V_0$}} 

Let us study the case $\mu>V_0$ where the mean densities far on both sides are
both positive, $\rho_{L,R} >0$. The square barrier thus acts as a perturbation inside 
the bulk of the Fermi gas. Using \eqref{e1} and \eqref{e3}, it is easy to see that the 
integration of Eq. (\ref{dmu})
leads to 
\bea \label{2eq}
&& K_\mu(x,y) = \frac{\sin(\sqrt{2\mu}|x-y|)}{\pi|x-y|} - \int_\mu^\infty d\mu' 
\frac{(\sqrt{\mu'} - \sqrt{\mu'-V_0})^2 \cos(\sqrt{2\mu'}\ (x+y))}{\pi V_0\sqrt{2\mu'}} \quad , \quad x, y < 0 \\
&& 
K_\mu(x,y) = \frac{\sin(\sqrt{2\mu-2V_0}\ (x-y))}{\pi(x-y)} + \int_\mu^\infty d\mu'  
\frac{(\sqrt{\mu'} - \sqrt{\mu'-V_0})^2 \cos(\sqrt{2\mu'-2V_0}\ (x+y))}{\pi V_0\sqrt{2\mu'-2V_0}} \quad , \quad x, y  > 0 \nonumber \;.
\eea 
The result in the region $x>0>y$ can be obtained from integration over $\mu'$ of
\eqref{vbigright} from $0$ to $V_0$, and of \eqref{e2} from $V_0$ to $\mu$, but
is not displayed here.

The results in \eqref{2eq} describes the deviations from the sine kernel forms,
which hold deep in the left and right bulks, due to the barrier at $x=0$. These deviations are written as integrals, which are convergent at large $\mu'$ since 
$(\sqrt{\mu'} - \sqrt{\mu'-V_0})^2   \simeq \frac{V_0^2}{4 \mu'}$ for $\mu' \gg V_0$,
and which decay far from the barrier.
The Eq.~\eqref{2eq} can be written in the scaling form \eqref{eq:kernel_scaling} where for $r<1$,
\begin{numcases}{\kappa_{r}\left(a,b\right)=}
\frac{\sin\left(\pi\left|a-b\right|\right)}{\pi\left|a-b\right|}-\sqrt{r}\int_{1/r}^{\infty}dw\frac{\left(\sqrt{w}-\sqrt{w-1}\right)^{2}\cos\left(\pi\sqrt{rw}\left(a+b\right)\right)}{2\sqrt{w}}, & $a,b<0$\\% \label{a}\\
\frac{\sin\left(\pi\sqrt{1-r}\left(a-b\right)\right)}{\pi\left(a-b\right)}+\sqrt{r}\int_{1/r}^{\infty}dw\frac{\left(\sqrt{w}-\sqrt{w-1}\right)^{2}\cos\left(\pi\sqrt{r\left(w-1\right)}\ \left(a+b\right)\right)}{2\sqrt{w-1}} & $a,b>0.$ %\label{b}
\end{numcases}

We recall that the mean density is given by $\rho(x)=K_\mu(x,x)$. We find that the density at $x=0$ 
can be expressed in terms of the densities $\rho_L$ and $\rho_R$ far from the barrier
given in \eqref{densLR} as
\bea
\label{eq:rho_at_zero_supercritical}
&& \rho(0)= \frac{2}{3} \frac{\rho_L^2 + \rho_R^2 + \rho_L \rho_R}{\rho_L + \rho_R} 
 = \frac{\rho_L+\rho_R}{2} + \frac{1}{12 \rho_L} (\rho_L-\rho_R)^2 + O((\rho_L-\rho_R)^3) .
\eea 
Far from the barrier the density decays to its asymptotic values as
\bea \label{asymptdens} 
&& \rho(x) \simeq \rho_L + \frac{(\sqrt{\mu} - \sqrt{\mu-V_0})^2 \sin(\sqrt{2\mu}\ 2 |x|)}{2 \pi V_0  |x|} 
= \rho_L + \frac{\rho_L-\rho_R}{\rho_L+\rho_R} \frac{\sin(2 \pi \rho_L |x|)}{2 \pi |x|} \quad , \quad x \to - \infty \\
&& \rho(x) \simeq \rho_R - \frac{(\sqrt{\mu} - \sqrt{\mu-V_0})^2 \sin(\sqrt{2\mu-2 V_0}\ 2 x)}{2 \pi V_0  x} 
= \rho_R - \frac{\rho_L-\rho_R}{\rho_L+\rho_R} \frac{\sin(2 \pi \rho_R x)}{2 \pi x} \quad , \quad x \to + \infty  \;.
\eea

%
%
%We can put these results in a dimensionless form if we introduce the parameter
%\be
%v_0 = \frac{V_0}{\mu} 
%\ee 
%and perform the change of variable $\mu'=\mu w$. Then we find
%\be
%K_\mu(x,y) = \sqrt{2 \mu} k_s( \sqrt{2 \mu} |x-y| ) + \sqrt{2 \mu} \, k^L_{v_0}(\sqrt{2 \mu} (x+y) ) 
%\ee 
%where $k_s(z)=\sin z/(\pi z)$ is the reduced sine kernel which describes the bulk on the left side
%with density $\rho_L$, and the correction is
%\be
%k_{v_0}(z) = - \frac{1}{2 \pi v_0} \int_1^{+\infty} \frac{dw}{\sqrt{w}} (2 w - 2 \sqrt{w(w-v_0)}- v_0) \cos( z \sqrt{w})
%\ee 
%where we recall that in this section $v_0<1$. One has
%\be
%k_{v_0}(0)= \frac{3 v_0+2 (1-v_0)^{3/2} -2}{3 \pi 
%   v_0}
%\ee 
%
%and for $x,\ y >0$
%\begin{equation}
%K_\mu(x,y) = \frac{\sin(\sqrt{2\mu-2V_0}\ (x-y))}{\pi(x-y)} + \int_\mu^\infty d\mu'  \frac{(2\mu'-2\sqrt{\mu'(\mu'-V_0)}-V_0)\cos(\sqrt{2\mu'-2V_0}\ (x+y))}{\pi V_0\sqrt{2\mu'-2V_0}}.
%\end{equation}
%
%\bea
%K_\mu(0,0)= \frac{2 \sqrt{2} \left(\mu ^{3/2}-(\mu -v)^{3/2} \theta(\mu-v) \right)}{3 \pi  v} \theta(\mu)
%\eea 
%\bea
%&& \rho(0)=-\frac{2 \left(\rho _1^3-\left(2 \rho _1^2-\rho _2^2\right){}^{3/2}\right)}{3
%   \left(\rho _1^2-\rho _2^2\right)} \\
%   && = \frac{\rho_1+\rho_2}{2} - \frac{5}{12 \rho_1} (\rho_2-\rho_1)^2 + O((\rho_2-\rho_1)^3) 
%\eea 
%

\subsection{Kernel in the critical -- just overflowing -- case}

For the square barrier one obtains a new form of edge when the Fermi energy intersects the potential, for $\mu=\mu_c =V_0$, see Fig. \ref{square}. The picture is that the bulk of the Fermi gas on inside the potential well for $x<0$ is on the point of overflowing out of the well for $x>0$. 

Let us first study the region $x,y>0$. Setting $\mu=V_0$ 
in the second equation in \eqref{2eq}, we see that the first term vanishes and we find 
\begin{eqnarray}
K_{\mu=V_0}(x,y)\equiv K_c(x,y) &=& \int_\mu^\infty d\mu' \frac{(\sqrt{\mu'} - \sqrt{\mu'-\mu})^2 \cos(\sqrt{2\mu'-2\mu}\ (x+y))}{\pi \mu\sqrt{2\mu'-2\mu}}\nonumber \\
%&=& \int_0^\infty du \frac{(\mu +2u -2\sqrt{u(\mu+u)})\cos(\sqrt{2u}\ (x+y))}{\pi \mu\sqrt{2u}}\nonumber\\
&=& \frac{\sqrt{\mu}}{\pi}\int_0^\infty dv \frac{(\sqrt{1+v}-\sqrt{v})^2 \cos(\sqrt{2v}\sqrt{\mu} (x+y))}{\sqrt{2v}} 
\quad , \quad x,y >0 
\end{eqnarray}
where we have set $\mu'=\mu(1+v)$. This integral can be evaluated and one find that the kernel takes the scaling form \eqref{eq:kernel_scaling}
%\begin{equation} \label{critscaling} 
%K_c(x,y)=\frac{1}{\ell} \,\, \kappa_c(\frac{x}{\ell},\frac{y}{\ell})   \quad , \quad \frac{1}{\ell} = \rho_L = \frac{p_F^L}{\pi} = \frac{\sqrt{2 \mu}}{\pi}  \quad , \quad x,y >0 \
%\end{equation}
%where $\ell$ is the typical inter-particle distance far in the bulk on the left side. We have introduced 
%the reduced kernel
where the critical reduced kernel $\kappa_{c}(a,b)\equiv\kappa_{r=1}(a,b)$ is given by
\begin{equation} 
\kappa_c(a,b) = \frac{2}{3} + \frac{\pmb{L}_2(\pi (a+b))-I_2(\pi(a+b))}{(a+b)}
\label{crit+} \quad , \quad a,b >0 
\end{equation}
where $\pmb{L}_\alpha$ denotes the modified Struve function and $I_\alpha$ the modified 
Bessel function \cite{abr65}.

It is worth noticing that Eq. (\ref{crit+}) can also be obtained, using \eqref{bigmu2} and \eqref{dmu},
by integrating Eq. (\ref{g2}) between
$\mu'=0$ and $\mu'=\mu$, since there are no states for $\mu'<0$
hence ${\rm Im}\ G_{\mu'<0}(x,y)=0$. Upon inserting $V_0=\mu$ and setting $\mu'=\mu(1-v)$ 
we obtain again \eqref{eq:kernel_scaling} %\eqref{critscaling}
together with the alternative, more useful, representation of the kernel
\begin{equation} 
\kappa_c(a,b) = \int_0^1 dv\ \sqrt{1-v}\exp(-\sqrt{v}\pi (a+b))\label{rep2}  \quad , \quad a,b >0  \;.
\end{equation}

Consider now the region $x, y<0$. Setting $\mu=V_0$ 
in the first equation in \eqref{2eq}, we find
\begin{equation}
K_c(x,y) = \frac{\sin(\sqrt{2\mu}(x-y))}{\pi(x-y)} -
\int_\mu^\infty d\mu' \frac{
(\sqrt{\mu'} - \sqrt{\mu'-\mu})^2 \cos(\sqrt{2\mu'}\ (x+y))}{\pi \mu\sqrt{2\mu'}}.
\end{equation}
The integral can be evaluated and the result takes again the scaling form \eqref{eq:kernel_scaling} %\eqref{critscaling}
with now  
\bea \label{kcleft} 
 \kappa_c(a,b) &=& \frac{\sin(\pi(a-b))}{\pi(a-b)} +\frac{\sin(\pi(a+b))}{\pi(a+b)}  \\
&+&
  \frac{1}{\pi(a+b)}\left[4\ \frac{  \pi(a+b)\cos(\pi(a+b))-\sin(\pi(a+b)) }{\pi^2(a+b)^2} - \pi J_2(\pi(a+b))\right]
  \quad , \quad a,b < 0 \nonumber
\eea
where $J_\alpha$ is the Bessel function of the first kind \cite{abr65}. Finally the kernel for
$x>0>y$, not displayed here, is obtained upon integrating \eqref{vbigright} between
$\mu'=0$ and $\mu'=\mu$.

%and the integral above can be evaluated to give
%\begin{eqnarray}
%&&K_c(x,y) =\nonumber \\&& \frac{\sin(\sqrt{2\mu}(x-y))}{\pi(x-y)} +\frac{\sin(\sqrt{2\mu}(x+y))}{\pi(x+y)}+
% \frac{1}{\pi(x+y)}\left[\frac{ -2\sin(\sqrt{2\mu}(x+y)) +2 \sqrt{2\mu}(x+y)\cos(\sqrt{2\mu}(x+y))}{\mu(x+y)^2}-\pi J_2(\sqrt{2\mu}(x+y))\right]\nonumber \\
%\end{eqnarray}

The mean fermion density is thus given by Eq.~\eqref{eq:density_scaling} with $n_{c}\left(a\right)\equiv n_{r=1}\left(a\right)=\kappa_{c}\left(a,a\right)$.
%\begin{equation}
%\rho(x) = \frac{1}{\ell} \, n_c(\frac{x}{\ell}) \quad , \quad \frac{1}{\ell} = \rho_L = \frac{\sqrt{2 \mu}}{\pi} \quad , \quad n_c(a)= \kappa_c(a,a) .\label{denscal} 
%\end{equation}
On the left side $x<0$ it is obtained from \eqref{kcleft} and it reaches its uniform
limit $\rho(x)=\rho_L$ for $x \to - \infty$, i.e. for $a \to - \infty$, by oscillating as
\be
n_{c}(a)=1+\frac{\sin(2\pi a)}{2\pi a}+O\left(\frac{1}{|a|^{3/2}}\right)\quad,\quad a\to-\infty\;.
\ee 
On the right side $x>0$ the density is obtained from \eqref{rep2}
and we find that it decays to zero as a power law
\be \label{alg1} 
n_{c}(a)=\frac{1}{2\pi^{2}a^{2}}-\frac{3}{8\pi^{4}a^{4}}+O\left(\frac{1}{a^{6}}\right).
\ee 
%The bulk density for $x,\ y\ll 0$ is obtained from 
%\begin{equation}
%\rho_c(x) = K_c(x,x)
%\end{equation}
%which leads to
%\begin{equation}
%\rho_b = \frac{\sqrt{2\mu}}{\pi},
%\end{equation}
%and this sets a length scale $\ell = 1/\rho_b$. 
%Using this we can write the above results as 
%\begin{equation}
%K_c(x,y)=\frac{1}{\ell}\kappa_c(\frac{x}{\ell},\frac{y}{\ell}),
%\end{equation}
%with 
%\begin{equation}
%\kappa_c(a,b) = \frac{2}{3} + \frac{\pmb{L}_2(\pi (a+b))-I_2(\pi(a+b))}{a+b}
%\end{equation}
%for $a,\ b >0$ and 
%
%\begin{equation}
%\kappa_c(a,b) = \frac{\sin(\pi(a-b))}{\pi(a-b)} +\frac{\sin(\pi(a+b))}{\pi(a+b)} +
%  \frac{1}{\pi(a+b)}\left[4\ \frac{  \pi(a+b)\cos(\pi(a+b))-\sin(\pi(a+b)) }{\pi^2(a+b)^2}-\pi J_2(\pi(a+b))\right],
%\end{equation}
%for $a,\ b<0$.
%It is worth noticing that Eq. (\ref{crit+}) can also be obtained by integrating Eq. (\ref{g2}) between
%$\mu'=0$ and $\mu'=\mu$, this is because there are no states for $\mu'<0$. This gives the alternative, more useful, representation of the kernel
%\begin{equation}
%\kappa_c(a,b) = \int_0^1 dv\ \sqrt{1-v}\exp(-\sqrt{v}\pi (a+b))\label{rep2}
%\end{equation}
%
%With the above notation we find that the local density relative to the bulk is given by
%\begin{equation}
%\frac{\rho(x)}{\rho_b}= n_c(\frac{x}{\ell}),
%\end{equation}
%where $n(a)=\kappa_c(a,a)$. 

The behavior of $n_c(a)$ is shown in Fig. (\ref{densc}). At the barrier we find that $n_c(a)$ and its derivative are continuous (as they must be). In addition we have $n_c(0)=2/3$. 

It is interesting to study the total number of fermions $N_R$ outside the well, i.e. on the right side $x>0$, together
with its fluctuations. Using Eq. (\ref{ns}) we see that its average is given by
\begin{equation}
\langle N_R \rangle  = \int_0^{+\infty} \rho(x) dx = \int_0^\infty n_c(a) da = 
\int_0^1 \frac{dv}{2 \pi \sqrt{v}} \sqrt{1-v} 
= \frac{1}{4},\label{avenout}
\end{equation}
where we used the representation \eqref{rep2} of the kernel. 
It is finite, and independent of $\mu$, which may be surprising a priori. This independence on $\mu$ is an interesting consequence 
of the scaling form \eqref{eq:density_scaling} %\eqref{denscal},
together with the convergence of the integral of $n_c(a)$ at large $a$. Recall that as soon as $\mu>V_0$, the mean number of fermions on the right, 
$\langle N_R \rangle$, becomes infinite (with a density $\rho_R = \sqrt{2\mu-2V_0}/\pi$).

Using Eq. (\ref{vs}), and again the kernel representation Eq. (\ref{rep2}), we find that the variance of $N_R$ is given by
\begin{eqnarray}
\label{varnout}
{\rm Var}(N_{R}) &=& \langle  N_{R}\rangle -\int_0^\infty da db \  \kappa_c(a,b)^2
= \frac{1}{4} - \frac{1}{\pi^2} \int_0^1 dv \int_0^1 du \frac{\sqrt{1-v}\sqrt{1-u}}{(\sqrt{u}+ \sqrt{v})^2}  = \frac{2}{\pi^2},
\end{eqnarray}
and where computation of the last  integral above is explained in Appendix \ref{integral}.
Given the small value of $\langle N_R\rangle$ it would be tempting to assume that the number of fermions outside the well has a Bernoulli distribution, {\em i.e.} $N_{R}=1$ with probability $p$ and $N_{R}=0$ with probability $q=1-p$. This would imply from Eq. (\ref{avenout}) that $p=1/4$ and so ${\rm Var}(N_{R})=pq=3/16=0.1875$. However our exact calculation finds 
${\rm Var}(N_{R})=\frac{2}{\pi^2} = 0.202642$. Hence the actual random variable $N_R$ fluctuates more than a Bernoulli random variable, and the above result shows that the probability that $N_R \geq 2$ is strictly nonzero. {A similar calculation for the third cumulant (see Appendix \ref{appendixThirdCumulant} for the details) gives $\left\langle N_{R}^{3}\right\rangle _{c}=0.1281169\dots$.  Here too one can see the difference from the Bernoulli distribution, whose third cumulant is $p\left(1-p\right)\left(1-2p\right)$, which for $p=1/4$ would give $3/32=0.09375$.}

\begin{figure}[t!]
\begin{center}
  \includegraphics[scale=0.50]{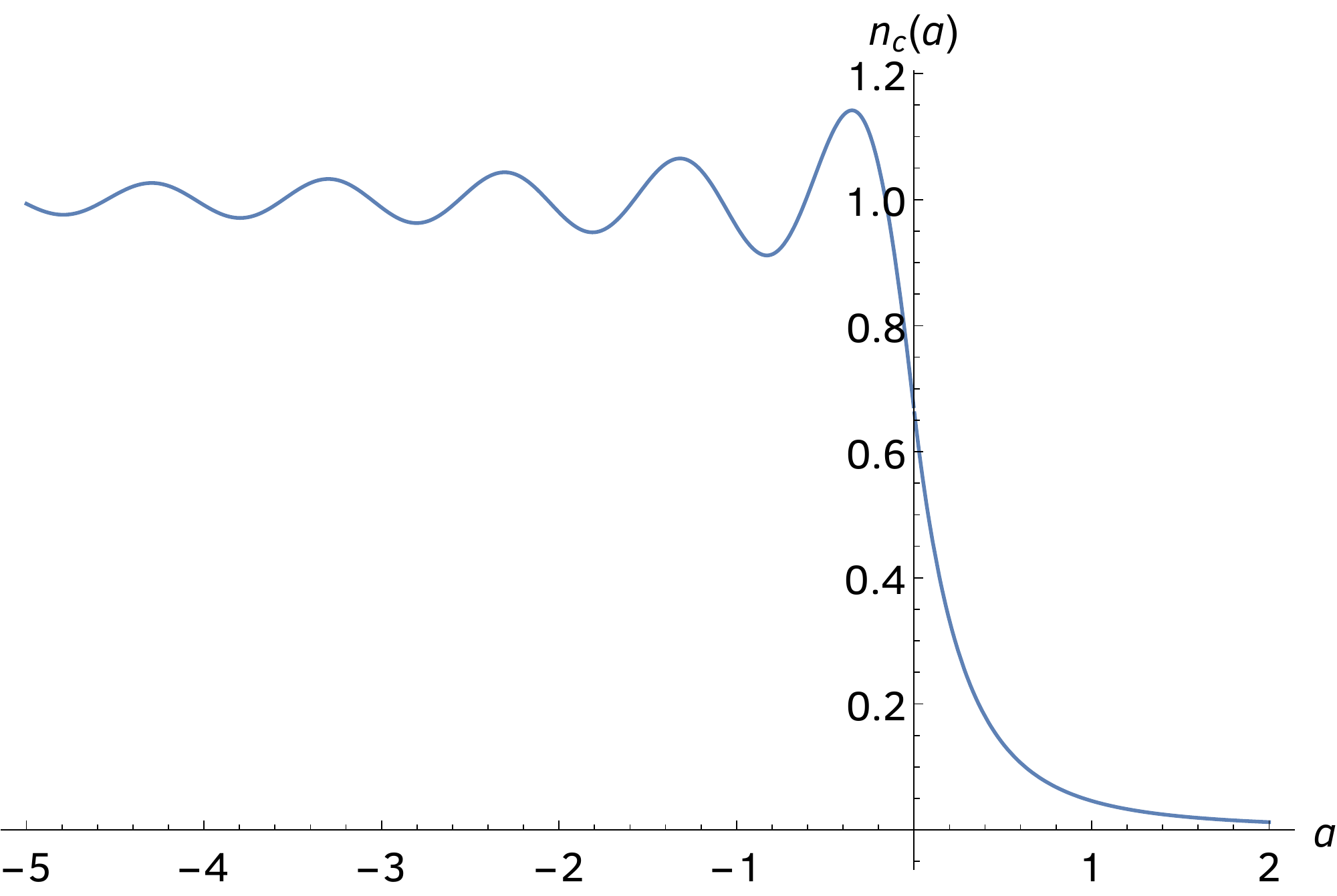} 
 \caption{The behavior of  $n_c(a)$ vs $a$. The density normalised by the bulk density, at the critical point where $\mu=\mu_c=V_0$.}\label{densc}
\end{center}
\end{figure}

\subsection{Kernel in the sub-critical regime}

We now consider the case where $\mu<V_0$. The physics here has similarities with the critical case, with additional off-critical features near the transition, which we now analyse. We will  
focus on the region $x,\ y>0$.
We integrate Eq. (\ref{g2}) over $\mu'$ from $\mu'=0$ to $\mu'=\mu$ and obtain
that the kernel takes the scaling form \eqref{eq:kernel_scaling}
%\begin{equation}
%K_\mu(x,y) = \frac{1}{\ell} \kappa_r(\frac{x}{\ell},\frac{y}{\ell}) \quad , \quad \frac{1}{\ell} = \rho_L = \frac{\sqrt{2 \mu}}{\pi}  \quad , \quad x,y >0 \
%\end{equation}
in terms of the reduced kernel, for $r = \frac{V_0}{\mu}>1$
\begin{equation} \label{kappar} 
\kappa_r(a,b) = \frac{1}{r}\int_0^1 dv\  \sqrt{v}\exp\left(-\pi\sqrt{r-v}(a+b)\right)  \quad , \quad a,b >0 .
\end{equation}
The critical kernel $\kappa_c(a,b)$ in \eqref{rep2} is recovered setting $r=1$. 
%The density takes again the scaling form 
%\begin{equation}
%\rho(x) = \frac{1}{\ell} \, n_r(\frac{x}{\ell}) \quad , \quad \frac{1}{\ell} = \rho_L = \frac{\sqrt{2 \mu}}{\pi} \quad , \quad n_r(a)= \kappa_r(a,a). \label{denscal2} 
%\end{equation}
The density takes the scaling form \eqref{eq:density_scaling} where
\begin{equation}
n_r(a)= \kappa_r(a,a) = n_{r}\left(a\right)=\kappa_{r}\left(a,a\right)=\frac{1}{r}\int_{0}^{1}dv\ \sqrt{v}\exp\left(-2\pi a\sqrt{r-v}\right). %\label{denscal2} 
\end{equation}
Remarkably, $\kappa_r(a,b)$ is only a function of the sum $a+b$. In particular, it satisfies 
$\kappa_r(a,b) = n_r((a+b)/2)$.
The rescaled density at $a=0$ is given by $n_r(0)=\frac{2}{3r}$. 
One can write this result together with Eq.~\eqref{eq:rho_at_zero_supercritical}, in the form 
\be
\label{eq:g_of_r}
\rho\left(0\right)=\frac{2\rho_{L}g\left(r\right)}{3}, \qquad
g\left(r\right)=\begin{cases}
1/r & r>1\\
\frac{2-r+\sqrt{1-r}}{1+\sqrt{1-r}} & 0<r<1
\end{cases}. 
\ee
In the vicinity of $r=1$, the function $g(r)$ exhibits the singular behavior
\be \label{gofr}
g\left(r\right)=\begin{cases}
1+\left(1-r\right)+\left(1-r\right)^{2}+\dots & r-1\ll1\\[0.1cm]
1+\left(1-r\right)-\left(1-r\right)^{3/2}+\left(1-r\right)^{2}+\dots & 1-r\ll1
\end{cases}.
\ee
The function $g(r)$ is plotted in Fig.~\ref{fig:g_of_r}.
\begin{figure}[t!]
\begin{center}
  \includegraphics[scale=0.50]{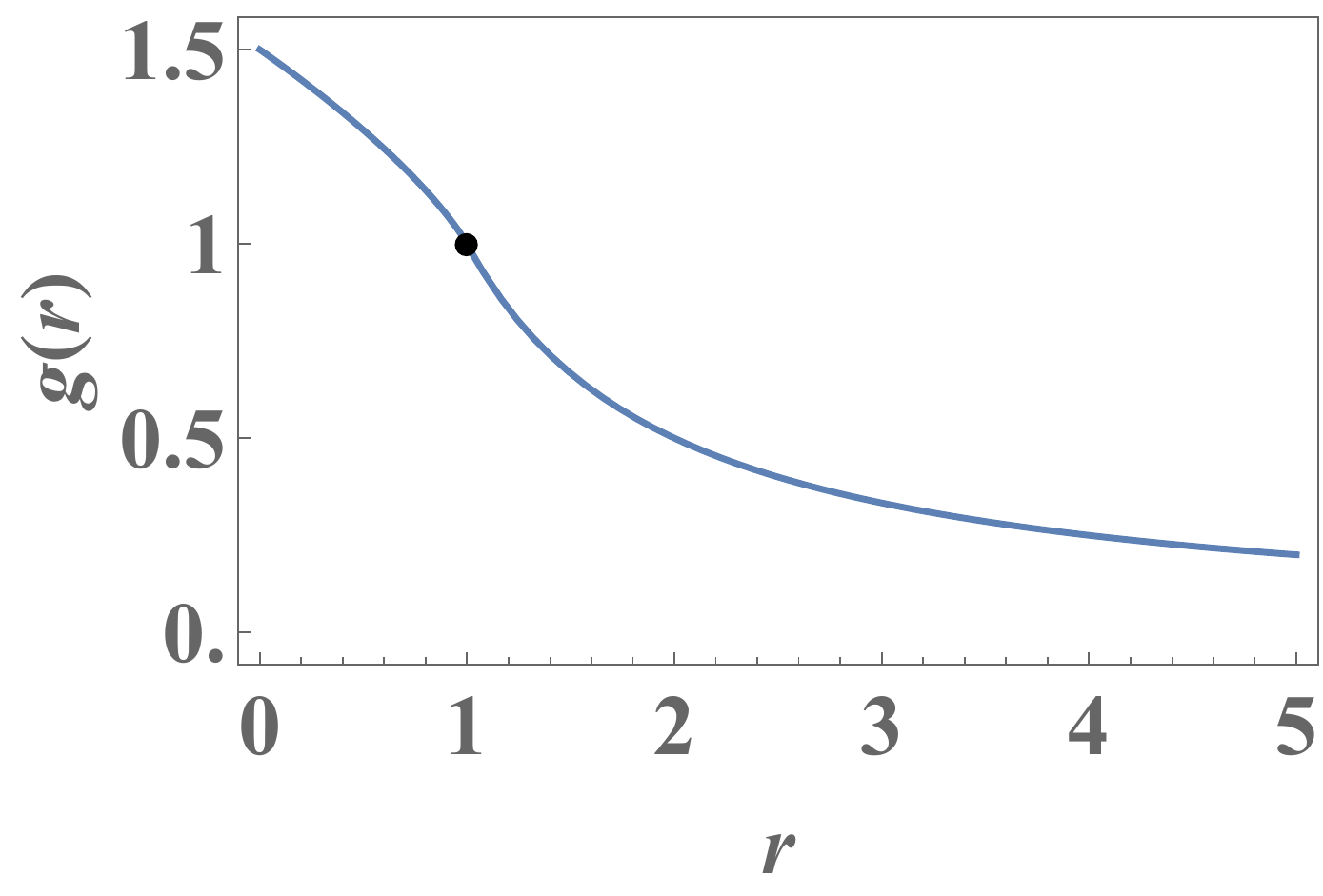} 
 \caption{The function $g(r)$ that describes the density at $x=0$, see Eq.~\eqref{eq:g_of_r}. The marked dot is the point $g(r=1)=1$ which corresponds to the critical case. {Note that $g(r)$ is non-analytic at this point $r=1$ [see Eq. (\ref{gofr})].}}\label{fig:g_of_r}
\end{center}
\end{figure}
For large $a>0$ and fixed $r>1$ one finds that the density decays exponentially
\be \label{decay1} 
n_{r}(a)\simeq\left(\frac{\sqrt{r-1}}{\pi ar}+O\left(\frac{1}{a^{2}}\right)\right)\exp\left(-2\pi a\sqrt{r-1}\right)\quad,\quad a\to+\infty.
\ee 
Hence for $r=V_0/\mu>1$ the density $\rho(x)$ decays exponentially for $x>0$, with a decay length $\xi_r=\ell/(2 \pi \sqrt{r-1})$ which diverges at the transition
with a square root singularity. Around the transition, in the double limit $r \to 1$ and $a \to +\infty$, with
$a \sqrt{1-r}$ fixed, it is easy to see by performing the change of variable
$v = 1 - (r-1) w$ in the integral \eqref{kappar}, that the reduced density takes the scaling form 
\be \label{nu} 
n_r(a) \simeq (r-1) \, \nu(2 \pi a \sqrt{1-r}) \quad , \quad \nu(\tilde a)= \frac{2}{\tilde a^2} (1+\tilde a) \exp\left(- \tilde a\right)
\ee
which describes the crossover between the exponential decay \eqref{decay1} and the
algebraic decay \eqref{alg1} at criticality. 

In this regime $V_0=r \mu$, $r>1$, we find that the average number of particles outside the well is given by
\begin{equation}
\label{eq:NR_subcritical}
\langle N_{R}\rangle=\int_{0}^{\infty}n_{r}(a)da=\frac{1}{2\pi r}\left[r\sin^{-1}\left(\frac{1}{\sqrt{r}}\right)-\sqrt{r-1}\right]=\begin{cases}
\frac{1}{4}-\frac{\sqrt{r-1}}{\pi}+O((r-1)^{3/2})\quad, & r\to1^{+}\\[0.2cm]
\frac{1}{3\pi r^{3/2}}+O(r^{-5/2})\quad, & r\to+\infty
\end{cases}
\end{equation}
which tends to zero as a power law for large $r$ and to $1/4$ as $r\to1$. 
Eq.~\eqref{eq:NR_subcritical} is plotted in Fig. \ref{fig:mean_and_var_NR} together with its asymptotics.
The variance is given 
by 
\be
\label{eq:var_NR_subcritical}
{\rm Var}\left(N_{R}\right)=\left\langle N_{R}\right\rangle -\frac{1}{\pi^{2}r^{2}}\int_{0}^{1}dv\int_{0}^{1}du\frac{\sqrt{v}\sqrt{u}}{\left(\sqrt{r-u}+\sqrt{r-v}\right)^{2}} =\left\langle N_{R}\right\rangle -F\left(\frac{r-1}{r}\right)
\ee

where 
\be
F\left(A\right)=\frac{\cos^{-1}\left(\sqrt{A}\right)\left[2\left(1-2A\right)\sqrt{A\left(1-A\right)}+\cos^{-1}\left(\sqrt{A}\right)\right]-\left(A-1\right)\left(3A+2A\ln A-2\right)}{\pi^{2}}\label{fint},
\ee
and the computation is explained in Appendix \ref{integral}.

As $r \to +\infty$ the second term {in Eq. (\ref{eq:var_NR_subcritical})} behaves as $\simeq -1/(9 \pi^2 r^3)$ and is thus small
compared to the first one. Consequently we see that $ N_R$ becomes a Bernoulli random variable in this limit, because  it satisfies ${\rm Var}\left(N_{R}\right)\simeq\left\langle N_{R}\right\rangle -\left\langle N_{R}\right\rangle ^{2}$ {[see the second line of Eq. (\ref{eq:NR_subcritical})]}.
\begin{figure}[t!]
\begin{center}
\includegraphics[scale=0.48]{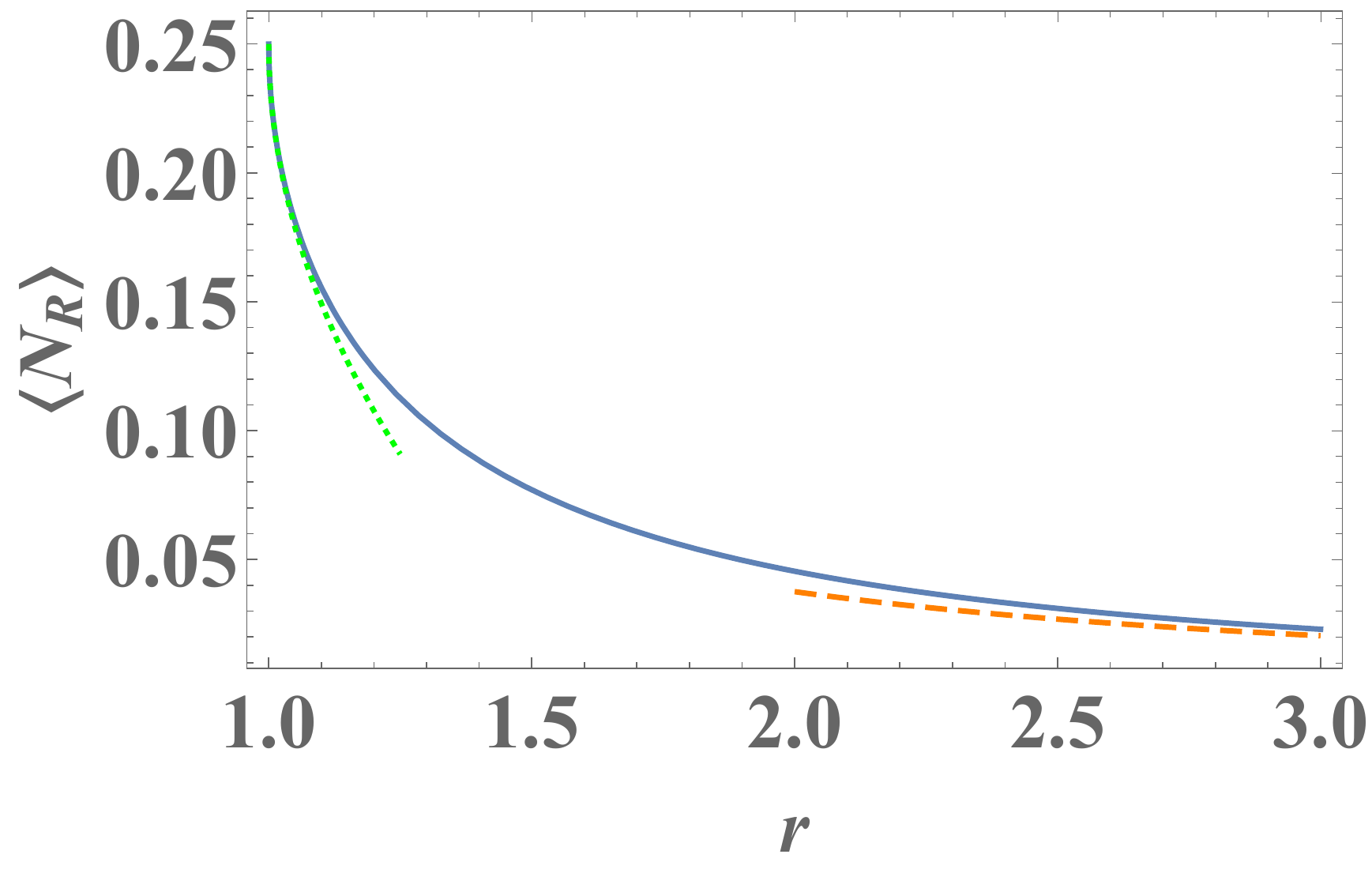} 
\includegraphics[scale=0.48]{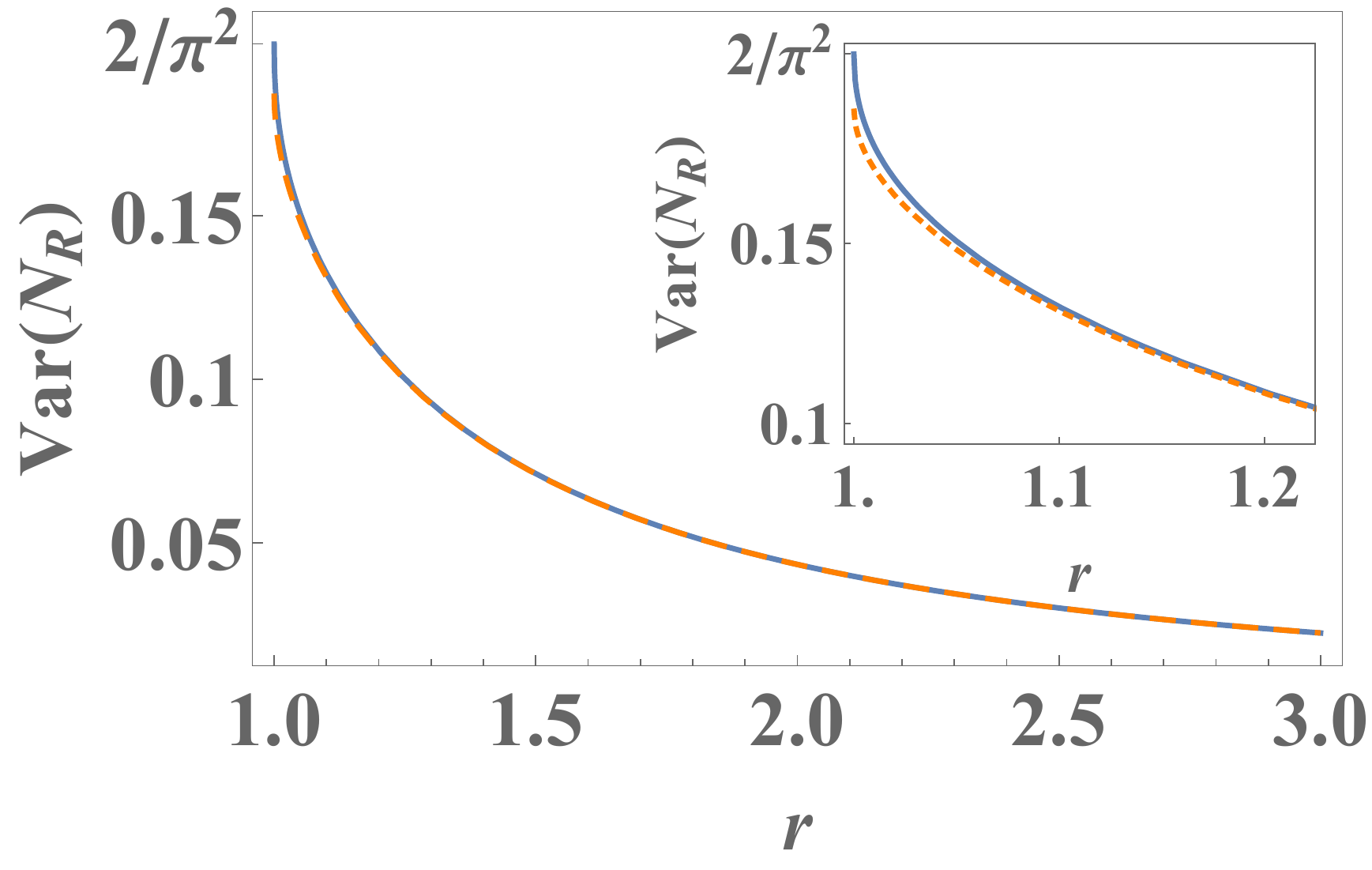} 
 \caption{ (a) The mean number of particles on the right side of the step $N_R$, see Eq.~\eqref{eq:NR_subcritical} (solid line), plotted together with its $r\to1^+$ and $r\gg1$ asymptotic behaviors (dotted and dashed lines respectively).
 (b) The variance of $N_R$, see Eq. \eqref{eq:var_NR_subcritical} (solid line), together with its $r\gg1$ approximation $\text{Var}\left(N_{R}\right)\simeq\left\langle N_{R}\right\rangle -\left\langle N_{R}\right\rangle ^{2}$ (dashed line) that corresponds to a Bernoulli random variable. Inset is a zoom in on the regime $r\simeq1$ where the result clearly deviates from the large-$r$ approximation.
At $r=1$ the mean and variance are $1/4$ and $2/\pi^2$ respectively.}
 \label{fig:mean_and_var_NR}
\end{center}
\end{figure}

The kernel $K_\mu(x,y)$ in the other regions $x,y<0$ and $x>0>y$, not displayed here, are obtained 
by integrating \eqref{vbigleft}  and \eqref{vbigright}, respectively, for $\mu'$ from
$0$ to $\mu$, and can be studied similarly.

%\newpage

\section{Smooth barrier, universality and scattering amplitudes}
\label{smooth}
{In this section we  consider  potentials which become asymptotically constant far from the origin. Without loss of generality we write 
\begin{equation}
V(x)=0 \ {\rm as}\ x\to -\infty; \ V(x) = V_0\  {\rm as}\ x\to \infty,
\end{equation}
where the convergence occurs beyond a typical scale which we call the {\em barrier width}. The discontinuous step potential is thus a special case which corresponds to zero barrier width. In what follows we show how the kernel can be obtained from scattering solutions, first via the Green's function method and then by a direct summation of the eigenfunctions.
We obtain formulas for the kernel valid at distances larger than the barrier width,
and for a general barrier potential, in terms of the coefficients of the scattering solutions. 
These formula recover the exact result in the case of the discontinuous barrier. 
In a second part we give the exact solution for a special {smoothened} step potential, known as the Woods-Saxon potential in the context of nuclear physics, and show the convergence to the aforementioned large distance 
formula. In that part we identify which features of the transition at $\mu=V_0$
are universal, i.e. independent of the details of the shape of the barrier.}

\subsection{General representation for a barrier in terms of scattering solutions}\label{gfg}

Here we determine the Green's function for a general barrier, and from it we derive the kernel.
In general in one dimension the Green's function can be written as
\begin{equation}
G_{\mu'}(x,y) = A_{R{\mu'}}(y)\phi_{R{\mu'}}(x) \ {\rm for }\ x>y,\ \ \ \  G_{\mu'}(x,y) = A_{L{\mu'}}(y)\phi_{L{\mu'}}(x) \ {\rm for }\ y>x;
\end{equation}
where $\phi_{R{\mu'}}(x)$ and $\phi_{L{\mu'}}(x)$ and  are solutions to the homogeneous equation $({\mu'}-i0^+-H)\phi_{{\mu'}}(x)=0$ respecting the boundary conditions $\phi_{R/L{\mu'}}(x)\to 0$ as $x\to-\infty$ (L) and as $x\to+\infty$ (R).
Now matching the solutions at $x=y$ gives the result
\begin{equation}
G_{\mu'}(x,y) =-\frac{ 2\phi_{R{\mu'}}(x)\phi_{L{\mu'}}(y)}{W} \ {\rm for }\ x>y,\ \ \ \  G_{\mu'}(x,y)=-\frac{ 2\phi_{R{\mu'}}(y)\phi_{L{\mu'}}(x)}{W} \ {\rm for }\ y>x, \label{match} 
\end{equation}
where $W \equiv W[\phi_{R{\mu'}},\phi_{L{\mu'}}]= \phi_{R{\mu'}}(x)\phi'_{L{\mu'}}(x)- \phi_{L{\mu'}}(x)\phi'_{R{\mu'}}(x)$ is the Wronskian and is constant for differential equations having no first derivative term as is the case here.
However as the potential becomes constant away from the barrier, we must find the bulk solutions which are given up to a constant prefactor by 
 \begin{equation}
\phi_{L{\mu'}}(x)\sim\exp\left(i\sqrt{2{\mu'}-i0^{+}}\ x\right),\ {\rm as}\ x\to-\infty,\label{as0}
\end{equation}
and 
\begin{equation}
\phi_{R{\mu'}}(x)\sim\exp\left(-i\sqrt{2{\mu'}-i0^{+}-2V_{0}}\ x\right),\ {\rm as}\ {x\to+\infty,} \label{as1} 
\end{equation}
in the case ${\mu'}>V_0$ and 
\begin{equation}
\phi_{R{\mu'}}(x)\sim\exp\left(-\sqrt{2V_{0}-2{\mu'}+i0^{+}}\ x\right),{\rm as}\ {x\to+\infty,}
\end{equation}
in the case { ${\mu'}<V_0$.} This means that if we  write $k=\sqrt{2{\mu'}-i0^+-2V_0}$, the region $2{\mu'}-i0^+-2V_0>0$ gives values of $k$ along the positive real axis whereas when $2{\mu'}-2V_0<0$ we obtain $k=-ik'$ where $k'>0$.\\

{\noindent \bf The  Green's function for ${\mu'}>V_0$ and the supercritical case}. 
\vspace*{.3cm}

In the case ${\mu'}>V_0$, we can use the solutions $\psi_{k_1}(x)$ of the Schr\"odinger equation $(k_1^2/2 - H)\psi_{k_1} = 0$  corresponding to the
energy $\mu'=k_1^2/2$ which have the asymptotic form of plane waves far away from the barrier. For potential barriers, eigenstates are often expressed in terms of the scattering of a plane wave coming from the left of the barrier. The incoming momentum to the left of the barrier is $k_1=\sqrt{2{\mu'}-i0^+}$ and the outgoing momentum to the right is $k_2=\sqrt{2{\mu'}-i0^+-2V_0}$ (we added the $-i 0^+$
for future convenience when discussing the Green's function). Such a plane wave is partially transmitted and one writes
\bea \psi_{k_1}(x) = 
\begin{cases}
&\exp\left(i k_1 x\right)+ \frac{C_2(k_1,k_2)}{C_1(k_1,k_2)} \exp\left(-ik_1x\right) \;, \;  x \to - \infty \\
& \\
& \frac{1}{C_1(k_1,k_2)} \exp\left(i k_2 x\right) \;, \;\hspace*{1.1cm} x \to +\infty  \;,
\label{asleft}\end{cases}
\eea
with ${\mu'}=\frac{ k_1^2}{2}= \frac{ k_2^2}{2}+V_0$. Here $\frac{C_2(k_1,k_2)}{C_1(k_1,k_2)}$ is the reflection amplitude and $\frac{1}{C_1(k_1,k_2)}$ the transmission amplitude, both of which depend on the precise form of the barrier. The reflection probability is then given by
\begin{equation}
R(k_1,k_2) =\left|\frac{C_2(k_1,k_2)}{C_1(k_1,k_2)} \right|^2,
\end{equation}
and the transmission probability is $T(k_1,k_2)=1-R(k_1,k_2)$.

The above wave function clearly does not satisfy the correct boundary conditions for either $\phi_{R{\mu'}}$ or $\phi_{L{\mu'}}$ however, comparing with Eq. \eqref{as1}, we see that we can write
\begin{equation}
\phi_{R{\mu'}}(x)=\psi_{k_1}^*(x),
\end{equation}
as, if $\psi_{k_1}(x)$ is an eigenfunction then so is $\psi^*_{k_1}(x)$, one just has to verify that it is not the same eigenfunction. Now if we write $\phi_{L{\mu'}}(x)= \psi_{k_1}(x)
+B \psi^*_{k_1}(x)$, the asymptotic condition given in Eq. \eqref{as0}, i.e. $\phi_{L{\mu'}}(x)\sim \exp\left(i k_1 x\right)$, as $x\to-\infty$ then gives
\begin{equation}
\left[\exp\left(ik_{1}x\right)+\frac{C_{2}(k_{1},k_{2})}{C_{1}(k_{1},k_{2})}\exp\left(-ik_{1}x\right)\right]+B\left[\exp\left(-ik_{1}x\right)+\frac{C_{2}^{*}(k_{1},k_{2})}{C_{1}^{*}(k_{1},k_{2})}\exp\left(ik_{1}x\right)\right]=A\exp\left(ik_{1}x\right),
\end{equation}
where $A$ is a constant. Note that since $k_1$ and $k_2$ are real we denote $C^*_j(k_1,k_2) := ( C_j(k_1,k_2) )^*$ [see the discussion around Eq. (\ref{conjugate})]. This yields $B=-C_2(k_1,k_2)/C_1(k_1,k_2)$ and thus
\begin{equation}
\phi_{L{\mu'}}(x)=\psi_{k_1}(x)-\frac{C_2(k_1,k_2)}{C_1(k_1,k_2)}\psi^*_{k_1}(x).
\end{equation}
From this we find that the Wronskian is given by $W[\phi_{{\mu'} R},\phi_{{\mu'} L}] = W[\psi^*_{k_1},\psi_{k_1}]$. However as this is a constant we can evaluate it in the regions $x\to+\infty$ where its is known, i.e., 
\begin{equation}
W[\psi^*_{k_1},\psi_{k_1}]= \frac{2ik_2}{C_1(k_1,k_2)C_1^*(k_1,k_2)}.
\end{equation}
On the other hand, the evaluation of the Wronskian as $x\to-\infty$ yields
\begin{equation}
W[\psi_{k_{1}}^{*},\psi_{k_{1}}]=2ik_{1}\left[1-\frac{C_{2}(k_{1},k_{2})C_{2}^{*}(k_{1},k_{2})}{C_{1}(k_{1},k_{2})C_{1}^{*}(k_{1},k_{2})}\right] \, .
\end{equation}
Equating the two expressions for the Wronskians yields the Wronskian identity, which is the well known formula corresponding the the conservation of current,
\begin{equation}
\frac{k_{2}}{|C_{1}(k_{1},k_{2})|^{2}}=k_{1}\left[1-\frac{|C_{2}(k_{1},k_{2})|^{2}}{|C_{1}(k_{1},k_{2})|^{2}}\right]\, ,\label{wid1}
\end{equation}
which can be written as 
\begin{equation}
T(k_1,k_2)=1-R(k_1,k_2) = \frac{k_2}{k_1|C_1(k_1,k_2)|^2} \, .\label{tr}
\end{equation}
From the above relation we see that
when $k_2$ vanishes $\psi_{k_1}^*$ and $\psi_{k_2}$ correspond to the same wave function as the Wronskian vanishes - physically this is due to total reflection of the incoming wave. The case $k_2^2<0$ must thus be treated separately and will be considered at the end of this section.

Putting all the above results together and using Eq. \eqref{match}, we obtain, for $x>y$,
\begin{equation}
G_{\mu'}(x,y) = \frac{i C_1(k_1,k_2) C^*_1(k_1,k_2)}{k_2}\psi_{k_1}^*(x)\left[\psi_{k_1}(y) - \frac{C_2(k_1,k_2)}{C_1(k_1,k_2)}\psi^*_{k_1}(y)\right],\label{gsoft}
\end{equation}
where we emphasize again that here, for $\mu'>V_0$, $k_1,k_2$ are both real and positive.
This is an explicit formula for the Green's function in terms of the scattering eigenstates.

Although the function {$\psi_{k_1}(x)$} is not necessarily known everywhere, its asymptotics can be read of from
Eq. (\ref{asleft}).  As $x,\ y\to +\infty$, with $x>y$, we find 
\begin{equation} \label{Gapprox} 
 \ G_{\mu'}(x,y)\approx \frac{i}{k_2}\exp(-ik_2(x-y))-i\frac{C_2(k_1,k_2)}{k_2C^*_1(k_1,k_2)}\exp(-ik_2(x+y)) \;.
\end{equation}
This gives, using Eq. (\ref{repi})
\be \label{Kmuprime}
K_\mu(x,y) \approx \frac{\sin(k_{2F}(x-y))}{\pi (x-y)}+\frac{1}{\pi}{\rm Im}\int_{k_{2F}}^{\infty} idk_2 \ \frac{C_2(\sqrt{k_2^2+2V_0},k_2)}{C^*_1(\sqrt{k_2^2+ 2V_0},k_2)}\exp(-ik_2(x+y))
\ee 
where $k_{2F}= \sqrt{2\mu-2V_0}$ and where we have used $d \mu' = k_2 dk_2$. 
%Note that {\red Naftali: It seems strange to me to refer here to this equation here (because it is in the appendix). Should we refer to \eqref{Kmuprime} instead?} 
\eqref{Kmuprime} should be valid for $x,y>0$ much larger than the {\em barrier width} when the asymptotics for above the wave-functions hold. The notion of a {\em barrier width} will be quantified in the next Section (and called $\lambda$) in the concrete example of the Woods-Saxon potential. In addition, one can perform a second asymptotics if furthermore $x+y \gg 1/k_{2F}$
(a scale which can become much larger than the barrier width near criticality, here the integral is dominated by the vicinity of $k_2 \approx k_{2F}$ and we obtain (see Appendix \ref{contsec} for details)
\begin{equation} \label{Kmu_asympt}
K_\mu(x,y) \approx \frac{\sin(k_{2F}(x-y))}{\pi (x-y)}-\frac{|C_2(k_{1F},k_{2F})|}{|C_1(k_{1F},k_{2F})|}\frac{\sin(k_{2F}(x+y)-\phi_{12})}{\pi(x+y)},
\end{equation}
where $\phi_{12}=\arg(C_2(k_{1F},k_{2F})/C^*_1(k_{1F}, k_{2F}))$. We thus see that at large distances from the barrier, the kernel takes the bulk sine-kernel form, plus an oscillatory correction with an amplitude that decays algebraically with distance from the barrier.

{
In the region to the left of the barrier as $x,\ y\to-\infty$ one finds
\begin{eqnarray}
G_{\mu'}(x,y) &=& i\frac{C_1^*(k_1,k_2) C_1(k_1,k_2)}{k_2}\left[1 -\frac{C_2^*(k_1,k_2) C_2(k_1,k_2)}{C_1^*(k_1,k_2) C_1(k_1,k_2)}\right]\left(\exp(-ik_1(x-y)) + \frac{C_2^*(k_1,k_2)}{C_1^*(k_1,k_2)}\exp(ik_1(x+y))\right)\nonumber \\
&=&
\frac{i}{k_1}\left(\exp(-ik_1(x-y)) + \frac{C_2^*(k_1,k_2)}{C_1^*(k_1,k_2)}\exp(ik_1(x+y))\right),\label{negxy}
\end{eqnarray}
where we have used the Wronskian identity Eq. (\ref{wid1}). Using Eq. (\ref{repi}), it leads to
\be \label{Kmuprime2}
K_\mu(x,y) \approx \frac{\sin(k_{1F}(x-y))}{\pi (x-y)}- \frac{1}{\pi}{\rm Im}\int_{k_{1F}}^{\infty} idk_1 \ \frac{C^*_2(k_1,\sqrt{k_1^2 + 2 V_0})}{C^*_1(k_1,\sqrt{k_1^2+ 2V_0})}\exp(ik_1(x+y))
\ee 
where $k_{1F}=\sqrt{2 \mu}$. A similar calculation as above (see Appendix \ref{contsec}) then yields
\begin{equation}
K_\mu(x,y) \approx \frac{\sin(k_{1F}(x-y))}{\pi (x-y)}- \frac{|C_2(k_{1F},k_{2F})|}{|C_1(k_{1F},k_{2F})|}\frac{\sin(k_{1F}(x+y)-\phi'_{12})}{\pi|x+y|}.\label{kas1}
\end{equation}
where $\phi'_{12}= \arg(C_2(k_{1F},k_{2F})/C_1(k_{1F},k_{2F}))$. 
The above formula agrees with Eq. 
\eqref{asymptdens} for the density of the square well when one sets $x=y$ and uses the corresponding scattering coefficients given in Eq. (\ref{G4}).}

\medskip

{\noindent \bf The Green's function for ${\mu'}<V_0$ and the subcritical case.}

\vspace*{0.3cm}

We now turn to the case where $k_2^2<0$, here two solutions can be found by analytic continuation of the previous solutions. Recalling that
$k_2=\sqrt{2{\mu'}-i0^+-2V_0}$ where the positive root is taken we see that when $V_0>{\mu'}$ the continuous branch of square root is $k_2= -i\kappa_2=-i\sqrt{ 2V_0 -2{\mu'} +i0^+}$ as this choice has a positive real part and 
negative imaginary part in the region where ${\mu'}\approx V_0$. The analytic continuation of the Green's function given in Eq. (\ref{gsoft}) is then, for $x>y$
\begin{equation}
G_{\mu'}(x,y) = \frac{- C^*_1(k_1,-i\kappa_2)C_1(k_1,-i\kappa_2)}{\kappa_2}\psi_{k_1}^{(2)}(x)\left[\psi_{k_1}^{(1)}(y) - \frac{C_2(k_1,-i\kappa_2)}{C_1(k_1,-i\kappa_2)}\psi^{(2)}_{k_1}(y)\right],\label{cgf}
\end{equation}
%where in the above formula a function $f(k_1,k_2)$ the complex conjugate $f^*(k_1,k_2)$ is computed at 
%$k_1$ and $k_2$ real and $f^*(k_1,-i\kappa_2)\equiv f^*(k_1,k_2=-i\kappa_2)$. However from this definition we see that one can write $f^*(k_1,-i\kappa_2)=\overline f(k_1,i\kappa_2)$, where $\overline\cdot$ denotes the complex conjugate. 
where here and below we denote the complex conjugate function 
\be \label{conjugate}
%f^{*}(k_{1},-i\kappa_{2}):=\left(f(k_{1},k_{2})|_{k_{1},k_{2}\in\mathbb{R}}\right)^{*}|_{k_{2}\to-i\kappa_{2}}\;.
f^{*}(k_{1},-i\kappa_{2}):=\left.\left(\left.f(k_{1},k_{2})\right|_{k_{1},k_{2}\in\mathbb{R}}\right)^{*}\right|_{k_{2}\to-i\kappa_{2}}\;.
\ee 
In other words $f^*(k_1,k_2)$ is calculated by taking first the complex conjugate of $f(k_1,k_2)$ with $k_1,k_2$ real, and then performing analytical continuation in $k_2$. Note that this is the
usual definition, which satisfies $( f(z,w) )^* = f^*(z^*,w^*)$, e.g if $f(z,w)=a z + b w$ then
$f^*(z,w)=a^* z + b^* w$.

In Eq. (\ref{cgf}) we have denoted the analytic continuation of  $\psi_{k_1}$ by $\psi_{k_1}^{(1)}$ and $\psi_{k_1}^*$ by  $\psi_{k_1}^{(2)}$. The analytic continuations have the asymptotic forms
%\bea
%\psi_{k_1}^{(1)}(x) = 
%\begin{cases}
%&\exp\left( i k_1 x\right)+ \frac{C_2(k_1,-i\kappa_2)}{C_1(k_1,-i\kappa_2)} \exp\left(-ik_1x\right) \;, \;  x \to - \infty \\
%& \\
%& \frac{1}{C_1(k_1,-i\kappa_2)} \exp\left(\kappa_2 x\right) \;, \;\hspace*{1.1cm} x \to +\infty  \;,
%\end{cases}
%\eea
\be
\psi_{k_{1}}^{(1)}(x)=\begin{cases}
\exp\left(ik_{1}x\right)+\frac{C_{2}(k_{1},-i\kappa_{2})}{C_{1}(k_{1},-i\kappa_{2})}\exp\left(-ik_{1}x\right)\;, & x\to-\infty\\[0.2cm]
%\\[0.1mm]
\frac{1}{C_{1}(k_{1},-i\kappa_{2})}\exp\left(\kappa_{2}x\right)\;, & x\to+\infty\;,
\end{cases}
\ee
and 
%\bea 
%\psi^{(2)}_{k_1}(x) = 
%\begin{cases}
%&\exp\left(-i k_1 x\right)+ \frac{C_2^*(k_1,- i\kappa_2)}{C_1^*(k_1,- i\kappa_2)} \exp\left(ik_1x\right) \;, \;  x \to - \infty \\
%& \\
%& \frac{1}{C_1^*(k_1,- i\kappa_2)} \exp\left(-\kappa_2 x\right) \;, \;\hspace*{1.1cm} x \to +\infty  \;.
%\end{cases}. \label{psi2} 
%\eea
\be
\psi_{k_{1}}^{(2)}(x)=\begin{cases}
\exp\left(-ik_{1}x\right)+\frac{C_{2}^{*}(k_{1},-i\kappa_{2})}{C_{1}^{*}(k_{1},-i\kappa_{2})}\exp\left(ik_{1}x\right)\;, & x\to-\infty\\[0.2cm]
%\\
\frac{1}{C_{1}^{*}(k_{1},-i\kappa_{2})}\exp\left(-\kappa_{2}x\right)\;, & x\to+\infty\;.
\end{cases}. \label{psi2} 
\ee
It is easy to see that these two functions used to construct the Green's function have the Wronskian 
identity 
\begin{equation}
W[\psi_{k_{1}}^{(2)},\psi_{k_{1}}^{(1)}]=\frac{2\kappa_{2}}{C_{1}(k_{1},-i\kappa_{2})C_{1}^{*}(k_{1},-i\kappa_{2})}=2ik_{1}\left[1-\frac{C_{2}(k_{1},-i\kappa_{2})C_{2}^{*}(k_{1},-i\kappa_{2})}{C_{1}(k_{1},-i\kappa_{2})C_{1}^{*}(k_{1},-i\kappa_{2})}\right].\label{w2}
\end{equation}

%{\tiny
%\begin{equation}
%\frac{i\kappa_2}{C_1(k_1,-i\kappa_2)\overline C_1(k_1,i\kappa_2)}= 2 i k_1(\frac{C_2(k_1,-i\kappa_2)\overline C_2(k_1,i\kappa_2)}{C_1(k_1,-i\kappa_2)\overline C_1(k_1,i\kappa_2)}-1).\label{w2}
%\end{equation}
%}

Now returning to the Green's function, in region $x\ ,y \to \infty$ we find, for $x>y$, 
\begin{equation}
G_{\mu'}(x,y) \approx -\frac{1}{\kappa_2}\exp(-\kappa_2(x-y))+ \frac{1}{\kappa_2}\frac{C_2(k_1,-i\kappa_2)}{C_1^*(\kappa_1, - i\kappa_2)}\exp(-\kappa_2(x+y)),
\end{equation}
%{\tiny \begin{equation}
%G_{\mu'}(x,y) \approx -\frac{1}{\kappa_2}\exp(-\kappa_2(x-y))+ \frac{1}{\kappa_2}\frac{C_2(k_1,-i\kappa_2)}{\overline C_1(\kappa_1, i\kappa_2)}\exp(-\kappa_2(x+y)),
%\end{equation}
%}
where we recall that here $k_1=\sqrt{2{\mu'}- i 0^+}$ and $\kappa_2=\sqrt{ 2V_0 -2{\mu'} +i0^+}$.
From this we find 
\begin{equation}
{\rm Im}\ G_{\mu'}(x,y)\approx \frac{\exp(-\kappa_2(x+y))}{\kappa_2}{\rm Im}\frac{C_2(k_1,-i\kappa_2)}{C_1^*(k_1, - i\kappa_2)} \, .
\end{equation}
%{\tiny
%\begin{equation}
%{\rm Im}\ G_{\mu'}(x,y)\approx \frac{\exp(-\kappa_2(x+y))}{\kappa_2}{\rm Im}\frac{C_2(k_1,-i\kappa_2)}{\overline C_1(\kappa_1, i\kappa_2)}
%\end{equation}
%}
This then gives the kernel as
\begin{equation}
K_\mu(x,y) \approx \frac{1}{\pi}\int_0^{k_{1F}} k_1 dk_1 \frac{\exp(-\kappa_2(x+y))}{\kappa_2}{\rm Im}\frac{ C_2(k_1,-i\kappa_2)}{C_1^*(k_1,- i\kappa_2)},\label{ksubplus}
\end{equation}
%{\tiny \begin{equation}
%K_\mu(x,y) \approx \frac{1}{\pi}\int_0^{k_{1F}} k_1 dk_1 \frac{\exp(-\kappa_2(x+y))}{\kappa_2}{\rm Im}\frac{ C_2(k_1,-i\kappa_2)}{\overline C_1(k_1, i\kappa_2)},\label{ksubplus2}
%\end{equation}
%}
 where $k_{1F}=\sqrt{2 \mu}$ and $\kappa_2=\sqrt{2V_0- k_1^2}$ in the integrand above. 
An important identity is derived in Appendix \ref{identityap}
\begin{equation}
{\rm Im}\frac{C_2(k_1,-i\kappa_2)}{C^*_1(k_1, - i\kappa_2)}= \frac{\kappa_2}{2k_1}\frac{1}{|C_1(k_1,i\kappa_2)|^2},\label{idformula}
\end{equation}
Hence we see that an alternative formula for the kernel (at distances much larger than the
barrier width) is 
\begin{equation}
K_\mu(x,y) \approx \int_0^{k_{1F}} \frac{dk_1}{2 \pi}
\frac{\exp(-\kappa_2(x+y))}{|C_1(k_1,i\kappa_2)|^2} \, .
\label{ksubplus2}
\end{equation}
This form is the one naturally obtained in the alternative method which uses the summation
over the eigenstates, as we will see in the next Section, see formula \eqref{KmuWood}.

 Further asymptotics can be performed if $x+y \gg 1/\kappa_{2F}$,
where $\kappa_{2F} = \sqrt{2 V_0- 2 \mu}$. Again this scale can be much
larger than the barrier width if one is near criticality $\mu \approx V_0$. In the
region $x+y \gg 1/\kappa_{2 F}$ the integral in Eq. (\ref{ksubplus}) is dominated by the region near $k_1=k_{1F}$ (equivalently $\kappa_2$ near $\kappa_{2F}$), due to the exponential decay of the integrand. Expanding the integral about $k_1=k_{1F}$  using
$k_1 dk_1 = \kappa_2 d\kappa_2$, yields the asymptotics for the kernel as $x,y \to +\infty$ as
\begin{equation} \label{asp} 
K_\mu(x,y) \approx \frac{\exp(-\kappa_{2F}(x+y))}{\pi (x+y)}{\rm Im} \frac{C_2(k_{1F},-i\kappa_{2F})}{C_1^*(k_{1F},- i\kappa_{2F})},
\end{equation}
%\begin{equation}
%K_\mu(x,y)= \frac{\exp(-\kappa_{2F}(x+y))}{\pi (x+y)}{\rm Im} \frac{C_2(k_{1F},-i\kappa_{2F})}{\overline C_1(k_{1F},i\kappa_{2F})},
%\end{equation}
 where we recall that $k_{1F} = \sqrt{2\mu}$ and $\kappa_{2F} = \sqrt{2V_0-k_{1F}^2}= \sqrt{2 V_0- 2 \mu}$. Let us emphasize again that the above calculation requires that $\kappa_{2F}>0$. The critical point where $\kappa_{2F}=0$ will be discussed below.

In the region $x\ ,y \to -\infty$, for $x>y$, we can  analytically continue Eq. (\ref{negxy}) to find
\begin{eqnarray}
G_{\mu'}(x,y) 
%&=& -\frac{\overline C_1(k_1,i\kappa_2) C_1(k_1,-i\kappa_2)}{\kappa_2}\left[1 -\frac{\overline C_2(k_1,i\kappa_2) C_2(k_1,-i\kappa_2)}{\overline C_1(k_1,i\kappa_2) C_1(k_1,-i\kappa k_2)}\right]\left(\exp(-ik_1(x-y)) + \frac{\overline C_2(k_1,i\kappa_2)}{\overline C_1(k_1,i\kappa_2)}\exp(ik_1(x+y))\right)\nonumber \\
&\approx&
\frac{i}{k_1}\left(\exp(-ik_1(x-y)) + \frac{C^*_2(k_1,- i\kappa_2)}{C^*_1(k_1,- i\kappa_2)}\exp(ik_1(x+y))\right).
\end{eqnarray}
Note that $C^*_j(k_1,- i\kappa_2)=( C_j(k_1, i \kappa_2) )^*$, $j=1,2$. Using Eq. (\ref{repii}), we obtain
\begin{equation} \label{asympt_subcrit}
K_\mu(x,y) \approx \frac{\sin(k_{1F}(x-y))}{\pi (x-y)} + \frac{1}{\pi}\int_0^{k_{1F}} dk_1 {\rm Im} \left(i
\frac{C^*_2(k_1,- i\kappa_2)}{C^*_1(k_1,- i\kappa_2)}\exp(ik_1(x+y)) 
\right)  \;,
\end{equation}
where $k_{1F} = \sqrt{2 \mu}$. This can  also be written using Eq. (\ref{repi}) (and analytically continuing he coefficients $C_1$ and $C_2$) as
\begin{equation} \label{asympt_subcrit2}
K_\mu(x,y) \approx \frac{\sin(k_{1F}(x-y))}{\pi (x-y)} - \frac{1}{\pi}\int_{k_{1F}}^\infty dk_1 {\rm Im} \left(i
\frac{C^*_2(k_1,- i\kappa_2)}{C^*_1(k_1,- i\kappa_2)}\exp(ik_1(x+y)) 
\right)  \;.
\end{equation}
Using the same method as in Appendix \ref{contsec}, the asymptotic behavior of $K_{\mu}(x,y)$ in (\ref{asympt_subcrit2}) for $x, y \to - \infty$ is given by
\begin{equation}
K_\mu(x,y) \approx \frac{\sin(k_{1F}(x-y))}{\pi (x-y)}+\frac{|C_2(k_{1F},-i\kappa_{2F})|}{|C_1(k_{1F},-i\kappa_{2F})|}\frac{\sin(k_{1F}(x+y)-\phi'_{12})}{\pi(x+y)}.\label{kas2}
\end{equation}
where $\phi'_{12}= \arg(C_2(k_{1F},i \kappa_{2F})/C_1(k_{1F}, i \kappa_{2F}))$.

%$\phi'_{12}= \arg(C_2(k_{1F},-i\kappa_{2F})/C_1(k_{1F},-i\kappa_{2F}))$
%which is simply the analytic continuation of Eq. (\ref{kas1}) (that is to say with $k_2\to-i\kappa_2$).
%{\red We find }.
%
%

\vspace*{0.3cm}

{\bf \noindent The critical case.}

\vspace*{0.3cm}

Here the critical case corresponds to $\kappa_{2F}=0$. If we use the representation in Eq. 
\eqref{ksubplus2}
%(\ref{idformula}) 
the kernel at the critical point,  for $x,\ y>0$ much larger than the barrier width, can be written as 
\begin{equation}
K_\mu(x,y) \approx \frac{1}{2\pi}\int_0^{k_{1F}} dk_1 \frac{1}{|C_1(k_1,i\kappa_2)|^2}\exp(-\kappa_2(x+y)),
\end{equation}
with $\kappa_2 = \sqrt{k_{1F}^2 - k_1^2}$ in the integrand. This kernel has the 
following asymptotics as $x+y \to +\infty$. The dominant contribution in that limit comes from
$\kappa_2\approx 0$ and so $\kappa_{1F} \approx k_{1F}$ and using this we find
\begin{equation} \label{GFres} 
K_\mu(x,y) \approx \frac{1}{2\pi |C_1(k_{1F},0)|^2 k_{1F}(x+y)^2},
\end{equation}
and so we see that the large distance $1/(x+y)^2$ decay of the kernel at the critical point is universal. This will be confirmed below from an exact solution for the Woods-Saxon potential \cite{woo54}.

\subsection{Direct summation of eigenfunctions and the Woods-Saxon potential}

We now analyze a fully solvable model of a smooth barrier, described by the Woods-Saxon potential \cite{woo54}
\be \label{smoothstep} 
V(x) = \frac{V_0}{1 + \exp(- \frac{x}{\lambda}) } 
\ee
such that $V(x) \to 0$ for $x \to - \infty$ and $V(x) \to V_0$ for $x \to + \infty$.
The length scale $\lambda$ thus controls the width of the step, and for $\lambda \to 0$ one
recovers the square step barrier. The method we use here is a direct summation
over the eigenstates. This will allow us to (i) connect with the previous
subsection where we obtained the asymptotics far from the barrier using the Green's function
method (we will see how these asymptotics emerge in this second method) (ii) explore the universality of the transition at $\mu=V_0$ with respect to the shape and width of the barrier. We will investigate here the situation where $\lambda$ and the typical inter-particle distance, $\ell$, are of the same order. 

%In the previous Section we have presented the exact solution for the square step barrier.
%It showed an interesting transition for $\mu=V_0$ with non trivial fermion number fluctuations.
%We also extracted some asymptotics for the kernel far from the barrier. To explore the universality of this transition with respect to the shape and width of the barrier, and of these asymptotics, 
%we now consider a smooth step barrier described by the Woods-Saxon potential \cite{woo54}
%\be \label{smoothstep} 
%V(x) = \frac{V_0}{1 + \exp(- \frac{x}{\lambda}) } 
%\ee
%such that $V(x) \to 0$ for $x \to - \infty$ and $V(x) \to V_0$ for $x \to + \infty$.
%The length scale $\lambda$ thus controls the width of the step, and for $\lambda \to 0$ one
%recovers the square step barrier. 
%
In this subsection we restrict to the case $\mu \leq V_0$, i.e. subcritical and critical.
We will only sketch the main results, the details are given in the Appendix \ref{app:smooth}. 
%The method used here is based on direct summation over the eigenstates. 
Using the standard solution \cite{Landau} for the eigenstates for energies $\epsilon_k< V_0$, one finds (see Appendix \ref{app:smooth}) the exact expression of the kernel for any $x,y$ and for $\mu \leq V_0$
\be \label{kernelsmooth} 
K_\mu(x,y) = \int_0^{\sqrt{2 \mu}} \frac{dk_1}{2 \pi} 
B(\lambda k_1, \lambda \kappa_2)  \phi^*_{\lambda k_1,\lambda \kappa_2}(x) \phi_{\lambda k_1,\lambda \kappa_2}(y)  \exp\left(- \kappa_2 (x+y)\right) \quad , \quad \kappa_2= \sqrt{2 V_0 - k_1^2} 
\ee
with
\be
\phi_{k_{1},\kappa_{2}}(x)={}_{2}F_{1}\left(ik_{1}+\kappa_{2},-ik_{1}+\kappa_{2},1+2\kappa_{2},-\exp\left(-\frac{x}{\lambda}\right)\right)~,~B(k_{1},\kappa_{2})=\left|\frac{\Gamma(-ik_{1}+\kappa_{2})\Gamma(1-ik_{1}+\kappa_{2})}{\Gamma(-2ik_{1})\Gamma(1+2\kappa_{2})}\right|^{2}
\ee
and ${}_2F_1$ is the standard hypergeometric function.
For $\lambda \to 0$ one has
$\phi_{\lambda k_1,\lambda \kappa_2}(x) \to 1$ and $B(\lambda k_1,\lambda \kappa_2)
\simeq 4 k_1^2/(k_1^2 + \kappa_2^2)$ and one recovers the result for the
square step barrier, in the form given in the appendix in \eqref{crit_kernel_app1}.

Let us first study the region $x,y>0$, i.e. the penetration of the fermions in the classically forbidden region.
We note that for $x \to +\infty$ the function
$\phi_{k_1,\kappa_2}(x)$ approaches unity exponentially fast, i.e. 
$\phi_{k_1,\kappa_2}(x) = 1 - \frac{2 V_0}{1 + 2 \lambda \kappa_2} \exp(- x/\lambda) + O(\exp({- 2 x/\lambda}))$. 
Hence for $x,y \gg \lambda$ the kernel takes the form
\bea \label{KmuWood} 
K_\mu(x,y) \simeq \int_0^{\sqrt{2 \mu}} \frac{dk_1}{2 \pi} 
B(\lambda k_1, \lambda \kappa_2)  \exp\left(- \kappa_2 (x+y)\right) \, .
\eea
Since $B(\lambda k_1, \lambda \kappa_2) = 1/|C_1(k_1,i \kappa_2)|^2$
in terms of the scattering amplitudes given in \eqref{scattWood},
the formula \eqref{KmuWood} is thus consistent with the result obtained in \eqref{ksubplus}
and \eqref{ksubplus2} by the Green's function method,
for a general barrier in terms of their associated scattering amplitudes.

At criticality $\mu=V_0$, one can shift to $\kappa_2=\sqrt{2 \mu-k_1^2}$ as integration
variable, and one sees that the kernel decays as a power law at large distance, as 
\be
\label{kernelsmoothTail} 
K_{\mu}(x,y)\simeq\int_{0}^{\sqrt{2\mu}}\frac{d\kappa_{2}}{2\pi}\frac{\kappa_{2}}{\sqrt{2\mu-\kappa_{2}^{2}}}B\left(\lambda\sqrt{2\mu-\kappa_{2}^{2}},\lambda\kappa_{2}\right)\exp\left(-\kappa_{2}(x+y)\right)\simeq\frac{2\lambda}{(x+y)^{2}}\coth\left(\pi\lambda\sqrt{2\mu}\right)+O\left(\frac{1}{(x+y)^{3}}\right)
\ee
since for large $x+y$ the integral is dominated by $\kappa_2 \approx 0$ and we used
that $B(\lambda \sqrt{2 \mu},0) = 4 \pi \lambda \sqrt{2 \mu} \coth(\pi \lambda \sqrt{2 \mu})$.
Hence, comparing with \eqref{alg1}, we see that the inverse square power law decay at large distance appears to be universal, but that the overall amplitude depends on the width of the barrier in units of inter-particle distance $\lambda \sqrt{2 \mu} = \pi \lambda/\ell$.  Again this is consistent
with the general result obtained by the Green's function method in \eqref{GFres}. The exact formula  Eq. \eqref{kernelsmooth} allows one to also obtain all sub-leading corrections.

We now show that, up to this overall amplitude, the scaling function $\nu$ defined in \eqref{nu} is 
universal.
It describes the decay of the kernel and of the density at large distance in the critical region,
i.e. $r-1 \ll 1$ and $x,y$ large of the order of the decay length 
$\xi_r = \ell/(2 \pi \sqrt{r-1})$ introduced above \eqref{nu}. Note that as $r \to 1$, $\xi$ becomes much larger than the
width of the barrier, hence it is natural to expect universality. 
Let us define $k_1=\sqrt{v} \sqrt{2 \mu}$, $\kappa_2=\sqrt{r-v} \sqrt{2 \mu}$ with $r=V_0/\mu \geq 1$ and
we see that the kernel can be put in the scaling form \eqref{eq:kernel_scaling}, with $\ell = \frac{\pi}{\sqrt{2\mu}}$ the typical inter particle distance, and $\tilde \lambda = \lambda \sqrt{2 \mu} = \pi \lambda/\ell$ 
\be
K_{\mu}(x,y) \simeq \frac{1}{\ell}\kappa_{r,\tilde \lambda}\left(\frac{x}{\ell},\frac{y}{\ell}\right) 
\quad , \quad 
\kappa_{r,\tilde \lambda}(a,b) = \int_0^1 \frac{dv}{4 \sqrt{v}} 
B(\tilde \lambda \sqrt{v}, \tilde \lambda \sqrt{r-v}) e^{- \pi \sqrt{r-v} (a+b) } 
  \quad , \quad a,b \gg \lambda/\ell
\end{equation}
If we take $\tilde \lambda \to 0$ we have $B(\tilde \lambda \sqrt{v}, \tilde \lambda \sqrt{r-v})\simeq \frac{4 v}{r}$ and one recovers \eqref{kappar}. Let us now write $v=1- (r-1) w$ and perform 
an expansion in $r-1$. We obtain
\bea
 \kappa_{r,\tilde \lambda}(a,b)   & \simeq &  (r-1)
%\frac{(r-1)}{\ell} \int_0^{1/(r-1)} \frac{dw}{4 \sqrt{1- (r-1) w}} 
%A(\sqrt{1- (r-1) w},\sqrt{r- 1} \sqrt{1+w}) 
%e^{- \pi \sqrt{r- 1} \sqrt{1+w}  (a+b)} \\
 \int_0^{1/(r-1)} dw 
\left( \pi \lambda \coth(\pi \lambda)  + O(\sqrt{r-1}) \sqrt{1+w} \right)
e^{- \pi \sqrt{r- 1} \sqrt{1+w}  (a+b)} \\
& \simeq & \pi \lambda \coth(\pi \lambda)  (r-1) \, 
\nu(\pi (a+b) \sqrt{r-1}) \quad , \quad \nu(\tilde a)= \frac{2}{\tilde a^2} (1+\tilde a) \exp\left(- \tilde a\right)
\label{nu2}
\eea 
hence in the critical region $x,y = O(\xi_r)$ the kernel and the density take the same
scaling form as in \eqref{nu}. Note that inside the subcritical phase the decay is exponential with a
rate predicted by \eqref{nu2} for any $r$. However, away from the critical region, that is
for large $a,b>0$ and fixed $r>1$, the amplitude of the
large distance decay is given by
\be
\kappa_{r,\tilde \lambda}(a,b) \simeq \frac{\sqrt{r-1}}{2 \pi (a+b)} B(\tilde \lambda, \tilde \lambda \sqrt{r-1}) 
\exp\left(-  \pi (a+b) \sqrt{r-1}\right) .
\ee
The above is the analog of 
\eqref{decay1}, but exhibits a non-universal $r$-dependent amplitude.

%{\red P: in conclusion there is some universality but probably not for $Var N_R$. Should we not say anything or say something but then positive and quantitative? }

Let us now discuss the region $x,y<0$ for $\mu \leq V_0$. From \eqref{psi_step22}
the wave functions are oscillating for $x,y \to -\infty$, and for $-x,-y \gg \lambda$,
the kernel takes the form
\be \label{left} 
K_\mu(x,y) \simeq \int_0^{\sqrt{2 \mu}} \frac{dk_1}{\pi} \left( \cos k_1 (x-y) 
+ {\rm Re} \beta_{k_1,\kappa_2}^* e^{i k_1 (x+y)} \right) 
= \frac{\sin \sqrt{2 \mu} (x-y)}{\pi (x-y)} 
+ \int_0^{\sqrt{2 \mu}} \frac{dk_1}{\pi}  {\rm Re} \beta_{k_1,\kappa_2} e^{- i k_1 (x+y)} 
\ee 
where we recall that $\kappa_2= \sqrt{2 V_0 - k_1^2}$  and
\be
\beta_{k_1,\kappa_2} = \frac{C_2(k_1,i\kappa_2)}{C_1(k_1,i\kappa_2)}
\ee
is the reflection amplitude (whose modulus square $|\beta|^2$ is the reflection coefficient of the barrier,
here equal to unity). In \eqref{left} the second term (the reflected kernel) gives the far away 
correction to the sine kernel of the bulk due to the presence of the barrier. 
Note that the asymptotic form \eqref{left} for $x,y \to -\infty$, and the expression of the reflected part of the kernel, is very general for any barrier. It is in perfect agreement with the formula
\eqref{asympt_subcrit} obtained by the Green's function method.
In the case of the Woods-Saxon potential
\be
\beta_{k_1,\kappa_2} = \frac{ \Gamma(2 i \lambda k_1) \Gamma(- i \lambda k_1+ \lambda \kappa_2 ) \Gamma(1- i \lambda k_1+ \lambda \kappa_2)  }{\Gamma(-2 i \lambda k_1) \Gamma(i \lambda k_1 + \lambda \kappa_2) \Gamma(1+ i \lambda k_1 + \lambda \kappa_2)}
\ee 
The limit of the square barrier is recovered for $\lambda \to 0$
in which case $\beta_{k_1,\kappa_2} = \frac{k_1- i \kappa_2}{k_1 + i \kappa_2}$.
Note that for any $\lambda$ in the limit $V_0 \to +\infty$ one has $\beta_{k_1,\kappa_2} \to -1$
and one recovers the infinite wall reflected kernel \eqref{reflectedkernel}. From \eqref{left}
we can extract the asymptotics of the mean fermion density as $x \to - \infty$. In that limit the integral
is dominated by the vicinity of $k_1=k_F=\sqrt{2 \mu}$ leading to the general formula for $\mu \leq V_0$
\be
\rho(x) \simeq \rho_L + \frac{1}{2 \pi |x|}  {\rm Im} \left[ \beta_{\sqrt{2 \mu},\sqrt{2 (V_0- \mu)}}\exp(2 i k_F |x|) \right]
\quad , \quad x \to - \infty
\ee 
which, for the square barrier gives 
\be
\rho(x)\simeq\rho_{L}+\frac{1}{2\pi V_{0}|x|}{\rm Im}\left[\left(\sqrt{\mu}-i\sqrt{V_{0}-\mu}\right)^{2}\exp\left(2ik_{F}|x|\right)\right]\quad,\quad x\to-\infty
\ee 
i.e. the continuation for $\mu \leq V_0$ of the formula \eqref{asymptdens} (which was
valid for $\mu \geq V_0$). The similar asymptotics were derived for a general barrier in terms of scattering amplitudes for
$\mu > V_0$ using the Green's function method in the previous subsection. 

{
The density $\rho(x)=K_\mu\left(x,x\right)$ for the Woods-Saxon potential in the critical case, with $\mu=V_0=1$, is plotted in Fig.~\ref{fig:woods_saxon_density} together with its $x \to \infty$ asymptotic behavior \eqref{kernelsmoothTail}. For a broad barrier (which, for the Woods-Saxon potential corresponds to large $\lambda$), the density is described correctly by the LDA, because the potential varies slowly in space. This is seen in the figure even for the moderately large value $\lambda=2$. Likewise, the density correlations are described by the sine kernel \eqref{sk}. Note that even for large $\lambda$, there is ultimately a power law decay of the density a large distance, as seen in the inset of Fig. \ref{fig:woods_saxon_density}.

%{\red Naftali: should I explain about the double structure of the tail of the density at large $\lambda$ or is this a little cumbersome? Should I plot also the $x \to -\infty$ tails?}
}

\begin{figure}[t!]
\begin{center}
\includegraphics[scale=0.6]{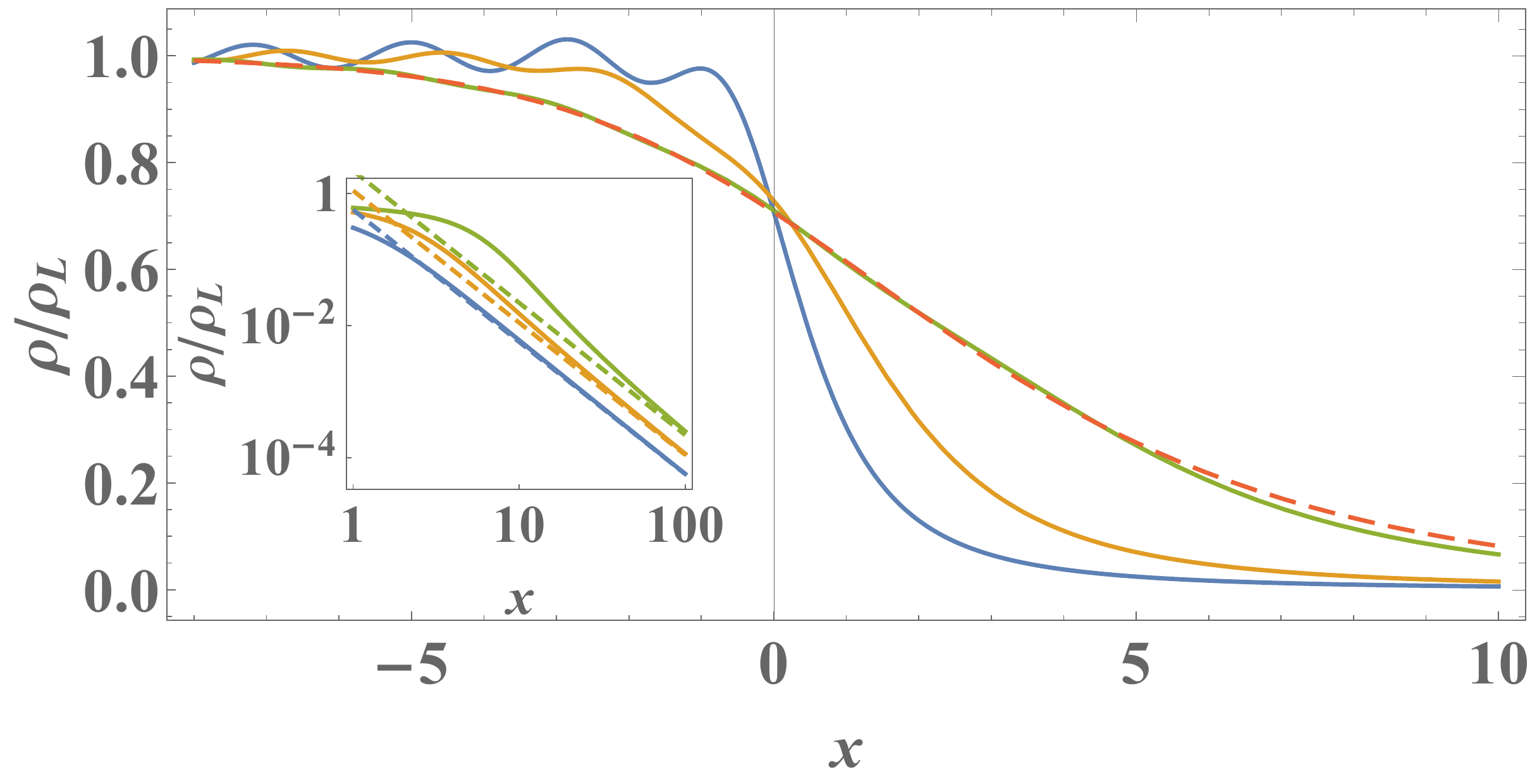} 
 \caption{Solid lines: The density $\rho(x)/\rho_{L}=K_\mu\left(x,x\right)/\rho_{L}$  vs.  $x$, for the Woods-Saxon potential \eqref{smoothstep} with $\mu=V_0=1$ (so $r=1$, the critical case) and three different values of $\lambda$: 1/2, 1 and 2, obtained via a numerical evaluation of Eq.~\eqref{kernelsmooth}. Dashed line: The prediction of the LDA, Eq.~\eqref{eq:densityLDA}, for $\lambda=2$. Inset: The large-$x$ tails of the density (solid lines) compared to the asymptotic \eqref{kernelsmoothTail} (dashed lines).}
 \label{fig:woods_saxon_density}
\end{center}
\end{figure}
%

%{\red P: Naftali there is another effect which may offset criticality if $L$ is not strictly
%infinite, due to the quantization of level, which may place the system slightly above of
%below criticality. Could you comment or give some quantitative criterion on that effect?} 

\section{Conclusions}\label{sec:conclusion}

%In this paper we have shown how the single particle Green's function can be used to compute the kernel for systems of trapped, spinless, noninteracting fermions. 

In this paper we studied the quantum correlations of spinless non-interacting fermions 
in their ground state. We
introduced an alternative Green's function method to compute the kernel
which does not rely on an explicit summation over eigenstates. 
We first showed how it allows to recover the known results for smooth potentials.
It allows one to derive, in a compact way, the kernel in the bulk, which is given by the LDA approximation, and to ascertain the validity of this approximation. 
We also showed how the same basic method can be adapted 
to study the properties of the Airy gas at the edges of the Fermi gas.

%and obtained some new results for a non-smooth ones, such as the step potential. 
%
%
%In particular, for the critical kernel case,
%we have also computed the kernel by the direct summation of the eigenfunctions up to the 
%Fermi level in Appendix B. As shown in this Appendix, one has to be ultra-careful in choosing
%the right set of eigenfunctions to sum over, while this is automatically taken into account
%in the Green�s function method. This simple example demonstrates the power of our approach. 
%Furthermore, in a companion paper, this method will be applied to treat the 
%case of delta impurities for which again it turns out to be more convenient.
%

The method is particularly useful when one has exact results for the single particle Green's function. 
This is the case for a system in the presence of a finite step in the potential for which we have 
obtained the kernel and the fermion density analytically. We have analyzed the cases where the Fermi energy is below the height of the step (the sub-critical case) and  above the height of the trap (the super-critical case). Of particular interest is the critical case where the step height coincides with the Fermi energy. Here the kernel takes a particularly simple form and one can show that the number of fermions $N_R$  to the right of the edge is of order $1$, even though the system itself is macroscopic. The analysis of the second moment shows that the distribution is not Bernoulli in most cases, showing that more that than one fermion may {\em leak} over the edge. However as the Fermi energy is lowered below the step height, we find that the distribution of the number of fermions becomes Bernoulli, but with a probability of presence $p$ that becomes very small. For completeness, we have shown in the Appendix B how to recover the kernel for the step potential from a direct summation over eigenstates, focusing for illustration on the simplest case $\mu \leq V_0$. This method, which also allowed us to analyze the case of a step of finite width, has its advantages, but it requires a careful treatment of  the normalization factors and selection of the proper eigenstates which contribute, technical details which are automatically taken into account in the Green's function method. Furthermore, in a companion paper, this Green's function method will be applied to treat the 
case of delta impurities for which again it turns out to be particularly well adapted.

Next we considered the case of a general smoothed potential and showed how the asymptotics of the  Green's function (far from the region where the potential varies) can be written in terms of generic scattering coefficients of plane waves arriving from the left. From this we were able to derive asymptotic results for the kernel and density  far from the step in the potential, in particular we were able to show that the algebraic decay $1/x^2$ of the density, far to the right of the step, is universal at the critical point $\mu=V_0$. The behavior of the density close to the step does however depend on the shape of the step. For the Woods-Saxon potential we have given an integral expression for the kernel and density in the subcritical and critical regimes and explicitly verified that for this potential, at distances greater than the step width $\lambda$, the general asymptotic results derived here hold.

This study opens up a number of perspectives for further research. In particular achieving close to zero temperatures in experiments is still an on going challenge. The effect of finite temperature can be incorporated 
in a straight forward manner at finite temperature and in the grand canonical ensemble. Here the kernel is given by \cite{dea15b,dea16,dea19}
\begin{equation}
K_{\tilde \mu}(x,y)=\sum_k \frac{1}{1 + \exp\left(\beta( \tilde \mu-\epsilon_k)\right)}\psi_k^*(x)\psi_k(y),
\end{equation}
where $\tilde \mu$ is the chemical potential, which becomes the Fermi energy in the zero temperature limit.
Applying the results presented here it is straightforward to see that 
\be
K_{\tilde \mu}(x,y)=
 \frac{1}{\pi}  \int d\mu' \frac{1}{1 + \exp\left(\beta (\mu'-\tilde \mu)\right)} { \rm Im} \, G_{\mu'}(x,y),
\ee
From this formula the effect of a finite temperature for a step potential can be analysed, in particular one can ask how the distribution of fermions to the right of the step in the critical and sub-critical regimes
will depend on the temperature. 

It would also be interesting to investigate the properties of the Wigner function \cite{wig32,cas08} for stepwise potentials. Existing methods based on the extraction of the Wigner function from the kernel have revealed interesting and universal properties at the edges of trapped systems both for statics \cite{dea18} and dynamics \cite{dea19b} and the methods proposed here might facilitate further studies.  

In a similar vein, the determinantal properties of trapped fermionic systems allow one to study extreme value statistics, typically answering questions such as what is the distribution of the farthest fermion from the center of a trap \cite{dea16,dea17,dea19}. For one dimensional smooth traps these statistics are given by the celebrated Tracy-Widom \cite{tra94} distribution  \cite{dea16,dea19}, while in higher dimensional systems with eigenstate degeneracy Gumbel type distributions are found \cite{dea17}. In principle  extreme value statistics can be determined from knowledge of a Fredholm determinant involving the kernel \cite{bor11,dea16,dea19}, however the computation of the Fredholm determinant presents a daunting mathematical task. It is however possible that the simple form of the critical kernel given by Eq. (\ref{rep2}) may allow further analytical progress.

%For the square step potential studied in the present work, in the critical case $\mu=V_0$ one can obtain by using dimensional analysis the scaling form $\mathcal{P}\left(x\right)=\left(1/\ell\right)f\left(x/\ell\right)$
%for the PDF $\mathcal{P}\left(x\right)$ of the position of the rightmost particle, where $f$ is universal. The right tail of the distribution is expected to coincide, $f\left(a\right)\simeq n_{c}\left(a\right)$, $a\gg1$, with the tail \eqref{alg1} of the density in the leading order.

\vspace*{0.5cm}

{\it Acknowledgments:} 
NRS acknowledges support from the Yad Hanadiv fund (Rothschild fellowship). This research was supported by ANR grant ANR-17-CE30-0027-01 RaMaTraF.

\appendix

%\newpage

{ \section{Alternative integral representations and asymptotic analysis of the kernel }\label{contsec}

\subsection{Integral representations of the kernel}

In this Appendix, we derive alternative integral representations for the kernel which are useful to study the asymptotics discussed in the text. 
When considering Green's functions for systems which have variations in their potentials in localised regions, for instance step like potentials, we will see that the Green's function for a Hamiltonian $H$ takes a generic form
\begin{equation}
G_{\mu'}(x,y) = G_{0\mu'}(x,y) + \Delta G_{\mu'}(x,y) ,
\end{equation}
where $G_{0\mu'}(x,y)$ is the Greens' function for a bulk system with constant potential, with Hamiltonian $H_0$. 
Here $\Delta G_{\mu'}(x,y)$ represents the change in the Green's function due to the variation of the potential.
The function $G_{0\mu'}(x,y)$ has poles at $\mu'=\epsilon^{(0)}_k+ i 0^+$, where $\epsilon^{(0)}_k$ are the energy levels of $H_0$, and $G_{\mu'}(x,y)$ has poles at $\mu'=\epsilon_k+ i 0^+$,where $\epsilon_k$ are the energy levels of $H$. We thus see that  $\Delta G_{\mu'}(x,y)$ can generally have poles 
at $\mu'=\epsilon^{(0)}_k+ i 0^+$ and $\mu'=\epsilon_k +i 0^+$. The important point is that the poles of  $\Delta G_{\mu'}(x,y) $ lie infinitesimally above the real axis - as shown in Fig. \ref{contours1} by the symbols $\times$'s. If $K_{0\mu'}(x,y)$ represents the bulk kernel then we have two representations of the kernel. The first (i) is obtained from Eq. (\ref{bigmu}) and reads 
\begin{equation}
K_\mu(x,y) = K_{0\mu}(x,y) -\int_{\mu}^{\infty}d\mu'  \frac{1}{\pi} { \rm Im} \,\Delta G_{\mu'}(x,y).
\end{equation}
\begin{figure}[t!]
\begin{center}
  \includegraphics[scale=0.5]{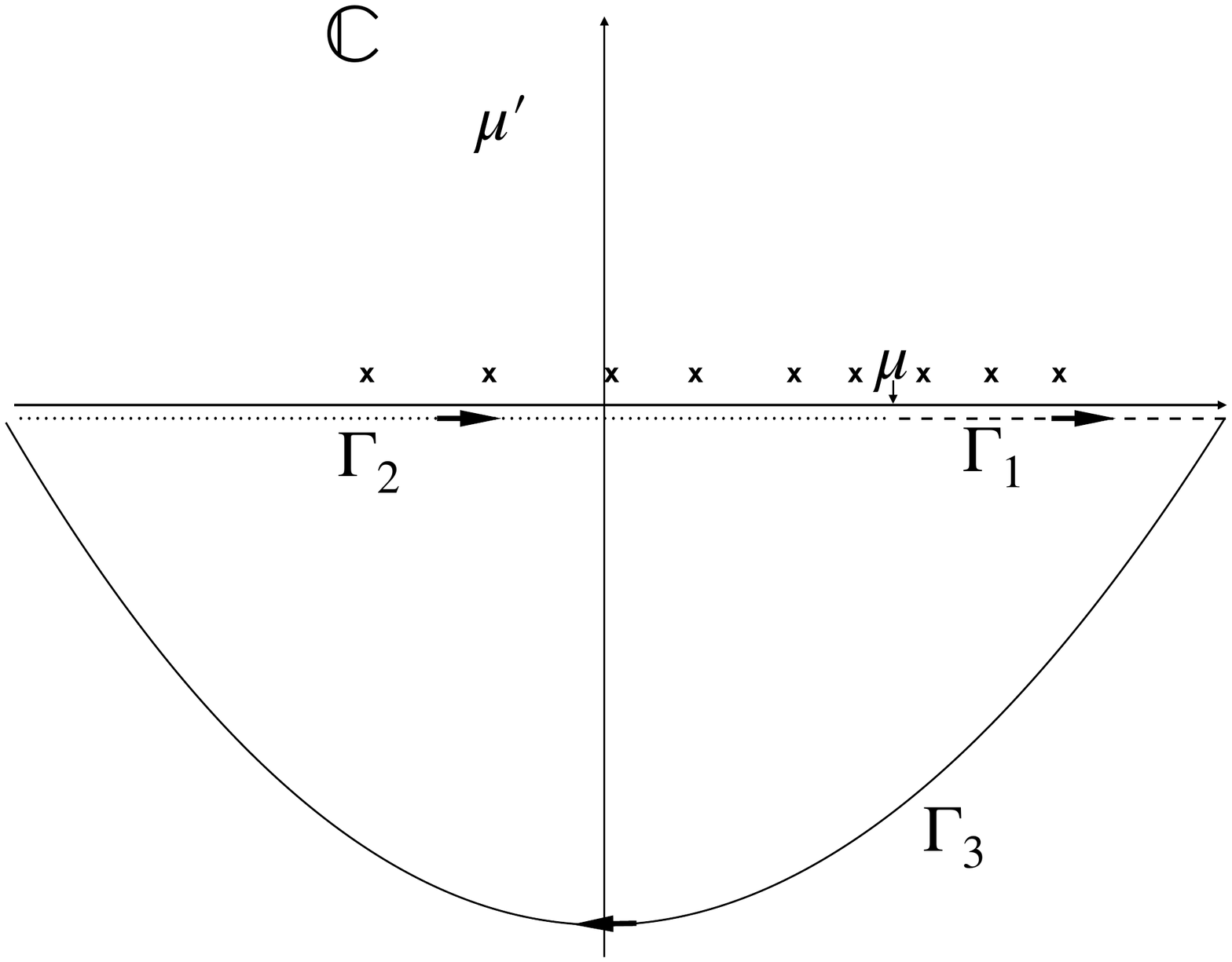} 
 \caption{Contour integrals used in the integral representations of the kernel. Crosses, $\times$,  indicate the poles of the Green's function which are just above the real axis.  $\Gamma_1=[\mu,\infty]$ is the contour used in representation (i) and $\Gamma_2=[-\infty,\mu]$ that used for representation (ii). The contour $\Gamma_3$ is used to close the contour $\Gamma_1\cup\Gamma_2$ and is taken to be a semi-circle in the lower half of the complex plane whose radius is taken to $\infty$. }\label{contours1}
\end{center}
\end{figure}}

The second (ii), obtained from Eq. (\ref{bigmu2}), gives
\begin{equation}
K_\mu(x,y) = K_{0\mu}(x,y)+  \frac{1}{\pi}\int_{-\infty}^{\mu}d\mu'  \,{ \rm Im}\ \Delta G_{\mu'}(x,y).
\end{equation}
This means that the change in the kernel due to the variation of the potential 
\begin{equation}
\Delta K_\mu(x,y) = K_\mu(x,y) - K_{0\mu}(x,y),
\end{equation}
has two representations
\begin{eqnarray}
{\rm Representation\ (i)}:\  \Delta K_\mu(x,y)&=&-\frac{1}{\pi} { \rm Im}\int_{\mu}^{\infty}d\mu'   \,\Delta G_{\mu'}(x,y).\label{repi}\\
{\rm Representation\ (ii)}:\  \Delta K_\mu(x,y)&=&\frac{1}{\pi} { \rm Im}\int_{-\infty}^{\mu} d\mu'  \,\Delta G_{\mu'}(x,y)\label{repii},
\end{eqnarray}
and note that {the imaginary part} can be taken outside of the integral as the integration range is real.
%and note that the taking of the imaginary part can be taken outside of the integral as the integration range is real.
The integration ranges are shown in Fig. (\ref{contours1}) as contours $\Gamma_1$ and $\Gamma_2$. If we denote by
\begin{equation}
I_\Gamma = - \int_\Gamma d\mu' \,\ \Delta G_{\mu'}(x,y),
\end{equation}
where $\Gamma$ is an arbitrary contour in the complex $\mu'$ plane, then representation (i) is equivalent to 
$\Delta K_\mu(x,y)=\frac{1}{\pi} {\rm Im}\ I_{\Gamma_1}$ and representation (ii) is equivalent to $\Delta K_\mu(x,y)=-\frac{1}{\pi} {\rm Im}\ I_{\Gamma_2}$.
The equivalence of the representations corresponds to ${\rm Im}[I_{\Gamma_1}+ I_{\Gamma_2}]=0$. This can be seen 
by applying Cauchy's theorem to the contour $\Gamma_2\cup\Gamma_1\cup\Gamma_3$, in the limit where $\Gamma_3$ shown in Fig. (\ref{contours1}) is extended to an infinite semi-circle. Due to the absence of poles in the lower half of the complex plane we find $I_{\Gamma_2}+ I_{\Gamma_2}+ I_{\Gamma_3}=0$. However we can formally write, for $\mu' \in \Gamma_3$
\begin{equation}
\Delta G_{\mu'} = (\mu'-H)^{-1}- (\mu'-H_0)^{-1}\approx  \frac{(H-H_0)}{ \mu'^2} 
\end{equation}
and thus we see that $I_{\Gamma_3}=0$, as $\Delta G_{\mu'}$ decays as $1/\mu'^2$ on $\Gamma_3$. Furthermore the integral over any subset of $I_{\Gamma_3}$ is also clearly zero. We thus recover the equivalence of representations (i) and (ii) from the, stronger, identity $I_{\Gamma_1}+ I_{\Gamma_2}=0$.

However $H-H_0$ expressed in terms of the eigenfunctions of $H$ and $H_0$ is an infinite sum, for instance $H=\sum_k \epsilon_k \psi^*_k(x)\psi_k(y)$ is clearly not convergent as by definition $\epsilon_k$ increases with $k$ and the number of states is not bounded. It is perhaps possible to use the above argument using a finite dimensional lattice model, however in what follows we propose a solution in the continuum setting.  One proceeds by  examining the behavior of the Green's function which for large $|\mu'|$ in the complex plane.  It is approximately  given by the free Green's function, see Eq. (\ref{gflocal}) with $V=0$, (as we can ignore the potential for $|\mu'|\to\infty$) 
\begin{equation}
G_{\mu'}(x,y) = \frac{i\exp\left(-i\sqrt{2\mu'}|x-y|)\right)}{\sqrt{2\mu'} }.
\end{equation}
The point is now that the branch of the square root must be chosen such that ${\rm Im}\  \sqrt{2\mu'}<0$ as the Greens function must decay for large $|x-y|$ (this is a guiding principle throughout the paper). 
%Furthermore we note that this choice  of the square root is continuous in the lower complex plane.
However as $|\mu'| \to \infty$ along $\Gamma_3$ we thus have ${\rm Im}\  \sqrt{2\mu'}\to -\infty$ and so clearly $G_{0\mu'}(x,y) \to 0$ as $|\mu'| \to \infty$ exponentially quickly, but only so long as $x\neq y$ ! So  we are obliged to treat the case $x=y$ specifically. Note that from Eq. (\ref{gflocal}) we can write the Greens function with a potential $V$ as (again when $|\mu'| >> |V(x)|$
)
\begin{equation}
G_{\mu'}(x,x) = \frac{i}{\sqrt{2\mu' -2V(x)} }.
\end{equation}
From this we see that
\begin{equation}
\Delta G_{\mu'}(x,x) \approx \frac{i}{\sqrt{2\mu' -2V(x)} }- \frac{i}{\sqrt{2\mu'} }
\end{equation}
and so on the contour $\Gamma_3$ we find that $\Delta G_{\mu'}(x,x)$ decays as $V(x)/(\mu')^{\frac{3}{2}}$, so slower than $1/\mu^2$ but nonetheless ensures that the integral along $\Gamma_3$ is zero.

\begin{figure}[t!]
\begin{center}
  \includegraphics[scale=0.2]{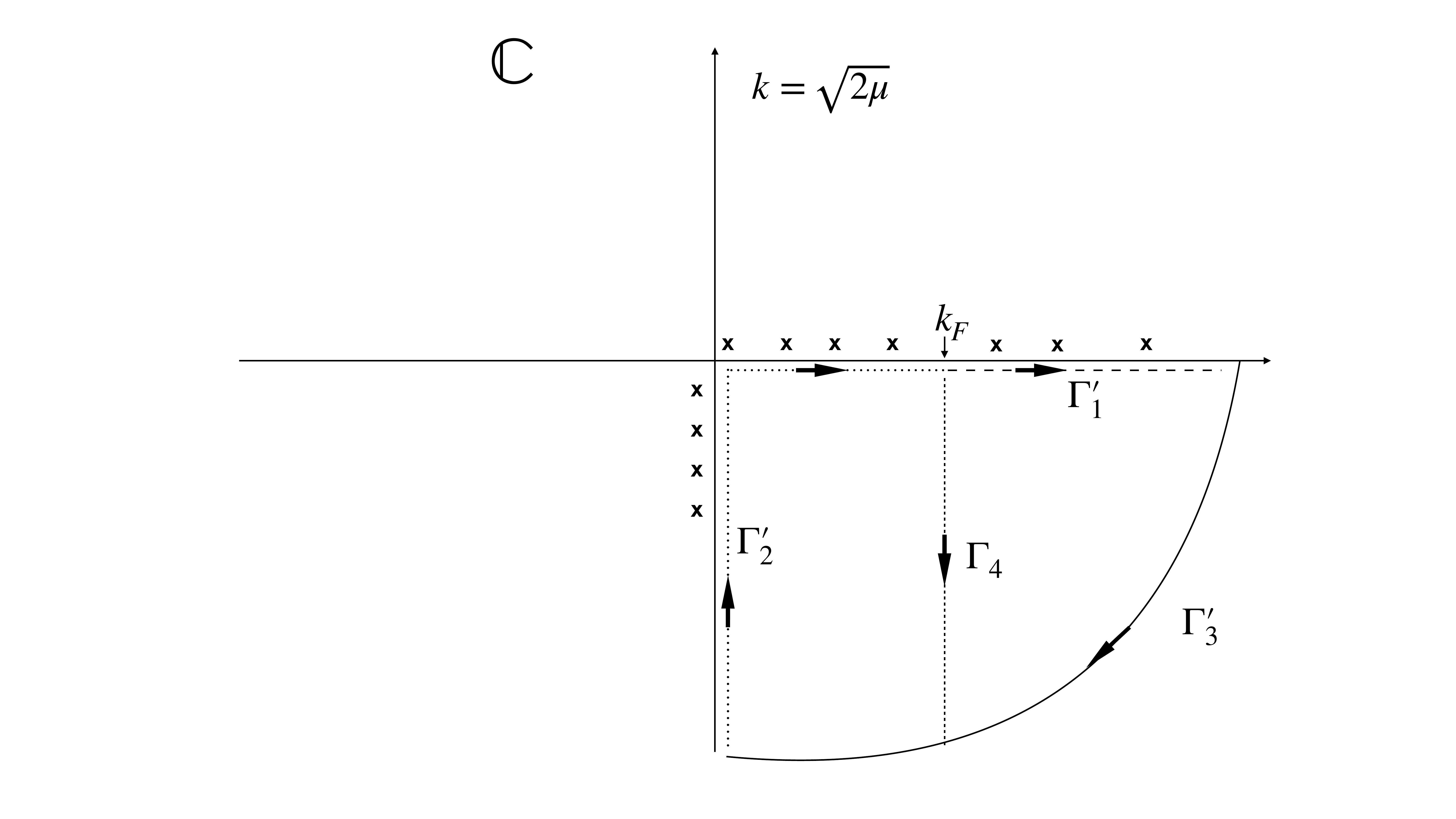} 
 \caption{Contour integrals used in the integral representations of the kernel in terms of the variable $k=\sqrt{2\mu'}$. Crosses indicate the poles of the Green's function in the complex plane of $k$.  The contours 
 $\Gamma'_{i}$ for $i=1,\ 2,\ 3$ correspond to the contours $\Gamma_{i}$ when mapped into the $k$ plane. The contour  $\Gamma_4=[k_F, k_F-i\infty]$, which is useful for asymptotic analysis, is also shown.
 }\label{contours2}
\end{center}
\end{figure}

In the analysis of an arbitrary barrier the Green's function is conveniently expressed in terms of wave vectors 
$k=\sqrt{2\mu}$. For $\mu'<0$, the fact that the Green's function must decay at large distances will be shown to imply that $k=-ik'$ where $k'>0$, thus determining the branch of the square root on the negative real axis. The contours $\Gamma$  in the $\mu'$ plane are mapped to their images $\Gamma'$ in the complex $k$ plane and are shown in Fig. (\ref{contours2}), as well as the positions of the poles of the Green's function after this transformation, again marked by a $\times$.

The values of the integrals are of course not changed, that is to say $I_{\Gamma_i}=I_{\Gamma'_i}$ (the integration measures being related by $kdk=d\mu'$). In particular in the limit where the contour $\Gamma_3$ is extended to infinity we have $I_{\Gamma_3}= I_{\Gamma'_3}=0$. This means that the representation (i) which now reads
\begin{equation}
\Delta K_\mu(x,y)= \frac{1}{\pi} {\rm Im}\  I_{\Gamma'_1},
\end{equation}
can be rewritten using Cauchy's theorem $I_{\Gamma'_1}+I_{\Gamma_3^{'*}}-I_{\Gamma_4}=0$ where 
$\Gamma^{'*}_3$ is the subset of the contour $\Gamma'_3$ which starts where $I'_{\Gamma_3}$
intersects the real axis and terminates where $I'_{\Gamma_3}$ intersects $I_{\Gamma_4}$. However as the integral along $I_{\Gamma^*_3}$ is zero for any subinterval  of $\Gamma^*_3$ as it moves out to infinity, we find
\begin{equation}
\Delta K_\mu(x,y)=\frac{1}{\pi} {\rm Im}\ I_{\Gamma_1}= \frac{1}{\pi} {\rm Im}\ I'_{\Gamma_1}=\frac{1}{\pi} {\rm Im}\ I_{\Gamma_4}.
\end{equation}
Now writing $\Delta G_\mu(x,y)= \Delta G(k,x,y)$ (where recall  $k=\sqrt{2\mu'}$ and so $d\mu'=kdk$) we can write
\begin{equation}
\Delta K_\mu(x,y)=-\frac{1}{\pi}{\rm Im} \int_{\Gamma_4} kdk\   \Delta G(k,x,y),\label{asky}
\end{equation}
along the contour $\Gamma_4=[k_F,k_F-i\infty)$. This result will turn out to be useful to explore the properties of general step like potentials in terms of their scattering coefficients in section (\ref{gfg}) as the integral decays exponentially as one moves along the contour $\Gamma_4$.

\subsection{Asymptotic analysis}

Starting from Eq. (\ref{Kmuprime}) in the text and using the representation of $\Delta K_\mu$ (with a rotated contour) given by Eq. (\ref{asky}), we can also write an alternative formula which is more amenable to an asymptotic analysis
\begin{eqnarray} 
K_\mu(x,y) &\approx& \frac{\sin(k_{2F}(x-y))}{\pi (x-y)}+\frac{1}{\pi}{\rm Im}\int_{k_{2F}}^{k_{2F}-i\infty} idk_2 \ \frac{C_2(\sqrt{k_2^2+2V_0},k_2)}{C^*_1(\sqrt{k_2^2+ 2V_0},k_2)}\exp(-ik_2(x+y))\label{approx1} \\
&\approx& \frac{\sin(k_{2F}(x-y))}{\pi (x-y)}+\frac{1}{\pi}{\rm Im}\int_{0}^{\infty} du \ \frac{C_2(\sqrt{(k_{2F}-iu)^2+2V_0},k_{2F}-iu)}{C^*_1(\sqrt{(k_{2F}-iu)^2+ 2V_0},k_{2F}-iu)}\exp((-ik_{2F}-u)(x+y)) \label{approx1a}
\end{eqnarray}
with $k_{2F}= \sqrt{2\mu-2V_0}$. Note, with reference to the discussion in section \ref{contsec}, that first term in Eq. \eqref{Gapprox} corresponds to what we referred to as $G_{\mu'}(x,y)$ and yields  the sine kernel for free fermions in a constant potential $V_0$, while the second terms corresponds to what we referred to as
$\Delta G_{\mu'}(x,y)$.  

We now perform the asymptotics for $x+y \gg 1/k_{2F}$. Here the integral in Eq. (\ref{approx1a}) is dominated by the region $u=0$, as the integrand clearly decays exponentially with $u$, and find Eq. (\ref{kas2}).
%
%\begin{equation} \label{Kmu_asympt}
%K_\mu(x,y) \approx \frac{\sin(k_{2F}(x-y))}{\pi (x-y)}-\frac{|C_2(k_{1F},k_{2F})|}{|C_1(k_{1F},k_{2F})|}\frac{\sin(k_{2F}(x+y)-\phi_{12})}{\pi(x+y)},
%\end{equation}
%where $\phi_{12}=\arg(C_2(k_{1F},k_{2F})/C^*_1(k_{1F}, k_{2F}))$. We thus see that at large distances from the barrier, the kernel takes the bulk sine-kernel form, plus an oscillatory correction with an amplitude that decays algebraically with distance from the barrier. 

%\section{Smooth potentials}
%
%{Before applying this method to obtain new exact solutions (for any $\mu$) 
%for discontinuous potentials in the next section, we show how the method can be applied to
%analyse the well studied case of smooth potentials. In the bulk we recover the prediction
%of the LDA (which is exact for potentials constant in space) and identify the validity of the LDA via this method. We then examine, the again well known, edge {\em Airy gas} behavior, based on a local linear approximation to the potential
%(the method being again exact for purely linear potentials).}

\section{Derivation of the Airy Green's function}
\label{app:airygreen} 

Here we provide a derivation of the solution of \eqref{airyeq} given in \eqref{resg}.
We look for a solution of \eqref{airyeq} in the form 
%\begin{eqnarray}
%g(\zeta,\zeta') &=& A_+ Y_+(\zeta) \ {\rm for } \ \zeta>\zeta' \nonumber \\
%                        &=& A_- Y_-(\zeta)   \ {\rm for }    \                \zeta<\zeta' ,
%\end{eqnarray}
\be
g(\zeta,\zeta')=\begin{cases}
A_{+} U_{+}(\zeta) & {\rm for}\ \zeta>\zeta'\\[0.1cm]
A_{-} U_{-}(\zeta) & {\rm for}\ \zeta<\zeta',
\end{cases}
\ee
where $U_\pm(\zeta)$ are solutions of the homogeneous equation, i.e. the Airy equation 
associated with the Green's function 
\begin{equation}
\frac{\partial^2}{\partial \zeta^2}U(\zeta) - \zeta U(\zeta) = 0. \label{hom}
\end{equation} 
%and we look for a solution of the form
This leads to 
%\begin{eqnarray}
%g(\zeta,\zeta') &=& -\frac{Y_+(\zeta)Y_-(\zeta')}{\hat W(Y_+,Y_-)}  \ {\rm for } \ \zeta>\zeta' \\
%&=&-\frac{Y_-(\zeta)Y_+(\zeta')}{\hat W(Y_+,Y_-)} \ {\rm for } \ \zeta<\zeta'
%\end{eqnarray}
{
\be
g(\zeta,\zeta')=\begin{cases}
-\frac{U_{+}(\zeta)U_{-}(\zeta')}{\hat{W}(U_{+},U_{-})} & {\rm for}\ \zeta>\zeta'\\[0.25cm]
-\frac{U_{-}(\zeta)U_{+}(\zeta')}{\hat{W}(U_{+},U_{-})} & {\rm for}\ \zeta<\zeta',
\end{cases}
\ee}
where ${ \hat W[f,g]} = f(z)g'(z)-g(z)f'(z)$ denotes the Wronskian between the functions $f$ and $g$, which is 
a constant here, {\em i.e.} independent of $\zeta$. 

The two linearly independent solutions of the Eq. (\ref{hom}) are given by the Airy
functions ${\rm Ai}(\zeta)$ and ${\rm Bi}(\zeta)$. For $\zeta>\zeta'$ we must have $U_+(\zeta) = {\rm Ai}(\zeta)$ as the wave functions vanish outside the bulk region. For $\zeta<\zeta'$ we note that both the solutions ${\rm Ai}$ and ${\rm Bi}$ decay as $\zeta\to-\infty$ to leading order as \cite{abr65} 
%{\red Naftali: There is surely a more compact way to write these asymptotics. e.g., in my quantum mechanics TA session notes I found the formula $\text{Ai}\left(\zeta\to-\infty\right)\simeq\frac{\sin\left(\frac{2}{3}\left(-\zeta\right)^{3/2}+\frac{\pi}{4}\right)}{\sqrt{\pi}\,\left(-\zeta\right)^{1/4}}$.}
\begin{eqnarray}
{\rm Ai}(\zeta) &\simeq&   \frac{\sin(\frac{2}{3}(-\zeta)^{\frac{3}{2}} +\frac{\pi}{4})}{\sqrt{\pi}(-\zeta)^\frac{1}{4}}\ 
\frac{\Gamma(\frac{5}{6})\Gamma(\frac{1}{6})}{2\pi} - \frac{\cos(\frac{2}{3}(-\zeta)^{\frac{3}{2}} -\frac{\pi}{4})}{\sqrt{\pi}(-\zeta)^\frac{1}{4}}\  \frac{\Gamma(\frac{11}{6})\Gamma(\frac{7}{6})}{2\pi} \\
{\rm Bi}(\zeta) &\simeq& \frac{\cos(\frac{2}{3}(-\zeta)^{\frac{3}{2}} +\frac{\pi}{4})}{\sqrt{\pi}(-\zeta)^\frac{1}{4}}
\  \frac{\Gamma(\frac{5}{6})\Gamma(\frac{1}{6})}{{2\pi}} + \frac{\sin(\frac{2}{3}(-\zeta)^{\frac{3}{2}} -\frac{\pi}{4})}{\sqrt{\pi}(-\zeta)^\frac{1}{4}}\
 \frac{\Gamma(\frac{11}{6})\Gamma(\frac{7}{6})}{2\pi} \, .
 \label{airyas}
\end{eqnarray} 
However from  Eq. (\ref{bcag}) we see that, up to an overall constant $c$, we must have
for $\zeta \to - \infty$,  i.e. in the bulk region 
\begin{equation}
Y_-(z) \simeq \frac{c\exp\left(-i\frac{1}{3V'(x_e)}(2\mu'-2V(x_e)-2V'(x_e)z)^{\frac{3}{2}}\right)}{\left(2\mu' -2V(x_e)-2V'(x_e)z\right)^{\frac{1}{4}}}.\nonumber \\
\label{bcag2}
\end{equation}
Now, recalling that  $\mu' - V(x_e) -z V'(x_e) = -\alpha \zeta$, matching with the solution in the bulk implies that for $\zeta \to - \infty$ one must have
\begin{equation}
U_{-}(\zeta)=\frac{c'}{(-\zeta)^{\frac{1}{4}}}\exp\left(-i\frac{2}{3}(-\zeta)^{\frac{3}{2}}\right),\label{y-as}
\end{equation}
where $c'$ is again a constant independent of $\zeta$.
Then using  Eq. (\ref{airyas}) and Eq. (\ref{y-as}) it is easy to show that 
\begin{equation}
 U_-(\zeta) =- i{\rm Ai}(\zeta) +{\rm  Bi}(\zeta),
\end{equation}
and $c'=\exp\left(- i \pi/4\right)/\sqrt{\pi}$.
Now using $\hat W[{\rm Ai},{\rm Bi}] = 1/\pi$ \cite{abr65}, we find
\begin{equation}
\hat W[U_+,U_-] = \frac{1}{\pi}.
\end{equation}
This leads to the result for the Airy Green's function given in \eqref{resg}.

\section{Green's function for the square barrier}
\label{app:green} 

It is interesting to relate the Green's function $G_\mu(x,y)$ defined in the text to the Euclidean propagator in the time domain (which we
will denote by another letter) $\hat G(x,y,t)$, solution of  
\bea
\partial_t \hat G = -  H \hat G = \frac{1}{2} \frac{\partial^2}{\partial x^2} \hat G - V(x) \hat G \quad , \quad \hat G(x,y,t=0)=\delta(x-y)
\eea 
This propagator was extensively discussed in \cite{dea16} where it was shown that the kernel
$K_\mu(x,y)$ can be obtained as the inverse Laplace transform of $\frac{1}{t} \hat G(x,y,t)$,
i.e. $K_\mu(x,y) = \int_C \frac{dt}{2 i \pi t } \hat G(x,y,t)$, where $C$ is the Bromwich contour. 
For the square barrier that we consider now, i.e. $V(x)= V_0 \theta(x)$ it has also a nice application to Brownian motion. Let us define $T_{x,y}(t)= \int_0^t d\tau\  \theta(x(\tau))$ the total time spent on the positive axis $x>0$ between time $0$ and $t$ by a Brownian motion $x(\tau)$, started at $x(0)=x$ and ending at $x(t)=y$. Using the Feynman-Kac formula, one easily sees that the Laplace transform w.r.t. the parameter $V_0$ of 
the probability distribution $P_{t,x,y}(T)$ of the random variable $T_{x,y}(t)$ is precisely the propagator
\cite{kac49,maj05}
\be
\hat G(x,y,t) = \mathbb E( \exp\left(- V_0 T(t)\right) ) = \int_0^{+\infty} dT  \exp\left(- V_0 T\right) P_{t,x,y}(T)
\ee 
leading e.g. to the famous arcsine law \cite{lev39} for $P_{t,x,x}(T)$. For an
explicit expression for $\hat G$ in real time for a step potential see e.g. \cite{Carvalho}. 

For the square barrier the propagator is most easily studied via its Laplace transform w.r.t. time $\tilde G(x,y;s)= \int_0^{+\infty} dt \exp\left(-s t\right) \hat G(x,y;t)$, which satisfies 
\bea \label{Gt} 
- \delta(x-y) + s  \tilde G = \frac{1}{2} \frac{\partial^2}{\partial x^2} \tilde G - V_0 \theta(x) G 
\eea
with proper decay at infinity in space. Thus one has $\tilde G = (s+H)^{-1}$ and the relation to
the Green's function in the text is 
\be \label{relGG} 
G_\mu(x,y) = - \tilde G(x,y, s= -\mu + i 0^+) 
\ee 

The equation \eqref{Gt} is easily solved in terms of plane waves. Since it
must decay for $|x,y| \to +\infty$, and it must be symmetric $\tilde G(x,y,s)=\tilde G(y,x,s)$,
one looks for a solution in the form 
\be
\tilde{G}(x,y,s)=\begin{cases}
A_{-}(s)\exp\left(-\sqrt{2s}|x-y|\right)+B_{-}(s)\exp\left(\sqrt{2s}(x+y)\right)\quad, & x,y<0\\[0.1cm]
C(s)\exp\left(\sqrt{2s}y-\sqrt{2(s+V_{0})}x\right)\quad, & y<0<x\\[0.1cm]
A_{+}(s)\exp\left(-\sqrt{2(s+V_{0})}|x-y|\right)+B_{+}(s)\exp\left(-\sqrt{2(s+V_{0})}(x+y)\right)\quad, & x,y>0
\end{cases}
\ee
The unknown functions $A_\pm(s)$, $B_\pm(s)$ and $C(s)$ are determined as follows,
from \eqref{Gt}. The continuity in $x=y=0$ requires that $A_-(s) + B_-(s) = A_+(s) + B_+(s)= C(s)$,
the continuity of $\partial_x G$ at $x=0$ requires that 
$\sqrt{2 s} (A_-(s) - B_-(s))  = \sqrt{2 (s+V_0)}  C(s)$ and
$\sqrt{2 (s+V_0)} (A_+(s) - B_+(s))  = \sqrt{2 s}  C(s)$, and 
finally the jump condition for the first derivative (from the delta function),
$[\partial_x \tilde G]_{y^-}^{y^+}=-2$, gives $A_-(s) \sqrt{2 s} = 1$ and $A_+(s) \sqrt{2(s+V_0)} = 1$.
These conditions are compatible and lead to 
\bea
&& A_-(s) = \frac{1}{\sqrt{2 s}} \quad , \quad  B_-(s)=  \frac{1}{V_0 \sqrt{2 s}} (2 \sqrt{s} \sqrt{s+V_0} - V_0 - 2 s)
\eea
\bea
&& A_+(s)=
\frac{1}{\sqrt{2(s+V_0)}} \quad , \quad 
 B_+(s)= \frac{2 s + V_0 -2 \sqrt{s} \sqrt{s+V_0}}{\sqrt{2} V_0 \sqrt{s+V_0}} \quad, \quad C(s)=\frac{1}{V_0} \sqrt{2}  (\sqrt{s+V_0}-\sqrt{s}) 
\eea 

We now use \eqref{relGG} to obtain $G_\mu$ and its imaginary part. 

For $\mu'<0$, $\sqrt{s}= \sqrt{- \mu'- i 0^+}$ and $\sqrt{s+V_0}= \sqrt{- \mu'- i 0^+ + V_0}$
and one finds that $G_{\mu' <0}$ has a vanishing imaginary part.

For  $\mu'>0$ we must replace everywhere $\sqrt{s} = i \sqrt{\mu' - i 0^+}$.
As in the text we must distinguish two cases. If $\mu' < V_0$ we must replace 
$\sqrt{s+V_0} = \sqrt{V_0-\mu' + i 0^+}$, while if $\mu'>V_0$ we must 
replace $\sqrt{s+V_0} = i \sqrt{\mu' - V_0 - i 0^+}$. This leads to the following
expressions.\\

(i) For $0< \mu'<V_0$ (where, on the l.h.s. we denote $\mu' - i 0^+$ simply by $\mu'$)
\bea
&&G_{\mu'}(x,y)  %-  \frac{1}{\sqrt{2 s}} (\exp\left(- \sqrt{2} \sqrt{s} y} + \frac{\sqrt{s} - \sqrt{s+v}}{\sqrt{s} + \sqrt{s+v}} 
% \exp\left(\sqrt{2} \sqrt{s} y})   \exp\left(\sqrt{2 s} x} = 
%-  \frac{1}{\sqrt{2 s}} (\exp\left(- \sqrt{2} \sqrt{s} y} - \frac{1}{v} (v -2 \mu' - 2 \sqrt{s} \sqrt{s+v})
% \exp\left(\sqrt{2} \sqrt{s} y})   \exp\left(\sqrt{2 s} x} \nonumber  \\
%&& 
= \frac{i}{\sqrt{2 \mu'}} (\exp\left(- i \sqrt{2 \mu'} |x-y|\right) - \frac{1}{V_0} (V_0-2 \mu' - 2 i \sqrt{\mu'} \sqrt{V_0-\mu'})
 \exp\left(i \sqrt{2 \mu'} (x+y)\right))  \quad , \quad x , y < 0
\\
&& G_{\mu'}(x,y)
%=  - \frac{1}{v} \sqrt{2}  (\sqrt{s+v}-\sqrt{s}) \exp\left(\sqrt{2} \sqrt{s} y}
%\exp\left(- \sqrt{2 (s+v)} x}  
= - \frac{1}{V_0} \sqrt{2}  (\sqrt{V_0-\mu'}-i \sqrt{\mu'}) \exp\left(i \sqrt{2 \mu'} y\right)
\exp\left(- \sqrt{2 (V_0-\mu')} x\right)  \quad , \quad x > 0 > y \\
&& G_{\mu'}(x,y)
%=  -  \frac{\exp\left(-\sqrt{2} y \sqrt{s+v}} \left(v \exp\left(2 \sqrt{2} y \sqrt{s+v}}-2
%   \sqrt{s} \sqrt{s+v}+2 s+v\right)}{\sqrt{2} v \sqrt{s+v}} \exp\left(- \sqrt{2 (s+v)} x}  \\
%&& 
= -  \frac{ V_0 \exp\left(- \sqrt{2(V_0-\mu')} |x-y|\right)
+  \exp\left(-\sqrt{2(V_0-\mu')} (x+y)\right) (
-2
  i  \sqrt{\mu'} \sqrt{V_0-\mu'}+V_0-2 \mu' )}{\sqrt{2} V_0 \sqrt{V_0-\mu'}} 
  \quad , \quad x , y > 0\nonumber \\
\eea

(ii) For $\mu'>V_0$ (where, on the l.h.s. we denote $\mu' - i 0^+$ simply by $\mu'$)
\bea
&&G_{\mu'}(x,y)
%=  -  \frac{1}{\sqrt{2 s}} (\exp\left(- \sqrt{2} \sqrt{s} y} - \frac{1}{v} (v -2 \mu' - 2 \sqrt{s} \sqrt{s+v})
% \exp\left(\sqrt{2} \sqrt{s} y})   \exp\left(\sqrt{2 s} x} \\
%&& = 
 = \frac{i}{\sqrt{2 \mu'}} (\exp\left(- i \sqrt{2 \mu'}  |x-y|\right) - \frac{1}{V_0} (V_0 -2 \mu' + 2 \sqrt{\mu'} \sqrt{\mu'-v})
 \exp\left(i  \sqrt{2 \mu'} (x+y)\right))    \quad , \quad x , y < 0 \\
 && G_{\mu'}(x,y)
%=  - \frac{1}{v} \sqrt{2}  (\sqrt{s+v}-\sqrt{s}) \exp\left(\sqrt{2} \sqrt{s} y}
%\exp\left(- \sqrt{2 (s+v)} x}  
= 
- \frac{i}{V_0} \sqrt{2}  (\sqrt{\mu'-V_0}-\sqrt{\mu'}) \exp\left(i \sqrt{2 \mu'} y\right)
\exp\left(- i \sqrt{2 (\mu'-V_0)} x\right) \quad , \quad x > 0 > y \\
&& G_{\mu'}(x,y)=  i \frac{V_0 \exp\left( - i \sqrt{2} |x-y| \sqrt{\mu'-V_0}\right) + (2
   \sqrt{\mu'} \sqrt{\mu'-V_0}+V_0-2 \mu') \exp\left(-i  \sqrt{2(\mu'-V_0)} (x+y)\right)}{\sqrt{2} V_0  \sqrt{\mu'-V_0}} 
      \quad , \quad x, y > 0\nonumber \\
\eea
together with the formula obtained by the symmetry in $x,y$. Taking the imaginary parts leads to the formula for 
${\rm Im}\ G_{\mu'}(x,y)$ given in the text \eqref{vbigleft},\eqref{vbigright},\eqref{g2},\eqref{e1},\eqref{e2},\eqref{e3}.

\section{Evaluating the integral in Eq.~\eqref{eq:var_NR_subcritical}}\label{integral}

Changing variables
$$
u=r\left(1-\tilde{u}\right),\quad v=r\left(1-\tilde{v}\right)
$$ 
in the integral that appears in Eq.~\eqref{eq:var_NR_subcritical}, we obtain
\be
I(r) \equiv \frac{1}{\pi^{2}r^{2}}\int_{0}^{1}dv\int_{0}^{1}du\frac{\sqrt{vu}}{\left(\sqrt{r-u}+\sqrt{r-v}\right)^{2}}=\int_{1-1/r}^{1}d\tilde{v}\int_{1-1/r}^{1}d\tilde{u}f\left(\tilde{u},\tilde{v}\right),\quad f\left(\tilde{u},\tilde{v}\right)\equiv\frac{\sqrt{1-\tilde{v}}\sqrt{1-\tilde{u}}}{\pi^{2}\left(\sqrt{\tilde{u}}+\sqrt{\tilde{v}}\right)^{2}}.
\ee
Now using 
$\partial_{\tilde{u}}\partial_{\tilde{v}}\mathcal{F}\left(\tilde{u},\tilde{v}\right)=f\left(\tilde{u},\tilde{v}\right)$ 
where 
\bea
&&\mathcal{F}\left(\tilde{u},\tilde{v}\right)=-\frac{1}{2\pi ^ 2}\left\{ 2\text{arccos}\left(\sqrt{\tilde{u}}\right)\text{arcsin}\left(\sqrt{\tilde{v}}\right)-4\tilde{u}\left(1-\tilde{u}\right)\log\left(\frac{\sqrt{\tilde{u}}+\sqrt{\tilde{v}}}{\sqrt{\left(1-\tilde{u}\right)\left(1-\tilde{v}\right)}+\sqrt{\tilde{u}\tilde{v}}+1}\right)\right. \nn\\
&& +2\sqrt{\tilde{v}\left(1-\tilde{v}\right)}\left(1-2\tilde{v}\right)\text{arcsin}\left(\sqrt{\tilde{u}}\right)+2\sqrt{\tilde{u}\left(1-\tilde{u}\right)}\left(1-2\tilde{u}\right)\text{arcsin}\left(\sqrt{\tilde{v}}\right)+2\sqrt{\left(1-\tilde{u}\right)\left(1-\tilde{v}\right)}\left[2-2\left(\tilde{u}+\tilde{v}\right)+\sqrt{\tilde{u}\tilde{v}}\right] \nn\\
&& \left.+4\tilde{v}\left(1-\tilde{v}\right)\left[\log\left(1-\sqrt{\left(1-\tilde{u}\right)\left(1-\tilde{v}\right)}+\sqrt{\tilde{u}\tilde{v}}\right)+2\log\left(1+\sqrt{\left(1-\tilde{u}\right)\left(1-\tilde{v}\right)}+\sqrt{\tilde{u}\tilde{v}}\right)-3\log\left(\sqrt{\tilde{u}}+\sqrt{\tilde{v}}\right)\right]\right\} .
\eea
we obtain
\be
I\left(r\right)=F\left(\frac{r-1}{r}\right),\quad F\left(A\right)=\int_{A}^{1}d\tilde{v}\int_{A}^{1}d\tilde{u}f\left(\tilde{u},\tilde{v}\right)=\mathcal{F}\left(1,1\right)-\mathcal{F}\left(1,A\right)-\mathcal{F}\left(A,1\right)+\mathcal{F}\left(A,A\right),
\ee
which indeed yields the expression for $F\left(A\right)$ which is given in Eq.~\eqref{fint}.

%%Naftalis appendix - end

{
\section{Third cumulant}

\label{appendixThirdCumulant}

Here we calculate the third cumulant $\left\langle N_{R}^{3}\right\rangle _{c}$ in the critical case.
We begin from the general formula Eq.~(D.6) of~\cite{KrajenbrinkPLD2018} for (minus) the third cumulant of the distribution of linear statistics $A=\sum_{i=1}^{N}\varphi\left(x_{i}\right)$ for any function $\varphi$ and for a general determinantal point process with kernel $K(x,y)$, which we give here (up for the convenience of the reader
\be
\left\langle A^{3}\right\rangle _{c}=\text{Tr}\left(\varphi^{3}K\right)-3\text{Tr}\left(\varphi K\varphi^{2}K\right)+2\text{Tr}\left(\varphi K\varphi K\varphi K\right).
\ee
The particular case of counting statistics $\mathcal{N}_{\mathcal{I}}$ corresponds to an indicator function $\varphi=\chi_{\mathcal{I}}$, and then the formula simplifies to
\bea
\left\langle \mathcal{N}_{\mathcal{I}}^{3}\right\rangle _{c}	&=&\int_{\mathcal{I}}K\left(x,x\right)dx-3\int_{\mathcal{I}}\int_{\mathcal{I}}K\left(x,y\right)^{2}dxdy+2\int_{\mathcal{I}}\int_{\mathcal{I}}\int_{\mathcal{I}}K\left(x,y\right)K\left(y,z\right)K\left(z,x\right)dxdydz\nn\\
&=&-2\left\langle \mathcal{N}_{\mathcal{I}}\right\rangle +3\text{Var}\left(\mathcal{N}_{\mathcal{I}}\right)+2\int_{\mathcal{I}}\int_{\mathcal{I}}\int_{\mathcal{I}}K\left(x,y\right)K\left(y,z\right)K\left(z,x\right)dxdydz,
\eea
where we used Eq.~\eqref{vs} in the last equality.
Let us apply this to the particular case of the square barrier potential in the critical case, for the number of particles $N_{R}$ to the right of the barrier. Using the results \eqref{avenout} and \eqref{varnout} for the mean and variance (respectively) of $N_{R}$, and plugging in $K\left(x,y\right)=\frac{1}{\ell}\kappa_{c}\left(\frac{x}{\ell},\frac{y}{\ell}\right)$, we obtain $\left\langle N_{R}^{3}\right\rangle _{c}=-\frac{1}{2}+\frac{6}{\pi^{2}}+2\mathfrak{I}$ where $\mathfrak{I}=\int_{0}^{\infty}\int_{0}^{\infty}\int_{0}^{\infty}\kappa_{c}\left(a,b\right)\kappa_{c}\left(b,c\right)\kappa_{c}\left(c,a\right)da\,db\,dc$.
Finally, using the expression \eqref{rep2} for $\kappa_c$, we find
\bea
\mathfrak{I}&=&\int_{0}^{\infty}\!\!da\int_{0}^{\infty}\!\!db\int_{0}^{\infty}\!\!dc\int_{0}^{1}\!\!dv\int_{0}^{1}\!\!du\int_{0}^{1}\!\!dw\sqrt{\left(1-v\right)\left(1-u\right)\left(1-w\right)} \exp\left\{ -\pi\left[\sqrt{v}\,\left(a+b\right)+\sqrt{u}\,\left(b+c\right)+\sqrt{w}\,\left(a+c\right)\right]\right\} \nn\\
&=& \int_{0}^{1}dv\int_{0}^{1}du\int_{0}^{1}dw\frac{\sqrt{\left(1-v\right)\left(1-u\right)\left(1-w\right)}}{\pi^{3}\left(\sqrt{v}+\sqrt{w}\right)\left(\sqrt{v}+\sqrt{u}\right)\left(\sqrt{u}+\sqrt{w}\right)}=0.01009491\dots
\eea
where the last equality was obtained via a numerical integration. Altogether this yields the result for $\left\langle N_{R}^{3}\right\rangle _{c}$ given in the main text below Eq.~\eqref{varnout}.
}

{\section{An identity between scattering coefficients}\label{identityap}
In section \ref{gfg},  for $\mu<V_0$, we obtained two solutions of the Schr\"odinger equation denoted by $\psi_{k_1}^{(1)}$ and  
$\psi_{k_1}^{(2)}$ by analytic continuation of the solutions $\psi_{k_1}$ and $\psi_{k_1}^*$ respectively. The physical scattering solution is $\psi_{k_1}^{(2)}$ which decays exponentially as $x\to\infty$. Another way to obtain a decaying solution is to start with the solution $\psi_{k_1}^{(1)}$ but  choose the branch of the square root of $\kappa_2^2$ with the opposite sign. That is we take the solution $\psi_{k_1}^{(1)}$ with the substitution
$\kappa_2\to-\kappa_2$. This solution which we will call $\psi_{k_1}^{(3)}$ then has the asymptotic behavior
\bea 
\psi^{(3)}_{k_1}(x) = 
\begin{cases}
&\exp\left(i k_1 x\right)+ \frac{C_2(k_1,i\kappa_2)}{C_1(k_1,i\kappa_2)} \exp\left(- ik_1x\right) \;, \;  x \to - \infty \\[0.2cm]
%& \\
& \frac{1}{C_1(k_1,i\kappa_2)} \exp\left(-\kappa_2 x\right) \;, \;\hspace*{1.1cm} x \to +\infty  \;.
\end{cases}.
\eea
However as there are only two linearly independent solutions this solution must be proportional to $\psi^{(2)}_{k_1}(x)$. Hence there exists $B$ such that $\psi^{(3)}_{k_1}(x)=B\psi^{(2)}_{k_1}(x)$. Identifying
all the asymptotic amplitudes one finds that 
\be \label{BB} 
B=\frac{C_{1}^{*}(k_{1},-i\kappa_{2})}{C_{1}(k_{1},i\kappa_{2})}=\frac{C_{2}(k_{1},i\kappa_{2})}{C_{1}(k_{1},i\kappa_{2})}=\frac{C_{1}^{*}(k_{1},-i\kappa_{2})}{C_{2}^{*}(k_{1},-i\kappa_{2})}\equiv\left(\frac{C_{1}(k_{1},i\kappa_{2})}{C_{2}(k_{1},i\kappa_{2})}\right)^{*}
\ee 
The last two identities show that the following ratio has modulus unity
\be
\left|\frac{C_{2}(k_{1},i\kappa_{2})}{C_{1}(k_{1},i\kappa_{2})}\right|=\left|\frac{C_{2}^{*}(k_{1},-i\kappa_{2})}{C_{1}^{*}(k_{1},-i\kappa_{2})}\right|=1
\ee 
which simply expresses that the wave in \eqref{psi2} is totally reflected. In addition, from the first 
identity in \eqref{BB} we obtain
\begin{equation}
C_1^*(k_1,- i\kappa_2) =  C_2(k_1,i\kappa_2).
\end{equation}
Constructing a solution in a similar fashion from $\psi_{k_1}^{(1)}$ also yields
$C_1^*(k_1,i\kappa_2) =  C_2(k_1,- i\kappa_2)$. 

%The same identity shows that the reflection coefficient obeys
%\begin{equation}
%\frac{C_2(k,-i\kappa_2)}{C_1(k_1,-i\kappa_2)} = \frac{C_2(k,-i\kappa_2)}{\overline C_2(k_1,-i\kappa_2)} ,
%\end{equation}
%and thus has modulus $1$ which implies total reflection. 

Now, using these relations in the Wronskian identity Eq. (\ref{w2}) we obtain Eq.(\ref{idformula}). For instance, eliminating all $C_j^*$ functions, one can 
check that both relations are equivalent to 
\be
C_1(k_1,-i \kappa_2) C_2(k_1,i \kappa_2) - C_1(k_1,i \kappa_2) C_2(k_1,-i \kappa_2) = - i \frac{\kappa_2}{k_1}
\ee 
}
These identities may be verified for the square shoulder potential and the Woods-Saxon potential.

\section{Kernel for the step potential via the summation over the eigenstates}\label{App:direct}

In this Appendix we show how to compute the kernel for the step potential
by a direct summation over the eigenfunctions. We will restrict to the simpler case $\mu \leq V_0$,
for which the kernel for $x,y>0$ was obtained in Eqs.~~\eqref{eq:kernel_scaling}, (\ref{rep2}) and \eqref{kappar},
but the method can be extended to $\mu >V_0$. Let us recall the definition of the kernel 
\bea \label{def_kernel_app}
K_\mu(x,y) = \sum_{k_1} \theta(\mu - \epsilon_{k_1}) \psi_{k_1}^*(x) \psi_{k_1}(y) 
\eea
Here $\psi_{k_1}(x)$ denote the eigenfunctions of the 
Schr\"odinger equation for the step potential $V(x) = V_0 \,\theta(x)$
\bea \label{Schrod_step}
- \frac{1}{2} \frac{\partial^2}{\partial x^2}\psi_{k_1}(x) + V_0 \, \theta(x) \psi_{k_1}(x) = \epsilon_{k_1} \psi_{k_1}(x) \;,
\eea
with eigenvalues $\epsilon_{k_1}=k_1^2/2$. The eigenfunctions are superpositions of
incoming and outgoing plane waves with different coefficients on both
sides of the step (parametrized generically by four amplitudes). For $\mu \leq V_0$, we only need the eigenfunctions for $\epsilon_{k_1} \leq V_0$, in which case there is no incoming wave from the right, and there are only three non zero amplitudes. 
In this simpler situation it can be treated as a problem of reflection by a barrier of an incident wave coming from the left, for which one can write (from standard
textbook see e.g. \cite{Landau})
%The eigenfunctions are well known from standard textbooks of quantum mechanics and read
\bea \label{psi_step}
\psi_{k_1}(x) = \frac{1}{\sqrt{2 \pi}} 
\begin{cases}
&\exp\left(i {k_1} x\right)+ \frac{k_1-k_2}{k_1+k_2} \exp\left(-ik_1x\right) \;, \; x \leq 0 \\[0.2cm]
%& \\
& \frac{2k_1}{k_1+k_2} \exp\left(i k_2x\right) \;, \;\hspace*{1.1cm} x \geq 0  \;,
\end{cases}
\eea
where the incoming wave vector is $k_1$ and the outgoing wave vector is $k_2 = \sqrt{k_1^2 - 2 V_0}$. In the notation of section \ref{gfg} the scattering coefficients are thus given by 

\begin{equation} \label{G4} 
C_1(k_1,k_2) = \frac{k_1+k_2}{2k_1}, \ \ C_2(k_1,k_2) = \frac{k_1-k_2}{2k_1},
\end{equation}
In this scattering problem, the incident wave is thus reflected with a reflection coefficient
$R(k_1,k_2)=|\frac{k_1-k_2}{k_1+k_2}|^2$ and transmitted with a transmission coefficient $T(k_1,k_2) = 1-R(k_1,k_2)=\frac{k_2}{k_1}|\frac{2k_1}{k_1+k_2}|^2$. Note that the highest allowed value of $k_1$ is $k_F = \sqrt{2 \mu}$. For $\mu \leq V_0$ one 
thus needs only to consider $k_1 \leq  \sqrt{2 V_0}$, and 
$k_2= i \sqrt{2V_0-k_1^2}$ is purely imaginary. This means that the corresponding eigenfunction in (\ref{psi_step}) in the region $x>0$ is exponentially damped
and the reflexion coefficient is $R=1$. As we mentioned above, the solution corresponding to $-k_1$ with the same energy $\epsilon_{k_1} = k_1^2/2$ is not physically allowed since there is no 
incident plane wave from the right, for $V_0 \geq \mu$. 
Therefore the allowed range of $k_1$ is $k_1 \in [0,k_{1F}]$ where $k_{1F}=\sqrt{2\mu}$. 

Hence, to compute the kernel $K_\mu(x,y)$, we just %have to 
substitute this expression for the eigenfunction (\ref{psi_step}) in Eq. (\ref{def_kernel_app}), and, in the limit of an infinite system with
the continuum normalization chosen in \eqref{psi_step}, replace $\sum_{k_1} \to \int_0^{k_F} \frac{dk_1}{2 \pi}$
(see the remark below). 
For simplicity, we just give the expression of the kernel when both $x,y>0$. In this case, $\psi_{k_1}(x)$ is simply given by he second line of Eq. (\ref{psi_step}). Consequently the kernel reads, for $x,y>0$
and any $\mu \leq V_0$
\bea \label{crit_kernel_app1}
K_{\mu}(x,y) &=& \frac{1}{2\pi} \int_0^{k_F} dk_1 \, \frac{4k_1^2}{|k_1+ i \sqrt{2 V_0-k_1^2}|^2} \exp\left(- \sqrt{2V_0-k_1^2}(x+y)\right) \nonumber \\
&=& \frac{1}{\pi V_0} \int_0^{\sqrt{2 \mu}} dk_1\, k_1^2 \,\exp\left(- \sqrt{(2V_0-k_1^2)/(2\mu)} \pi (a+b)\right)
 = \frac{1}{\ell} \kappa_r(a,b) \quad , \quad r=V_0/\mu \geq 1
\eea
and the last equality shows that it agrees with $\kappa_r(a,b)$ is given in \eqref{kappar}. We have 
rescaled $a = x/\ell$ and $b = y/\ell$ with $\ell = \pi/\sqrt{2 \mu}$ as in Eq.~\eqref{eq:kernel_scaling}  %(\ref{critscaling})
 in the main text, and performed the change of variable $k_1 = \sqrt{2 \mu v}$. Alternatively, for
 $\mu=V_0$, performing the change of variable $k_1=\sqrt{2 V_0} \sqrt{1-v}$ one obtains the critical kernel 
 as %directly in the form
\bea \label{crit_kernel_app2}
K_c(x,y) = \frac{1}{\ell} \kappa_c(a,b) \;, \;\;\; {\rm where} \;\;\; \kappa_c(a,b) = \int_0^1 dv \, \sqrt{1-v} \, \exp\left(-\sqrt{v} \pi (a+b)\right) \;,
\eea  
Hence the results for $\mu = V_0$ and for $\mu < V_0$ coincide with the formula 
obtained via the Green's function derivation given in Eqs.~\eqref{eq:kernel_scaling} %(\ref{critscaling})
 and (\ref{rep2}), \eqref{kappar}.
 
The kernel in the others domains of $x,y$ can be obtained in the same way for $\mu \leq V_0$.
The case $\mu>V_0$ is slightly more involved as it involves four amplitudes, but
can be done similarly. 
%Note that this direct derivation of the kernel may look simpler than the Green's function derivation. 

{\bf Remark}.
To establish more carefully the normalization we place an infinite wall at $x=-L$ so that $\psi_{k_1}(-L)=0$,
and consider the limit $L \to +\infty$. For $\epsilon_{k_1} < V_0$ this is sufficient since the wave-function decays exponentially for $x>0$. In this case the wave functions can be written (with a 
normalization different to \eqref{psi_step})
$\psi_{k_1}(x)=A \psi_{k_1}^0(x)$ with $\psi_{k_1}^0(x)=(\exp\left(i {k_1} x\right) + \alpha \exp\left(-i k_1 x\right))$ for $x<0$, with $\alpha=\frac{k_1 - i \sqrt{2 V_0-k_1^2}}{k_1 - i \sqrt{2 V_0-k_1^2}}$ with $A$ a real normalization amplitude. 
One can then perform the integral of the norm on the negative axis, and one finds, after inserting
in the expression the boundary condition $\exp\left(2 i k_1 L\right) = -1/\alpha$, that it simplifies into
$\int_{-L}^0 |\psi_{k_1}(x)|^2 = A^2 ( L (1 + |\alpha|^2) + O(1) ) = A^2 (2 L + O(1))$ 
since $|\alpha|=1$ for the case
considered. Since the normalization integral for $x>0$ is $O(1)$ this gives $A=1/\sqrt{2 L}$ for large $L$. On the other hand the boundary condition
shows that the quantized values of $k_1$ are $k_1= \frac{\pi n}{L} + c/L$, with $n$ positive
integer. Putting this together we 
see that at large $L$ one has
\be
K_\mu(x,y) = \sum_{k_1} \theta(\mu - \epsilon_{k_1}) \psi_{k_1}^*(x) \psi_{k_1}(y) \simeq \frac{L}{\pi} \int_0^{k_{1F}}  dk_1 \frac{1}{2 L} 
\psi^0_{k_1}(x)^* \psi^0_{k_1}(y) = \int_0^{k_{1F}} \frac{dk_1}{2 \pi} 
\psi^0_{k_1}(x)^* \psi^0_{k_1}(y) 
\ee
which justifies the formula given above. If one prefers continuum normalizations, it is also
possible to check (using regulators) that, since $|\alpha|=1$, the eigenfunctions given in
\eqref{psi_step} satisfy $\int_{-\infty}^{+\infty} dx \psi_{k_1}(x) \psi^*_{k_1'}(x)= (2 \pi) (\delta(k_1-k_1')+\delta(k_1+k_1')) + O(1) = (2 \pi) \delta(k_1-k_1')$ since $k_1,k_1'>0$, and to argue that
it leads to the same conclusion. 

In conclusion we see that in this direct summation method one needs to 
be careful in choosing the eigenfunctions that contribute to the sum in the kernel in Eq. (\ref{crit_kernel_app1}) and in determining their proper normalization so that
the discrete sum over states can be given a meaning as an integral in the continuum. 
Interestingly, in the Green's function approach, this is automatically taken 
care of by imposing appropriate boundary conditions at $x,y \to \pm \infty$.
%and one does not need any selection principle for the right set of eigenfunctions.

\section{The smooth step}
\label{app:smooth}

Here we give some more details on the smooth step barrier potential 
$V(x) = \frac{V_0}{1 + \exp(- x/\lambda) }$ studied in Section \ref{smooth}. Following \cite{Landau} one looks for solutions of the eigenfunction equation $(-\frac{1}{2} \frac{\partial^2}{\partial x^2} + V(x)) \psi(x) = \epsilon \psi(x)$ 
of the form 
\be \label{exactpsi} 
\psi(x)=\psi_{k_1}(x) = A \exp(i k_2 x) f\left(- \exp(- \frac{x}{\lambda})\right )  \quad , \quad \epsilon = \frac{k_1^2}{2} = V_0 + \frac{k_2^2}{2} 
\ee 
One finds that $f(z)$ must obey the hypergeometric equation $z(1-z) f''(z) + (1-z) (1- 2 i \lambda k_2) f'(z) + \lambda^2( k_2^2 - k_1^2) f(z) =0$. The solution
\be \label{fz} 
f(z)={}_{2}F_{1}\left(i\lambda(k_{1}-k_{2}),-i\lambda(k_{1}+k_{2}),1-2i\lambda k_{2},z\right)
\ee 
is such that, upon choosing $A=\frac{1}{C_1(k_1,k_2)}$ one can explicitly verify that the eigenfunction has the following asymptotics
(using the asymptotics of $f(z)$ for $z \to - \infty$ and $f(0)=1$)
\bea \label{psi_step22}
\psi_{k_1}(x) = 
\begin{cases}
&\exp\left(i k_1 x\right)+ \frac{C_2(k_1,k_2)}{C_1(k_1,k_2)} \exp\left(-ik_1x\right) \;, \;  x \to - \infty \\[0.2cm]
%& \\
& \frac{1}{C_1(k_1,k_2)} \exp\left(i k_2 x\right) \;, \;\hspace*{1.1cm} x \to +\infty  \;,
\end{cases}
\eea
with scattering coefficients
\be \label{scattWood} 
C_1(k_1,k_2) = \frac{ \Gamma(-2 i \lambda k_1) \Gamma(1- 2 i \lambda k_2) }{\Gamma(- i \lambda (k_1+k_2)) \Gamma(1- i \lambda (k_1+k_2))}
\quad , \quad 
C_2(k_1,k_2) = \frac{ \Gamma(2 i \lambda k_1) \Gamma(1- 2 i \lambda k_2) }{\Gamma(i \lambda (k_1-k_2)) \Gamma(1+ i \lambda (k_1-k_2))}
\ee
Here we take  $k_1>0$, corresponding to a plane wave coming from the left, with a
reflection amplitude $\frac{C_2(k_1,k_2)}{C_1(k_1,k_2)}$ and a transmitted amplitude $\frac{1}{C_1(k_1,k_2)}$. 

Although \eqref{psi_step22} is an eigenfunction for any $\epsilon$, we focus here on the case $\epsilon< V_0$. In that case $k_2= i \kappa_2$ with $\kappa_2=\sqrt{2 V_0-k_1^2}$ and the
eigenfunction decays exponentially on the right. The reflection amplitude has modulus one
\be
\beta= \beta_{k_1,\kappa_2} = \frac{C_2(k_1,i\kappa_2)}{C_1(k_1,i\kappa_2)}  = \frac{ \Gamma(2 i \lambda k_1) \Gamma(- i \lambda k_1+ \lambda \kappa_2 ) \Gamma(1- i \lambda k_1+ \lambda \kappa_2)  }{\Gamma(-2 i \lambda k_1) \Gamma(i \lambda k_1 + \lambda \kappa_2) \Gamma(1+ i \lambda k_1 + \lambda \kappa_2)}
\ee 
To obtain the kernel we now need to perform the summation over the eigenstates, with a proper
normalization and counting of the states, which is possible using the asymptotics \eqref{psi_step22}
(with $k_2=i \kappa_2$). The argument given in the remark at the end of the Appendix \ref{App:direct} (using
a hard wall at $x=-L$ for $L \to +\infty$, extends to this case,
since the wave function is also totally reflected for $\epsilon < V_0$. Indeed the normalization integrals
differ from those of the square barrier only up to distances $|x| = O(\lambda)$,
hence their $O(L)$ amplitude is not affected. In the $L \to +\infty$ limit the kernel, for $\mu \leq V_0$
thus reads
\be \label{KMU} 
K_\mu(x,y) = \int_0^{\sqrt{2 \mu}} \frac{dk_1}{2 \pi} \psi_{k_1}^*(x) \psi_{k_1}(y) 
\ee
where the $\psi_{k_1}(x)$ are given in \eqref{exactpsi} with $k_2=i \kappa_2$, $A= 1/C_1(k_1,i \kappa_2)$ 
and $f(z)$ given in \eqref{fz}. This leads to the exact formula \eqref{kernelsmooth} given in the text for
where we denoted $B(\lambda k_1, \lambda \kappa_2) 
= 1/|C_1(k_1,i \kappa_2)|^2$. The formula \eqref{left} in the text is then obtained by inserting the asymptotics 
for $x,y<0$ in \eqref{psi_step22} into \eqref{KMU}.

{}


\begin{thebibliography}{}
\bibitem{com1}It is customary in \eqref{k1} to choose $\theta(0)=1$, so that $\mu$ can be chosen to
equal $\mu=\epsilon_N$, but this choice is not important.
\bibitem{comreal} 
The eigenfunctions $\psi_{k}\left(x\right)$ can be chosen to be real. This fact follows from the fact that if $\psi_{k}\left(x\right)$ is an eigenfunction then so are its real and imaginary parts, $\text{Re}\, \psi_{k}(x)$ and $\text{Im}\,\psi_{k} (x)$ respectively.
This means that the kernel is real and symmetric.
\end{thebibliography}
\end{document}